\documentclass[aps,reprint,superscriptaddress,twocolumn,longbibliography,prx,footnoteinbib]{revtex4-1}
\usepackage{amsmath,amssymb,amsfonts}
\usepackage{graphicx}
\usepackage{bm}
\usepackage{dsfont}
\usepackage{color}
\usepackage[colorlinks=True,linkcolor=red,citecolor=blue,urlcolor=blue]{hyperref}
\usepackage{hypcap}
\usepackage{comment}
\usepackage{ulem}

\newcommand{\mkadd}[1]{{\color{blue} #1}}

\newcommand{\vk}{\mathbf{k}}

\newcommand{\mcl}[1]{\mathcal{#1}}

\def\be{\begin{equation}}
\def\ee{\end{equation}}
\def\bea{\begin{eqnarray}}
\def\eea{\end{eqnarray}}

\newcommand{\bB}{{\bf B}}
\newcommand{\bE}{{\bf E}}

\begin{document}

\title{Tunable axial gauge fields in engineered Weyl semimetals: \\ Semiclassical analysis and optical lattice implementations}

\author{Sthitadhi Roy}
\affiliation{Max-Planck-Institut f{\"u}r Physik komplexer Systeme, N{\"o}thnitzer Stra{\ss}e 38, 01187 Dresden, Germany}
\author{Michael Kolodrubetz}
\affiliation{Materials Sciences Division, Lawrence Berkeley National Laboratory, Berkeley, CA 94720, USA}
\affiliation{Department of Physics, University of California, Berkeley, California 94720, USA}

\author{Nathan Goldman}
\affiliation{Center for Nonlinear Phenomena and Complex Systems,
Universit\'e Libre de Bruxelles, CP 231, Campus Plaine, B-1050 Brussels, Belgium}

\author{Adolfo G. Grushin}
\affiliation{Department of Physics, University of California, Berkeley, California 94720, USA}
\affiliation{Institut N\'eel, CNRS and UniversitŽ Grenoble Alpes, F-38042 Grenoble, France}

\begin{abstract}
In this work, we describe a toolbox to realize and probe synthetic axial gauge fields in engineered Weyl semimetals. These synthetic electromagnetic fields, which are sensitive to the chirality associated with Weyl nodes, emerge due to spatially and temporally dependent shifts of the corresponding Weyl momenta.  First, we introduce two realistic models, inspired by recent cold-atom developments, which are particularly suitable for the exploration of these synthetic axial gauge fields. Second, we describe how to realize and measure the effects of such axial fields through center-of-mass observables, based on semiclassical equations of motion and exact numerical simulations. In particular, we suggest realistic protocols to reveal an axial Hall response due to the axial electric field $\bE_5$, and also, the axial cyclotron orbits and chiral pseudo-magnetic effect due to the axial magnetic field $\bB_5$.
\end{abstract}
\maketitle

%------------------
%-- Introduction --
%------------------

\section{Introduction}

Gauge theories---theories that are invariant under a continuous group of local transformations---
span various subjects of modern physics. 
Artificial generation of gauge fields, such as those emanating from electromagnetism, can be realized and probed not only in condensed matter~\cite{PGL13}, but also in engineered lattice systems, such as cold atoms in optical lattices~\cite{GBZ16,dalibard2011colloquium,goldman2014light,aidelsburger2017artificial} and photonic crystals~\cite{Lu2014topological,hafezi2014synthetic}. 
A recent example in solid state physics was the realization and successful manipulation of artificial gauge fields in Graphene~\cite{amorim2016}.
In this case, the effective gauge fields emerge from a strain field that couples to the Dirac quasiparticles in this material.
Spatial derivatives of the strain field define a gauge field that, unlike the electromagnetic gauge field,
couples with opposite signs to each valley (Dirac node). 
In parallel, synthetic systems have also been successful in emulating 
and controlling effective electromagnetic gauge fields~\cite{dalibard2011colloquium,goldman2014light,GBZ16,Lu2014topological,hafezi2014synthetic}.
Examples include the recent optical lattice realization and characterization of the Hofstadter and Haldane models~\cite{aidelsburger2013realization,miyake2013realizing,aidelsburger2015measuring,KBCK2015,jotzu2014experimental,wu2016realization,flaschner2016observation,tai2017microscopy}, which open the possibility of 
probing exotic topological states in ways that are challenging to realize in condensed matter, such as monitoring ``heating"~\cite{TDG17}. Interestingly, effective magnetic fields could also be engineered by combining strain methods and optical lattice technologies, as was recently proposed in Ref.~\cite{Tian2015Landau}.

%---------
%Weyl semimetals

Engineered gauge fields not only provide an intriguing and promising avenue towards the control of electronic properties of two-dimensional materials~\cite{PGL13},
they also lie at the core of the recently discovered three-dimensional Weyl semimetals~\cite{CFL15,Pikulin2016,grushin2016inhomogeneous,Gorbar2017,Gorbar2017a,Gorbar2017b,Gorbar2017c,Gorbar2017d,Liu2017,Gorbar2017e,Gorbar2017f,Huang2017,Guan2017}. 
The band structure of Weyl semimetals hosts a set of band-touching points around which quasiparticles disperse as massless Weyl fermions~\cite{V03,Turner:2013tf,hosur2013recent,AMV17}.
This description assigns quasiparticles a chirality, a quantum number that reflects the parallel or anti-parallel orientation of the spin with respect to the momentum of massless particles.
Weyl fermions appear in the Brillouin zone in pairs of opposite chirality~\cite{NielNino81a,NielNino81b,NielNino81c,NielNino83}; 
they are separated in energy-momentum space by a four-vector $b_{\mu}=(b_0,\mathbf{b})$.
Unlike in Graphene, the gapless touching points, known as the Weyl nodes, do not necessarily coincide with a high-symmetry point of reciprocal space in the absence of strain.
However, similar to Graphene, it was recently realized~\cite{CFL15} that strain can promote $b_{\mu}$ to a local vector field that couples 
with opposite signs to opposite chiralities, termed the chiral or {\it axial} gauge field.
This identification enables one to define an axial magnetic field $\mathbf{B}_5 = \bm{\nabla}_\mathbf{r}\times \mathbf{b}(\mathbf{r},t)$ 
and an axial electric field $\mathbf{E}_5=-\bm{\nabla}_\mathbf{r}b_0(\mathbf{r},t) -\partial_t \mathbf{b}(\mathbf{r},t)$ in analogy with the usual electromagnetic fields.
%

%---------
% Response and previous predictions

Several groups have used this powerful analogy to predict novel phenomena and provide a different perspective on the properties of Weyl semimetals~\cite{CCG14,Jiang2015,Yang2015,Pikulin2016,grushin2016inhomogeneous,Schuster2016,cortijo2016strain,cortijo2016emergent,Gorbar2017,Gorbar2017b,Gorbar2017c,Gorbar2017d,Gorbar2017d,Liu2017,Huang2017} and Helium-3~\cite{bevan1997momentum,klinkhamer2005emergent,V03}.
Analogous to the positive magneto-conductivity proportional to magnetic field $\mathbf{B}$ that emerges from the chiral anomaly~\cite{son2013chiral}, 
a strain field that creates a constant $\mathbf{B}_5$ is predicted to result in a strain enhanced conductivity\cite{Pikulin2016,grushin2016inhomogeneous}.
Additionally, a strain induced $\mathbf{B}_5$ can lead to pseudo-magnetic oscillations~\cite{Liu2017}. 
Moreover, the surface states of Weyl semimetals can be reinterpreted as the zeroth pseudo-Landau levels due to a $\mathbf{B}_5$ field 
localised at the boundary~\cite{grushin2016inhomogeneous,Tchoumakov2017b}. 
This is a fruitful re-interpretation of the origin of the surface states: it is a natural framework 
to treat smooth interfaces between topological states ~\cite{Grushin2015,grushin2016inhomogeneous,Lau2017,Inhofer2017,Tchoumakov2017}, as well as edge states in the strained Haldane model~\cite{Ho2017}.
Finally, through this identification, the contribution of axial gauge fields to 3+1 dimensional anomalies known in the high-energy literature~\cite{bertlmann2000anomalies}
could be in principle explored in full~\cite{CCG14,Land14,Pikulin2016,grushin2016inhomogeneous}.
Of particular interest is the difference between covariant and consistent versions of the anomaly~(see Ref.~\cite{Landsteiner2016} and references therein for a review) 
which have been shown to matter even at the kinetic theory level~\cite{Gorbar2017c}.
%

%---------
%This work:

Despite the intriguing predictions and exciting prospects above, the realization of controllable and sizeable axial gauge fields is challenging in condensed matter.
Although realistic proposals exist, they rely on interface strain effects, defects~\cite{grushin2016inhomogeneous} or bulk strain patterns that are so far difficult to engineer at will~\cite{Pikulin2016}.

In this work, we investigate a host of phenomena that are made possible by axial gauge fields in engineered Weyl semimetals. 
We specifically focus on ultra-cold atomic realizations, where we show that simple tuning of the lattice enables arbitrary control over the axial fields of interest. 
We argue that these axial fields are more readily controllable over a larger dynamic range than their condensed matter counterpart, 
which combined with the pristine nature of optical lattices for ultra-cold atoms suggests these systems as ideal platforms for studying Weyl physics. 
Furthermore, while it is not straightforward to perform conventional transport experiments in cold-atom experiments (see Ref.~\cite{BrantutReview} for a review), these synthetic systems are conducive to dynamical (in-situ) density measurements. 
In particular, wavepacket dynamics often serve as ideal probes for extracting electromagnetic responses~\cite{price2012mapping,dauphin2013extracting,duca2014aharonov,aidelsburger2015measuring,price2016measurement}. 
Therefore, using a semiclassical wavepacket formalism, 
we show how various responses to engineered gauge and axial gauge potentials 
give rise to measurable quantities in realistic experimental settings that are unique to axial electromagnetic fields.
Our work complements previous Kubo-based approaches on lattice models~\cite{Gorbar2017,Gorbar2017a,Liu2017} that compute transport properties arising due to the presence of axial fields.

In section~\ref{sec:models} we set the stage by introducing notation and discussing two realistic models that
allow to control axial-gauge fields in realistic setups.
In section~\ref{sec:semiclassics} we discuss signatures of the axial fields that can be probed in experiment monitoring
wavepacket dynamics.
Finally in section~\ref{sec:conclusions} we discuss our results and present some concluding remarks.

%%%%%%%%%%%%%%%%%%%%%%%%%%%%%%%%%%%%%%%%%%%%%%%%%%%%%%%%%%%%%%%%%%%%%%%%%%%%
%%%%%%%%%%%%%%%%%%%%%%%%%%%%%%%%%%%%%%%%%%%%%%%%%%%%%%%%%%%%%%%%%%%%%%%%%%%%
\section{Models and axial gauge fields \label{sec:models}}
%%%%%%%%%%%%%%%%%%%%%%%%%%%%%%%%%%%%%%%%%%%%%%%%%%%%%%%%%%%%%%%%%%%%%%%%%%%%
%%%%%%%%%%%%%%%%%%%%%%%%%%%%%%%%%%%%%%%%%%%%%%%%%%%%%%%%%%%%%%%%%%%%%%%%%%%%

In this section, we discuss the basic ingredients that are needed for the realization of axial gauge fields in (engineered) Weyl semimetals. Based on two experimentally-relevant lattice models, we describe how such axial gauge fields naturally emerge as one modulates the model parameters in space or in time; this analysis also highlights the tunability of these fields under realistic conditions. Finally, we discuss optical lattice realizations of these two toy models.

%%%%%%%%%%%%%%%%%%%%%%%%%%%%%%%%%%%%%%%%%%%%%%%%%%%%%%%%%%%%%%%%%%%%%%%%%%%%
%%%%%%%%%%%%%%%%%%%%%%%%%%%%%%%%%%%%%%%%%%%%%%%%%%%%%%%%%%%%%%%%%%%%%%%%%%%%
\subsection{Axial gauge fields \label{subsec:fields}}
%%%%%%%%%%%%%%%%%%%%%%%%%%%%%%%%%%%%%%%%%%%%%%%%%%%%%%%%%%%%%%%%%%%%%%%%%%%%
%%%%%%%%%%%%%%%%%%%%%%%%%%%%%%%%%%%%%%%%%%%%%%%%%%%%%%%%%%%%%%%%%%%%%%%%%%%%

%n the emergence of gauge fields A simple physical interpretation of the gauge fields can be obtained by 
As a starting point we review how, in the low-energy theory close to the Weyl nodes, the four-vector, $b_{\mu}=(b_0,\mathbf{b})$ denoting the separation between two Weyl points can be reinterpreted as a chiral or axial gauge field~\cite{CFL15}. 
The time-like component $b_0$ denotes the Weyl node separation in energy space, and the space-like component $\mathbf{b}$ denotes their separation in momentum space.
This identification naturally follows from the low-energy Hamiltonian for the two decoupled Weyl nodes, which  can be generically represented as 

\begin{equation}
\mathcal{H}_{\mathrm{WSM}}^{\text{eff}}  = \sum_{\eta=\pm 1} \left[\eta b_0\mathds{I}_2 + \sum_{i,j=x,y,z}[\mathcal{D}^{(\eta)}]_{i}^{j}\sigma^i(k_j-\eta b_j)\right],
\label{eq:hamweyllowenergy}
\end{equation}
where $\mathrm{sgn}(\mathrm{Det}[\mathcal{D}^{(\eta)}])=\eta~(=\pm1)$ denotes the chirality of the Weyl node located at momentum $\mathbf{k}=\eta\mathbf{b}$ and energy $\varepsilon=\eta b_0$. 
Comparing Eq.~\eqref{eq:hamweyllowenergy} with that of a fermion of charge $e$ minimally coupled to an external gauge field $A_\mu=(\Phi,\mathbf{A})$ via $\mathcal{H}(\mathbf{k})\rightarrow \Phi+\mathcal{H}(\mathbf{k}-e\mathbf{A})$
immediately leads to the identification of the Weyl node separation $b_{\mu}$ as an effective gauge potential experienced by the Weyl fermions. Henceforth we work in units where $e=\hbar=1$. Also, all lengths are in units of the lattice constant and all time scales are in units of inverse hopping amplitude of the lattice Hamiltonians, which we specify below.

A few remarks are in order.
First, the axial gauge potential $b_{\mu}$ couples to the Weyl fermions with different signs, depending on their chirality, and thus it resembles 
an axial gauge field $A^{\mu}_{5}$ for Weyl fermions employed in high-energy physics~\cite{bertlmann2000anomalies}.
However, it is important to keep in mind that, unlike the axial gauge field $A^{\mu}_{5}$ introduced in the high-energy context, 
a gauge configuration of $b^{\mu}$ is in fact observable in the present framework.
This seemingly innocent remark has key implications, such as the vanishing of the chiral magnetic effect in Weyl semimetals~\cite{Vazifeh:2013fe,Land14}.
Second, the effective model $\mathcal{H}_{\mathrm{WSM}}^{\text{eff}}$ breaks inversion symmetry ($\mcl{I}$) if $b_{0} \neq 0$ and time-reversal symmetry ($\mcl{T}$)
if ${\bf b}\neq 0$.
These symmetries can be restored by introducing a suitable number of copies of the Hamiltonian \eqref{eq:hamweyllowenergy}, as will be explained shortly.

In the following, we are interested in situations where the parameters of a microscopic lattice Hamiltonian $\mathcal{H}_{\mathrm{WSM}}$, which describes a Weyl semimetal, are varied in space and/or time. We will assume that such modulations are performed on timescales that are much slower than the intrinsic timescales of the system (e.g.,~as set by the bandwidth of the single-particle energy spectrum), and on length-scales that are much longer than the intrinsic length-scales (e.g.,~the lattice spacing). Under such assumptions, these modulations then directly affect the low-energy Hamiltonian \eqref{eq:hamweyllowenergy} through smooth spatiotemporal dependences, in particular $b_\mu \rightarrow b_\mu(\mathbf{r},t)$. This immediately allows for the definition of an axial electric and magnetic fields, 
\begin{subequations}
\begin{equation}
\mathbf{E}_5(\mathbf{r},t) = -\bm{\nabla}_\mathbf{r}b_0(\mathbf{r},t) -\partial_t \mathbf{b}(\mathbf{r},t),
\end{equation}
\begin{equation}
\mathbf{B}_5 (\mathbf{r},t)= \bm{\nabla}_\mathbf{r}\times \mathbf{b}(\mathbf{r},t),
\end{equation}
\label{eq:e5b5definition}
\end{subequations}
which, at low energies can be regarded as effective external fields, which have direct consequences on transport properties (see Section~\ref{sec:semiclassics}). In a lattice tight-binding model, the locations and energies of Weyl nodes would generically depend on all the parameters entering the microscopic Hamiltonian. However, the identification of realistic schemes realizing non-trivial $\mathbf{E}_5(\mathbf{r},t)$ or $\mathbf{B}_5 (\mathbf{r},t)$ requires a careful analysis of simple Weyl semimetal models, as we now illustrate.
%

%%%%%%%%%%%%%%%%%%%%%%%%%%%%%%%%%%%%%%%%%%%%%%%%%%%%%%%%%%%%%%%%%%%%%%%%%%%%
%%%%%%%%%%%%%%%%%%%%%%%%%%%%%%%%%%%%%%%%%%%%%%%%%%%%%%%%%%%%%%%%%%%%%%%%%%%%
\subsection{Axial gauge fields from a simple lattice model \label{subsec:lattice}}
%%%%%%%%%%%%%%%%%%%%%%%%%%%%%%%%%%%%%%%%%%%%%%%%%%%%%%%%%%%%%%%%%%%%%%%%%%%%
%%%%%%%%%%%%%%%%%%%%%%%%%%%%%%%%%%%%%%%%%%%%%%%%%%%%%%%%%%%%%%%%%%%%%%%%%%%%

\begin{figure*}[!]
\includegraphics[width=12cm]{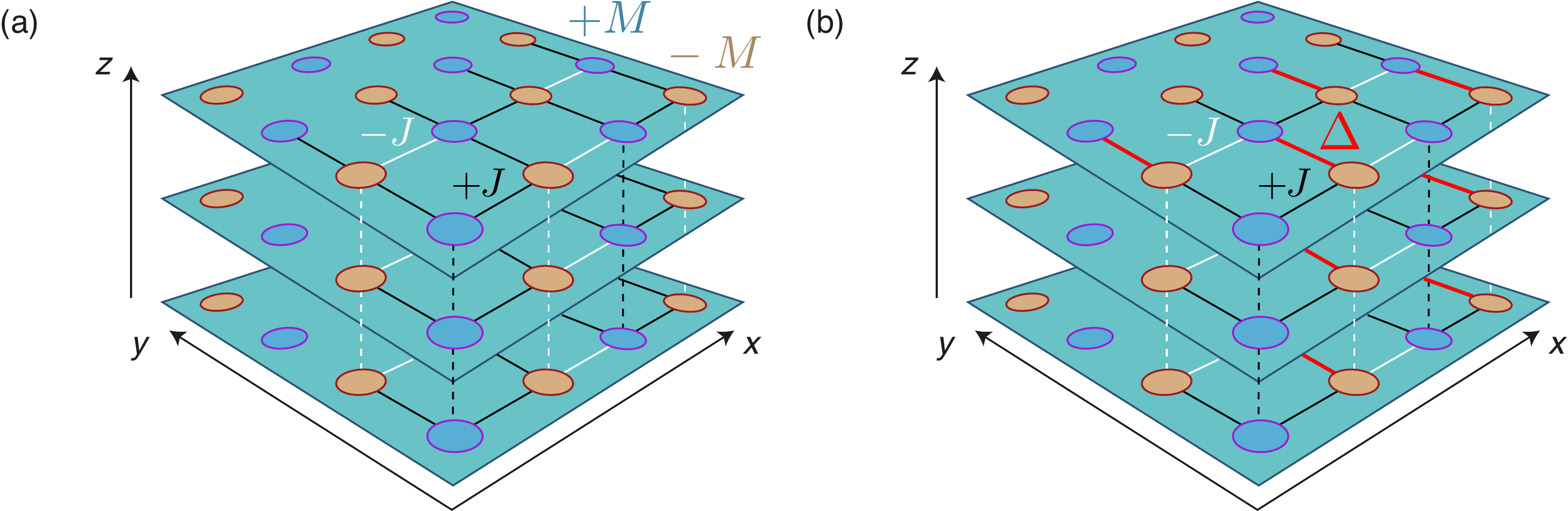}
\caption{{\bf{Schematic of the lattice models of Weyl semimetals on a cubic lattice.}} Depiction of (a) the staggered mass model \eqref{eq:hamwsm1r} 
and (b) the staggered hopping model \eqref{eq:hamwsm2r}.}
\label{fig:kett}
\end{figure*}

In this subsection, we discuss how arbitrary configurations of $\mathbf{E}_5$ and $\mathbf{B_5}$ [Eqs.~\eqref{eq:e5b5definition}] can be obtained by modifying a simple Weyl semimetal Hamiltonian, suitable for cold-atom implementations~\cite{dubeck2015weyl}.
The lattice model of Ref.~\cite{dubeck2015weyl} is defined on a cubic lattice, and is captured by the tight-binding Hamiltonian
\begin{align}
\mathcal{H}_{\mathrm{WSM}}=-J\sum_{\mathbf{r}} & \Big[ (-1)^{x+y}(c_\mathbf{r}^\dagger c_{\mathbf{r}+a\hat{x}}+c_\mathbf{r}^\dagger c_{\mathbf{r}+a\hat{z}}) \notag\\
&+c_\mathbf{r}^\dagger c_{\mathbf{r}+a\hat{y}}\Big] + \mathrm{h.c.},
\label{eq:hamwsmr}
\end{align}
where $c_\mathbf{r}^\dagger$ creates a particle at lattice site $\mathbf{r}$, $J$ denotes the hopping amplitude, and $\hat{x}$ denotes the unit vector along the $x$ direction. Importantly, the sign of the tunneling matrix elements alternates for hopping processes taking place along the $x$ and $z$ directions. This leads to a two-site sublattice structure, corresponding to sites with even or odd values of $x+y$, as illustrated in Fig.~\ref{fig:kett}(a). Diagonalizing the Hamiltonian in Eq.~\eqref{eq:hamwsmr} reveals two pairs of Weyl nodes in the first Brillouin zone. Specifically, this model realizes a Weyl semi-metal with time-reversal symmetry (see Ref.~\cite{dubeck2015weyl}). 

\subsubsection{ The staggered mass model $\mathcal{H}_{\mathrm{WSM}}^{M}$}

In order to realize tunable axial gauge fields, we must introduce additional ingredients to the model of Ref.~\cite{dubeck2015weyl}. As a first example, we propose to build on the sublattice structure of this model by adding a staggered potential that shifts the on-site energy of one sublattice with respect to the other. The corresponding Hamiltonian therefore takes the form 
\begin{eqnarray}
\mathcal{H}_{\mathrm{WSM}}^{M}&=-J&\sum_{\mathbf{r}}[(-1)^{x+y}(c_\mathbf{r}^\dagger c_{\mathbf{r}+a\hat{x}}+c_\mathbf{r}^\dagger c_{\mathbf{r}+a\hat{z}})\nonumber\\
&&+c_\mathbf{r}^\dagger c_{\mathbf{r}+a\hat{y}} + \mathrm{h.c.}]+M\sum_\mathbf{r}(-1)^{x+y}c_\mathbf{r}^\dagger c_\mathbf{r},
\label{eq:hamwsm1r}
\end{eqnarray}
with a staggered potential of strength $M$.

%%%%%%%%%%%%%%%%%%%%%%%%%%%%%%%%%%%%%%%%%%%%%%%%%%%%%%%%
%%%%%%%%%%%%%%%%%%%%%%%%%%%%%%%%%%%%%%%%%%%%%%%%%%%%%%%%

 For the sake of simplicity, we will analyze this model in a reference frame that is rotated by $\pi/4$ about the $z$-axis, such that each sublattice forms a cubic lattice with its primary lattice vectors given by $\sqrt{2}a\hat{x}$, $\sqrt{2}a\hat{y}$, and $\sqrt{2}a\hat{z}$ respectively; for the rest of the paper we will also set $a=1/\sqrt{2}$.
In this rotated frame, the Hamiltonian in Eq.~\eqref{eq:hamwsm1r} can now be expressed as
\begin{align}
\mathcal{H}_{\mathrm{WSM}}^M= -J\sum_{\mathbf{r}}&[\hat{c}_{A,\mathbf{r}}^\dagger (\hat{c}_{B,\mathbf{r}} -  \hat{c}_{B,{\mathbf{r}-\hat{x}}}+ \hat{c}_{B,{\mathbf{r}-\hat{y}}}\nonumber\\
&+ \hat{c}_{B,{\mathbf{r}-\hat{x}-\hat{y}}}+ \hat{c}_{A,{\mathbf{r}+\hat{z}}})-\hat{c}_{B,\mathbf{r}}^\dagger \hat{c}_{B,{\mathbf{r}+\hat{z}}} + \mathrm{h.c}]\nonumber\\
&+M\sum_\mathbf{r}[\hat{c}_{A,\mathbf{r}}^\dagger\hat{c}_{A,\mathbf{r}}-\hat{c}_{B,\mathbf{r}}^\dagger\hat{c}_{B,\mathbf{r}}],
\label{eq:hamwsm1rb}
\end{align}
where $c^\dagger_{A,\mathbf{r}}$ ($c^\dagger_{B,\mathbf{r}}$) denotes the creation operator on the $A$ ($B$) sublattice within the unit cell located at position $\mathbf{r}$.

Owing to the bipartite structure of Hamiltonian \eqref{eq:hamwsm1rb}, it may be represented in a simple form in reciprocal space, $\mathcal{H}_{\mathrm{WSM}}^M(\mathbf{k})\!=\!\mathbf{d} (\mathbf{k})\cdot\bm{\sigma}$, where $\sigma^{x,y,z}$ are Pauli matrices and the Pauli vector $\mathbf{d} (\mathbf{k})$ is
\begin{align} 
d_x &= J(1-\cos k_x+\cos k_y + \cos(k_x+k_y)),\nonumber\\
d_y &= J(-\sin k_x+\sin k_y + \sin(k_x+k_y)),\label{eq:hamwsm1rk}\\
d_z &= 2J\cos k_z+M.\nonumber
\end{align}

Analyzing the energy spectrum, $\varepsilon_\mathbf{k}=\pm\vert\mathbf{d}_\mathbf{k}\vert$, reveals that there remain two pairs of Weyl nodes in the spectrum, one pair being the time-reversed partner of the other~\cite{dubeck2015weyl}.
The locations of the four Weyl nodes in momentum-space are given by
\begin{equation}
\mathbf{k}_W=\pm(\pi/2,\pi/2,\pm \cos^{-1}(-M/2J)).
\label{eq:wsm1wps}
\end{equation}
Importantly, the Weyl node locations (and separations) are found to only depend on one dimensionless parameter: $M/J\!\equiv\!M$, setting $J\!=\!1$.
Hence, a spatiotemporal variation of the parameter $M$ should generate axial fields [Eqs.~\eqref{eq:e5b5definition}], as we will now show explicitly.

In the presence of multiple pairs of Weyl nodes, only those pair(s) for which the spatiotemporally varying parameter leads to a relative shift in the position of the two Weyl nodes can lead to axial fields. Thus, in this case, we can focus our analysis on the pair of Weyl points located at $\mathbf{k}_{W,\pm}=(\pi/2,\pi/2,\pm\cos^{-1}(-M/2))$~\footnote{Note that, for the pair of Weyl points at $\mathbf{k}_{W,\pm}=(-\pi/2,-\pi/2,\pm\cos^{-1}(-M/2))$, the analysis remains unchanged except for a flip in the chiralities. On the other hand, choosing the pair of Weyl points located at the same value of $k_z$, for instance the pair $\mathbf{k}_{W,\pm}=(\pm\pi/2,\pm\pi/2,\cos^{-1}(-M/2))$ does not result in axial fields since any variation in $M$ leads to an overall shift in the Weyl points and not a relative shift between the two. The effective low-energy Hamiltonian for such pair does not posses an axial gauge potential.}.
Around these two Weyl points, the Hamiltonian can be expanded to linear order in momentum as
\begin{align}
\mathcal{H}_\pm=&(k^\prime_x-k^\prime_y)\sigma^x-(k^\prime_x+k^\prime_y)\sigma^y\nonumber\\
&\mp\sqrt{4-M^2}[k_z\mp\cos^{-1}(-M/2)]\sigma^z,\label{eq:hamwsm1rlin}
\end{align}
where $k^\prime_{x(y)}\!=\!k_{x(y)}-\pi/2$.
Comparing the form of the linearized Hamiltonian \eqref{eq:hamwsm1rlin} with Eq.~\eqref{eq:hamweyllowenergy}, one can make the identification 
\begin{equation}
\mathcal{D}^{(\eta)}=\begin{pmatrix}1&-1&0\\-1&-1&0\\0&0&-\eta\sqrt{4-M^2}\end{pmatrix},
\label{eq:hamwsm1rlindmat}
\end{equation}
which shows that the two Weyl nodes considered above are indeed associated with opposite chiralities, since $\mathrm{Det}[\mathcal{D}^{(\eta)}]=2\eta\sqrt{4-M^2}$. 
Importantly, this comparison allows for the identification of the axial gauge potential
\begin{equation}
\mathbf{b}=(0,0,\cos^{-1}(-M/2)) ,
\end{equation}
which can also be directly identified through Eq.~\eqref{eq:wsm1wps}.
Hence, allowing $M(\mathbf{r},t)$ to depend parametrically on space and time, leads to the axial fields
\begin{subequations}
\begin{equation}
\mathbf{E}_5 (\mathbf{r},t)= \frac{\partial_t M(\mathbf{r},t)}{\sqrt{4-M(\mathbf{r},t)^2}}\hat{z},
\end{equation}
\begin{equation}
\mathbf{B}_5 (\mathbf{r},t)= \frac{\partial_y M(\mathbf{r},t)\hat{x}-\partial_x M(\mathbf{r},t)\hat{y}}{\sqrt{4-M(\mathbf{r},t)^2}}.
\end{equation}
\label{eq:hamwsm1re5b5}
\end{subequations}

One should note that the spatiotemporal variation of the parameter $M$ does not only generate axial fields in the equations of motion. It also modulates the shape of the band structure, and hence introduces spatiotemporal dependences in the band velocity~\cite{DeJuan2012}. This effect is apparent in our model, where from Eq.~\eqref{eq:hamwsm1rlin} we observe that the Fermi velocity along $z$ depends on $M$. When analyzing the semiclassical equations of motion in Section~\ref{sec:semiclassics}, we will neglect this spatiotemporal variation of the Fermi velocity, which is well justified when the variations of $M$ are taken to be small. 
Furthermore, within the linearized regime, the relative momentum will also be taken to be small, and hence any variation of the Fermi velocity will appear as a second order effect, which can hence be neglected.
Such spatiotemporal Fermi velocity effects have been considered previously in Refs.~\cite{Yang2015,Jiang2015}.

\subsubsection{The staggered hopping model $\mathcal{H}_{\mathrm{WSM}}^\Delta$ and gauge field tunability}

Inspection of Eq.~\eqref{eq:hamwsm1re5b5} reveals that a spatiotemporal variation of the staggered-potential strength $M(\mathbf{r},t)$ produces axial fields with highly-constrained orientations: $\mathbf{E}_5$ is necessarily directed along $\hat{z}$, while $\mathbf{B}_5$ lies in the $(x,y)$ plane. In particular, we find that $\mathbf{E}_5$ is restricted to the direction set by the Weyl node separation, while $\mathbf{B}_5$ is restricted to be perpendicular to this direction. 

In this paragraph, we propose an alternate modification of the model in Eq.~\eqref{eq:hamwsmr}, which allows for more flexibility over the pseudo-field orientation. Instead of introducing a staggered potential, we propose to modulate the tunneling matrix elements along the $y$ direction, as described by the following Hamiltonian
\begin{align}
\mathcal{H}_{\mathrm{WSM}}^\Delta=-\sum_{\mathbf{r}}&[(-1)^{x+y}J(c_\mathbf{r}^\dagger c_{\mathbf{r}+a\hat{x}}+c_\mathbf{r}^\dagger c_{\mathbf{r}+a\hat{z}})\nonumber\\
& + J_y(\mathbf{r},t) c_\mathbf{r}^\dagger c_{\mathbf{r}+a\hat{y}} + \mathrm{h.c.}],
\label{eq:hamwsm2r}
\end{align}
where 
\begin{equation}
J_y(\mathbf{r},t) = \frac{1}{2}[(J+\Delta(\mathbf{r},t))+(-1)^{x+y}(J-\Delta(\mathbf{r},t))].
\end{equation}
As illustrated in Fig.~\ref{fig:kett}(b), the alternating hopping amplitudes $(J, \Delta)$ along $y$ preserve the sublattice structure of the original model~\eqref{eq:hamwsmr}.
Therefore, as for the staggered mass model [Eq.~\eqref{eq:hamwsm1rb}], we write the Hamiltonian in a rotated frame, which reads
\begin{eqnarray}
\mathcal{H}_{\mathrm{WSM}}^\Delta&= -J&\sum_{\mathbf{r}}[\hat{c}_{A,\mathbf{r}}^\dagger (\hat{c}_{B,\mathbf{r}} -  \hat{c}_{B,{\mathbf{r}-\hat{x}}}+ \hat{c}_{B,{\mathbf{r}-\hat{y}}}+\hat{c}_{A,{\mathbf{r}+\hat{z}}}+\nonumber\\
&& (\Delta/J)\hat{c}_{B,{\mathbf{r}-\hat{x}-\hat{y}}})-\hat{c}_{B,\mathbf{r}}^\dagger \hat{c}_{B,{\mathbf{r}+\hat{z}}} + \mathrm{h.c}].
\label{eq:hamwsm2rb}
\end{eqnarray}

In reciprocal space, this Hamiltonian can be written in the form $\mathcal{H}_{\mathrm{WSM}}^\Delta(\mathbf{k})\!=\!\mathbf{d} (\mathbf{k})\cdot\bm{\sigma}$ with
\begin{align} 
d_x &= J(1-\cos k_x+\cos k_y + (\Delta/J)\cos(k_x+k_y)),\nonumber\\
d_y &= J(-\sin k_x+\sin k_y + (\Delta/J)\sin(k_x+k_y)),\label{eq:hamwsm2rk}\\
d_z &= 2J\cos k_z. \nonumber
\end{align}

Analyzing the energy spectrum of the Hamiltonian $\mathcal{H}_{\mathrm{WSM}}^\Delta(\mathbf{k})$ reveals that the Weyl nodes are now located at
\begin{equation}
\mathbf{k}_W = \pm(\mathcal{K},\pi-\mathcal{K},\pm\pi/2),
\end{equation}
where 
\begin{equation}
\mathcal{K} = \tan^{-1}\left[\frac{\sqrt{(3J-\Delta)(J+\Delta)}}{J-\Delta}\right].
\label{eq:frakK}
\end{equation}
Note that the original Weyl nodes, in the absence of perturbation ($\Delta\!=\!J$), are located at $\mathbf{k}_W = \pm(\pi/2,\pi/2,\pm\pi/2)$.

As for the staggered mass model, \eqref{eq:hamwsm1r}, we focus on the Weyl node pairs which will lead to emergent axial fields.
In this case, these include those for which the nodes are located at the same value of $k_z$, say $k_z\!=\!\pi/2$, denoted by
\begin{equation}
\mathbf{k}_{W,\pm} = (\pm\mathcal{K},\pm(\pi-\mathcal{K}),\pi/2).
\end{equation}
The axial gauge potential can then be identified as
\begin{equation}
\mathbf{b} = (\mathcal{K},\pi-\mathcal{K},0).\label{delta_potential}
\end{equation}
Hence, using Eqs.~\eqref{eq:e5b5definition}, \eqref{eq:frakK}, and \eqref{delta_potential}, we find that a spatiotemporal variation of the tunneling parameter $\Delta(\mathbf{r},t)$ can be used to produce the axial  fields $\mathbf{E}_5$ and $\mathbf{B}_5$, which can be expressed as
\begin{subequations}
\begin{equation}
\mathbf{E}_5(\mathbf{r},t)=-\partial_t\mathcal{K}(\mathbf{r},t)[\hat{x}-\hat{y}];
\end{equation}
\begin{equation}
\mathbf{B}_5(\mathbf{r},t) = \partial_z\mathcal{K}(\mathbf{r},t)[\hat{x}+\hat{y}]-[\partial_x\mathcal{K}(\mathbf{r},t)+\partial_y\mathcal{K}(\mathbf{r},t)]\hat{z}.
\end{equation}
\label{eq:e5b5wsm2}
\end{subequations}
In contrast with the staggered-mass model \eqref{eq:hamwsm1r}, this approach allows one to generate a field $\mathbf{E}_5(\mathbf{r},t)$ that is {\it perpendicular} to the direction set by the original separation between the Weyl nodes [$\mathbf{b}\propto (\hat{x}+\hat{y})$; see Eq.~\eqref{delta_potential} for $\Delta\!=\!J$]. Moreover, we note that the field $\mathbf{B}_5(\mathbf{r},t) $ now has components that are parallel to the Weyl node separation.\\ 

As a final remark, we note that one could combine the ingredients of the staggered-mass \eqref{eq:hamwsm1r} and staggered-hopping models \eqref{eq:hamwsm2r}, in order to generate axial fields of any arbitrary directions (with respect to the original Weyl node separation).

%%%%%%%%%%%%%%%%%%%%%%%%%%%%%%%%%%%%%%%%%%%%%%%%%%%%%%%%%%%%%%%%%%%%%%%%%%%%
%%%%%%%%%%%%%%%%%%%%%%%%%%%%%%%%%%%%%%%%%%%%%%%%%%%%%%%%%%%%%%%%%%%%%%%%%%%%
\subsection{Optical lattice implementation \label{subsec:amo}}
%%%%%%%%%%%%%%%%%%%%%%%%%%%%%%%%%%%%%%%%%%%%%%%%%%%%%%%%%%%%%%%%%%%%%%%%%%%%
%%%%%%%%%%%%%%%%%%%%%%%%%%%%%%%%%%%%%%%%%%%%%%%%%%%%%%%%%%%%%%%%%%%%%%%%%%%%

 In this section, we describe realistic schemes that realize the staggered mass and staggered hopping models, defined by Eq.~\eqref{eq:hamwsm1r} and Eq.~\eqref{eq:hamwsm2r} respectively, using accessible optical lattice technologies. \\

A promising scheme realizing the simple model in Eq.~\eqref{eq:hamwsmr} has already been carefully described in Ref.~\cite{dubeck2015weyl}. The scheme is based on the observation that, in this model, the sign of tunneling matrix elements alternates for hopping processes taking place along the $x$ and $z$ directions. In order to modify these tunneling matrix elements, modulation-induced tunneling~\cite{kolovsky2011creating,bermudez2011synthetic,goldman2015periodically,creffield2016realization} is performed along these two directions. This is realized by tilting a 3D optical lattice along the $x$ and $z$ direction, for instance by using magnetic field gradients, and then restoring tunneling using an external resonant time-modulation via, e.g., superimposing a moving optical lattice whose frequency is resonant with the energy offsets generated by the tilt. The phase of this time-modulation, which is typically space dependent, can then be tuned so as to generate the pattern of alternating hopping amplitudes along both $x$ and $z$; see Ref.~\cite{dubeck2015weyl} for details. 

It is important to note that the time-modulated optical lattice of Ref.~\cite{dubeck2015weyl} realizes synthetic $\pi$-fluxes in a set of plaquettes defined in the $x-y$ and $x-z$ planes; hence, this artificial magnetic field~\cite{dalibard2011colloquium,goldman2014light,aidelsburger2017artificial} will be present in the following discussion, independently of the \emph{axial} gauge fields (on which the focus will be set). As will become apparent below, the schemes realizing standard (non-axial) synthetic gauge fields (e.g.~through light-induced or shaking methods~\cite{dalibard2011colloquium,goldman2014light,aidelsburger2017artificial}) are compatible with those generating axial gauge fields.

\subsubsection{Realizing the staggered-mass model} 
 
The only difference between the staggered-mass model Hamiltonian in Eq.~\eqref{eq:hamwsm1r} and that of Ref.~\cite{dubeck2015weyl} in Eq.~\eqref{eq:hamwsmr} is the presence of a staggered mass term $M$, which will be required to depend on both time and space. A simple way to realize the staggered mass model, starting from the configuration laid out in Ref.~\cite{dubeck2015weyl}, would consist in adding an additional square lattice in the $xy$, with spacing $a\sqrt{2}$ and  aligned with the blue sites in Fig.~\ref{fig:kett}(a). This would generate the staggered potential of Eq.~\eqref{eq:hamwsm1r} by changing the on-site energy of the even sublattice only. Time dependence of $M$ is trivially generated by modulating the  intensity of the laser field that generates this extra lattice potential. Space dependence is most easily generated by either slightly detuning the wavelength of the additional $M$ lattice from the original one to yield long-distance changes in $M$, or offsetting the $M$ lattice from the main lattice beams such that spatial intensity variations towards the edge of the beam waist give rise to spatial dependence of the staggered mass.

Another option for implementing the staggered mass model is based on directly exploiting the checkerboard (sublattice) structure displayed in the $xy$ plane [Fig.~\ref{fig:kett}(a)]. Following Ref.~\cite{goldman2013realizing}, we note that such a 2D checkerboard configuration can be implemented by trapping two internal states of an atom (e.g.~$g$ and $e$) with an optical square lattice field that is set at an ``anti-magic" wavelength, i.e., a special wavelength such that the $g$ $(e)$ states are trapped at the potential's minima (maxima), leading to a checkerboard configuration of $g$ and $e$ states in the 2D plane. As shown in Ref.~\cite{goldman2013realizing}, tunneling between the neighboring sites of this checkerboard lattice can be activated by resonantly coupling the $g$ and $e$ states~\cite{jaksch2003creation}. Interestingly, the tunneling along the $x$ and $y$ directions can be activated and tuned independently by exploiting transitions between degenerate Zeeman sublevels of the $g$ and $e$ manifolds~\cite{osterloh2005cold,goldman2013realizing}. The latter feature can be exploited to generate the alternating tunneling matrix elements ($\pm J$) in the $xy$ plane, as required in Eq.~\eqref{eq:hamwsmr}. Then, the staggered potential of Eq.~\eqref{eq:hamwsm1r} can be simply obtained by detuning the $g-e$ transitions, i.e., by slightly shifting the state dependent potential out of resonance (which could be realized in a spatiotemporal manner). Finally, tunneling along the $z$-direction could be modulation-induced so as to realize the desired tunneling matrix elements $\pm J$. Here, one could use an optical lattice along the $z$ direction that is set at a ``magic" wavelength, meaning that both $g$ and $e$ states feel the same potential along $z$. Modulation-induced tunneling can then be implemented as explained above by tilting the lattice and activating the hopping through an additional moving optical lattice.

\subsubsection{Realizing the staggered-hopping model}

The staggered hopping model in Eq.~\eqref{eq:hamwsm2r} builds on a very specific ingredient:~the tunneling amplitudes should be ``dimerized" in the $xy$ plane, with amplitudes $J$ and $\Delta$ [Fig.~\ref{fig:kett}(b)]. In order to achieve such a configuration, one could start from the tunable optical lattice potential that was introduced by Tarruel {\it et al.} in Ref.~\cite{tarruell2012creating}:~through a proper adjustment of the optical potential, the lattice sites can be combined by pairs, hence leading to strong coupling within ``dimers" (with coupling $\Delta$) and weaker couplings between the dimers (with amplitudes $J_1$ and $J_2$, along $x$ and $y$ respectively); see Fig. 1(b) in Ref.~\cite{tarruell2012creating}. By tuning the optical potential such that $J_1\!\sim\!J_2$, one can generate the alternating pattern of hopping amplitudes in the $xy$ plane ($J$,$\Delta$), as depicted in Fig.~\ref{fig:kett}(b). In this scheme, $\Delta$ can be varied in space and time by modulating the optical potential of Ref.~\cite{tarruell2012creating}, as required for the generation of axial fields; see also Ref.~\cite{goldman2016creating} for a scheme realizing smooth spatial modulations of this tunable lattice potential. Tuning the sign of the tunneling matrix elements along $x$ and $z$ could then be realized through modulation-induced tunneling methods~\cite{dubeck2015weyl}, as explained above for the staggered mass model. We have verified that the difference $J_1\ne J_2$, which is usually present in the ``dimer" potential of Ref.~\cite{tarruell2012creating}, does not modify the properties of axial fields in a significant manner. Finally, we note that the staggered mass and staggered hopping models could be combined, through a fusion of the schemes proposed in this section.

%%%%%%%%%%%%%%%%%%%%%%%%%%%%%%%%%%%%%%
\section{Semiclassical probes for $\mathbf{E}_5$ and $\mathbf{B}_5$ \label{sec:semiclassics}}
%%%%%%%%%%%%%%%%%%%%%%%%%%%%%%%%%%%%%%

In this section, we discuss how dynamics of wavepackets analyzed semiclassically can serve as probes for the axial fields. As we noted earlier, while conventional transport experiments are difficult in the context of ultra-cold atoms, unconventional probes such as direct non-equilibrium measurement of wavepacket dynamics are much more feasible.
Apart from the corrections corresponding to the usual anomalous Hall velocity~\cite{xiao2010berry}, the semiclassical equations we use also contain corrections due to the slow and parametric spatiotemporal dependence of the Hamiltonian parameters, which are accounted for via gradient corrections~\cite{sundaram1999wavepacket}.

\subsection{General formalism and methodology}

Let us start by considering the behavior of a wavepacket, centered around the position $\mathbf{r}_c$  and momentum $\mathbf{k}_c$, and prepared in the Bloch band associated with an eigenstate $\vert u\rangle\!\equiv\!\vert u (\bm{k}; \bm{r} ,t)\rangle$ of a generic Hamiltonian $\mathcal{H}(\mathbf{k}; \bm{r} ,t)$; the generalized semiclassical equations then read~\cite{sundaram1999wavepacket}
\begin{align}
\dot{\mathbf{r}}_c =& \bm{\nabla}_\mathbf{k}\varepsilon_\mathbf{k} -{\Omega}_{\mathbf{kr}}\dot{\mathbf{r}}_c-{\Omega}_{\mathbf{kk}}\dot{\mathbf{k}}_c+\bm{\Omega}_{t\mathbf{k}},\label{eq:sceomr}\\
\dot{\mathbf{k}}_c =& -\bm{\nabla}_\mathbf{r}\varepsilon_\mathbf{k} +{\Omega}_{\mathbf{rr}}\dot{\mathbf{r}}_c+{\Omega}_{\mathbf{rk}}\dot{\mathbf{k}}_c-\bm{\Omega}_{t\mathbf{r}},\label{eq:sceomk}
\end{align}
where ${\Omega}_\mathbf{kk}$, ${\Omega}_\mathbf{kr}$, and ${\Omega}_\mathbf{rr}$ are generalized Berry curvature matrices with their elements given by
\begin{equation}
({\Omega}_{\mathbf{kk}})^{ij} = {\Omega}_{k_i k_j}=i\left[\left\langle\partial_{k_i}u\vert\partial_{k_j}u\right\rangle-\left\langle\partial_{k_j}u\vert\partial_{k_i}u\right\rangle\right],
\label{eq:omegakk}
\end{equation}
and similarly for ${\Omega}_{\mathbf{rk}}$ and ${\Omega}_{\mathbf{rr}}$, while the components of the vector $\bm{\Omega}_{t\mathbf{k}}$ are defined as
\begin{equation}
(\bm{\Omega}_{t\mathbf{k}})^i = i\left[\left\langle\partial_t u\vert\partial_{k_i}u\right\rangle-\left\langle\partial_{k_i}u\vert\partial_tu\right\rangle\right],
\label{eq:omegatk}
\end{equation}
and similarly for $\bm{\Omega}_{t\mathbf{r}}$.
Note that the energy dispersion $\varepsilon_\mathbf{k}$ can  also depend parametrically on $\mathbf{r}$ and $t$, through the spatiotemporally-varying Hamiltonian parameters.

Let us note some important technical remarks about the applicability of Eqs.~\eqref{eq:omegakk} and \eqref{eq:omegatk}. In order for these semiclassical approaches to be valid, the temporal variations of the lattice system should occur on a much slower time scale as compared to the intrinsic time scales of the system (e.g., the inverse of hopping energies). Similarly, the spatial variations should occur on much longer length scales than the intrinsic length scales (e.g., the lattice spacing). When this assumption holds\mkadd{,} Eqs.~\eqref{eq:omegakk} and \eqref{eq:omegatk} are meaningful attempts to ``coarse grain'' the dynamics. For cold-atom experiments, which are the main focus of this work, the wavepackets are typically much larger than the lattice spacing, and much smaller than the cyclotron orbits (which are much smaller than the total size of the optical lattice). It is thus not surprising that such semiclassical equations of motions are found to well describe recent cold-atom experiments on Berry curvature effects~\cite{aidelsburger2015measuring,duca2014aharonov}. Additionally, as explained in detail in Ref.~\cite{sundaram1999wavepacket}, this formalism naturally allows us to treat external magnetic and electric fields in the same footing as perturbations due to spatio-temporal variations of the Hamiltonian parameters, as they enter through the Berry curvatures defined above. In what follows, we assume that there are no synthetic electromagnetic fields acting on the models in section~\ref{sec:models}, focusing instead on perturbations that generate external axial fields as we now describe.
%%%
%%%
%%%

For the two-band models described in Section \ref{sec:models} and defined by Eqs.~\eqref{eq:hamwsm1r} and \eqref{eq:hamwsm2r},  Eqs. \eqref{eq:omegakk} and \eqref{eq:omegatk} can be simplified to 
\begin{equation}
\Omega_{k_i k_j} = \mathbf{d}\cdot[\partial_{k_i}\mathbf{d}\times\partial_{k_j}\mathbf{d}]/\vert\mathbf{d}\vert^3,
\label{eq:omegatls}
\end{equation}
where $\mathbf{d}$ is the Pauli vector representing the Hamiltonian in momentum space defined by Eq. \eqref{eq:hamwsm1rk} and Eq. \eqref{eq:hamwsm2rk} for models \eqref{eq:hamwsm1rb} and \eqref{eq:hamwsm2r}, respectively.

For concreteness, we shall use the model described in Eqs.~\eqref{eq:hamwsm1rb} and \eqref{eq:hamwsm1rk} throughout this section.
To generate the $\mathbf{E}_5$ and $\mathbf{B}_5$ fields, $M$ is taken to be spatiotemporally inhomogeneous and linear, namely
\begin{equation}
M(\mathbf{r},t) = M_0 + M_1^{(t)}t +M_1^{(x)}x.
\label{eq:mrt}
\end{equation}
As the semiclassical equations of motion solely depend on local gradients of the various fields and pseudo-fields, this linear approximation plays no important physical role, and is thus invoked to simplify the analysis. More general (slow) spatio-temporal dependence can easily be treated by locally linearizing the equations.

In the following subsections, we shall show that finite $\mathbf{E}_5$ and $\mathbf{B}_5$ fields can be detected via anomalous Hall-like drifts and cyclotron orbits of a wavepacket, as well as through a pseudo chiral magnetic effect. In doing so, we will obtain a more transparent expression for the equations of motion given in Eqs.~\eqref{eq:sceomr} and \eqref{eq:sceomk}.  
Indeed, we will show that the latter can be recast in the more standard form~\cite{xiao2010berry}
\begin{align}
\dot{\mathbf{r}}^{\eta}_c = & \bm{\nabla}_\mathbf{q}\varepsilon -{\Omega}_{\mathbf{qq}}\cdot\dot{\mathbf{q}}^{\eta}_c,\label{eq:sceomrbare}\\
\dot{\mathbf{q}}^{\eta}_c =& \eta\mathbf{E}_5 + \eta\dot{\mathbf{r}}_c\times\mathbf{B}_5,\label{eq:sceomkbare}
\end{align}
where the reference frame was changed so as to measure the momentum relative to the Weyl nodes:~Specifically, for each node labelled by $\eta\!=\!\pm$, we defined the relative momentum $\mathbf{q}^{\eta}\!=\!\mathbf{k}-\eta\mathbf{b}$, where $\eta\mathbf{b}$ is the location of the Weyl node with respect to the wavepacket's momentum. Note that this alternative form~\eqref{eq:sceomrbare}-\eqref{eq:sceomkbare} is reminiscent of the standard semiclassical equations of motion, as modified by conventional external electromagnetic fields~\cite{xiao2010berry}. In particular, $\mathbf{E}_5$ and $\mathbf{B}_5$ can indeed be directly detected and characterized through wavepacket dynamics, as we now investigate in more detail.

\subsection{Anomalous Hall drifts due to $\mathbf{E}_5$ \label{subsec:e5}}
\subsubsection{Analytical description}

We first consider the case where $M_1^{(x)}\!=\!0$ and $M_1^{(t)}\!\ne\!0$, which leads to a finite $\mathbf{E}_5$ but zero $\mathbf{B}_5$; see Eq.~\eqref{eq:hamwsm1re5b5}.
Since there is no spatial dependence on the Hamiltonian, one obtains from Eq.~\eqref{eq:sceomk} that $\dot{\mathbf{k}}_c=0$, i.e., translational symmetry implies that the momentum of the wavepacket is a constant of motion. 
Using this in Eq.~\eqref{eq:sceomr} results in $\dot{\mathbf{r}}_c = \bm{\nabla}_\mathbf{k}\varepsilon +\bm{\Omega}_{t\mathbf{k}}$
where the first term is the trivial group velocity and the second term is a Berry curvature correction.
We now show that the second term $\bm{\Omega}_{t\mathbf{k}}$ is precisely the anomalous Hall velocity due to $\mathbf{E}_5$, as suggested by Eq.~\eqref{eq:sceomrbare}. 
Using Eq.~\eqref{eq:omegatls}, together with Eqs.~\eqref{eq:hamwsm1rk} and \eqref{eq:mrt}, we obtain
%{}
\begin{align}
(\bm{\Omega}_{t\mathbf{k}})^i &= -\epsilon^{lmn}d_l (\partial_{k_i}d_m)(\partial_td_n)/\vert\mathbf{d}\vert^3\nonumber\\
&=-\epsilon^{lm z}d_l (\partial_{k_i}d_m)M_1^{(t)}/\vert\mathbf{d}\vert^3,
\label{eq:omegaktwsm1}
\end{align}
where $\epsilon^{lmn}$ is the three dimensional Levi-Civita symbol.
For concreteness, we explicitly use the specifics of the model from Eq.~\eqref{eq:hamwsm1rk}. A more general analysis in the linearized regime is presented in Appendix \ref{sec:app1}.

We now connect Eq.~\eqref{eq:omegaktwsm1} to an anomalous Hall response due to $\mathbf{E}_5$, following a two-steps approach:~We first evaluate $\mathbf{E}_5$ using Eq.~\eqref{eq:hamwsm1rk},  and then, using a coordinate change, we show that the anomalous Hall response resulting from Eqs.~\eqref{eq:sceomrbare}-\eqref{eq:sceomkbare} is precisely given by \eqref{eq:omegaktwsm1}. 
To this end we first expand $d_z$ in Eq.~\eqref{eq:hamwsm1rk} (as it is the only time dependent term) to linear order in momentum around the Weyl points $\eta \cos^{-1}(-M/2)$, where $0\le\cos^{-1}(-M/2)\le\pi$, 
to obtain
\begin{align}
d_z^{(\eta)}&\approx -\eta\sqrt{4-M^2}[k_z-\eta\cos^{-1}(-M/2)]\nonumber\\
&\approx -\eta\left(\sqrt{4-M_0^2}-\frac{M_0M_1^{(t)}t}{\sqrt{4-M_0^2}}\right)\times\nonumber\\
&~~~\left[k_z-\eta\left(\cos^{-1}\left(\frac{-M_0}{2}\right)+\frac{M_1^{(t)}t}{\sqrt{4-M^2_0}}\right)\right],
\label{eq:dzmt}
\end{align}
where in the second step we have also expanded the terms to leading order in $M_1^{(t)}$. 
However, since we consider $M_1^{(t)}$ small, and also the analysis stays valid as long as $k_z$ is close to the Weyl node, we ignore the dependence on $M_1^{(t)}$ of the Fermi velocity to obtain
\begin{equation}
d_z^{(\eta)}\approx-\eta\sqrt{4-M_0^2}\left[k_z-\eta\left(\cos^{-1}\left(\frac{-M_0}{2}\right)+\frac{M_1^{(t)}t}{\sqrt{4-M^2_0}}\right)\right].
\label{eq:dzmtlin}
\end{equation}
Hence $\mathbf{b}$ can be identified as 
\begin{equation}
\mathbf{b}=\left(0,0,\cos^{-1}\left(\frac{-M_0}{2}\right)+\frac{M_1^{(t)}t}{\sqrt{4-M^2_0}}\right),
\end{equation}
and consequently $\mathbf{E}_5$ as
\begin{equation}
\mathbf{E}_5=-\partial_t\mathbf{b}=-\frac{M_1^{(t)}}{\sqrt{4-M_0^2}}\hat{z}.
\label{eq:e5wsm1}
\end{equation}
We now make the change of reference for the momentum, such that $k_z$ is now measured relative to the time-dependent momentum of the Weyl node. The transformed momenta to linear order in $M_1^{(t)}$ takes the form
\begin{equation}
q_{x(y)} = k_{x(y)},~ q_z = k_z-\eta\left(\cos^{-1}\left(\frac{-M_0}{2}\right)+\frac{M_1^{(t)}t}{\sqrt{4-M^2_0}}\right)\label{qx:eq}
\end{equation}
Since the transformed momentum is time-dependent, via Eq.~\eqref{eq:sceomkbare} it can be regarded as coming from a net force acting on the wavepacket in this reference frame, 
which is given by the time-derivative of the momentum as
$\dot{\mathbf{q}} = -\eta M_1^{(t)}\hat{z}/\sqrt{4-M_0^2}$. 
Note that, the magnitude of the force is concomitant with $\mathbf{E}_5$ derived in Eq.~\eqref{eq:e5wsm1} and the factor of $\eta$ is due to the chiral nature of $\mathbf{E}_5$.

The anomalous Hall response due to this field can then be calculated from Eq.~\eqref{eq:sceomrbare} as 
\begin{align}
-({\Omega}_{\mathbf{qq}}\dot{\mathbf{q}}_c)^i &= -\Omega_{q_i q_j}\dot{q}_{c,j}\nonumber\\
&=-\epsilon^{lmn}d_l(\partial_{q_i} d_m)(\partial_{q_j} d_n)\dot{q}_{c,j}/\vert\mathbf{d}\vert^3\nonumber\\
&=-\epsilon^{lm z}d_l(\partial_{q_i} d_m)(\partial_{q_z} d_z)\dot{q}_{c,z}/\vert\mathbf{d}\vert^3,
\label{eq:e5omegaqq}
\end{align}
where we have used that only the $z$-component of $\mathbf{q}$ depends on time and only $d_z$ is dependent on $q_z$. 
Using $\partial_{q_z}d_z = -\eta\sqrt{4-M_0^2}$ from Eq.~\eqref{eq:dzmtlin} and the form of $\dot{\mathbf{q}}$ from Eq.~\eqref{qx:eq} in Eq.~\eqref{eq:e5omegaqq}, we finally obtain the form of the anomalous Hall response as
\begin{align}
-({\Omega}_{\mathbf{qq}}\dot{\mathbf{q}}_c)^i &= -\epsilon^{lm z}d_l(\partial_{q_i} d_m)M_1^{(t)}/\vert\mathbf{d}\vert^3,
\label{eq:e5omegaqqfin}
\end{align}
which is exactly equal to the gradient correction to the velocity due to the time-dependent term $\bm{\Omega}_{t\mathbf{k}}$ in Eq.~\eqref{eq:omegaktwsm1}.

Hence, we have shown that the interpretation of the $\mathbf{E}_5$ as an axial electric field leading to a Hall response is equivalent to the 
generalized Berry curvature-like correction $\bm\Omega_{t\mathbf{k}}$ in Eq.~\eqref{eq:sceomr}. It is interesting to note that this geometrical contribution is independent of the chirality of the Weyl node. 
This is a consequence of the axial nature of $\mathbf{E}_5$; the axial electric field and Berry curvature have different signs at the two Weyl nodes and thus conspire to give the same effective Hall drift.

\subsubsection{Numerical analysis}

\begin{figure}
\includegraphics[width=\columnwidth]{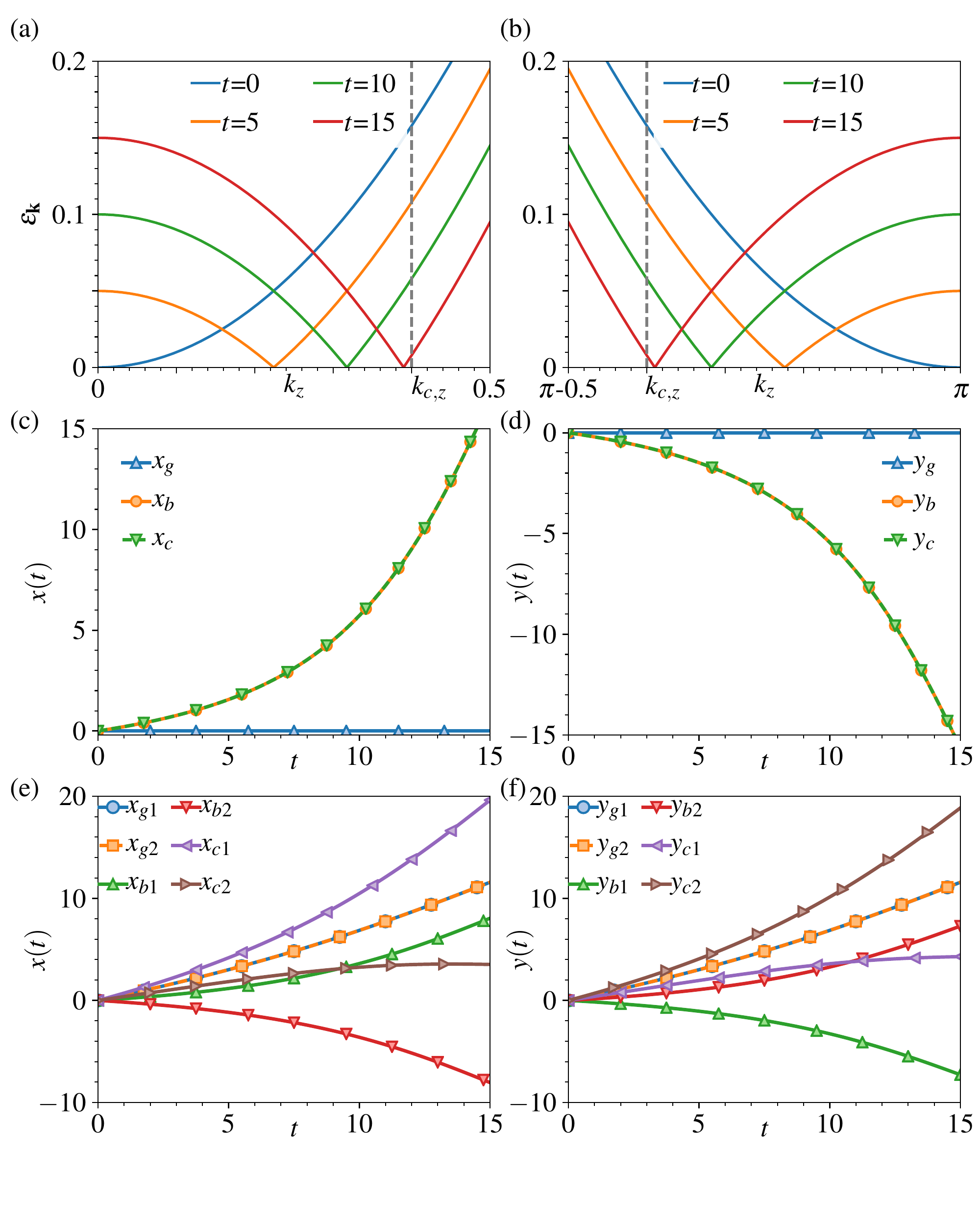}
\caption{{\bf{Differential measurement to extract the geometric contribution.}} (a) Parametric dependence of the spectrum and the Weyl node locations on time for $M_0=-2$, $M_1^{(t)}=0.01$, and $k_x=\pi/2=k_y$. The vertical dashed line corresponds to the wavepacket momentum $k_{c,z}$, which is a constant of motion. (b) Same as (a) but with $M_0=2$ and $M_1^{(t)}=-0.01$. (c) The different contributions to $x_{c}(t)$ as a function of $t$ for $k_x=\pi/2$, $k_y=\pi/2+0.05$ and $k_z=0.4$  with $M_0=-2$, $M_1^{(t)}=0.01$. (d) The same as (c) but for $y(t)$ with $k_x=\pi/2+0.05$ and $k_y=\pi/2$. (e)-(f) Robust extraction of the geometric contribution using the two protocols discussed in the main text. Note that the group velocity contribution for the two protocols is identical (blue circles and orange squares), where as the Berry velocity contribution is exactly opposite (green up and red down triangles). Hence the net displacements for the two protocols (purple left and brown right triangles) can be subtracted to obtain the geometric contribution. }
\label{fig:e5halldrift}
\end{figure}

We now corroborate these results by studying wavepacket trajectories obtained by solving the equations of motion \eqref{eq:sceomr} for a lattice model. 
The parametric dependence of the energy spectrum (including the Weyl node location) on time for $M_1^{(t)}\ne0$ is shown in Fig.~\ref{fig:e5halldrift}{(a)-(b)} \footnote{The parabolic spectrum at $t=0$ is an artifact of our initial condition $M_0=2$, and is inconsequential to our results as for any $t>0$, the system is a Weyl semimetal.}.
Without loss of generality, we consider the initial conditions $\mathbf{r}_c(t=0)=0$. 
The trajectories at any later time are then given by $\mathbf{r}_c(t) = \int_0^t dt'~[\dot{\mathbf{r}}_g(t')+\dot{\mathbf{r}}_b(t')]$, where $\dot{\mathbf{r}}_g(t)=\bm{\nabla}_\mathbf{k}\varepsilon_\mathbf{k}$ is the group velocity and $\dot{\mathbf{r}}_b(t)=\bm{\Omega}_{t\mathbf{k}}$.
Their explicit forms for the model \eqref{eq:hamwsm1rb} are
\begin{align}
\dot{x}_{g} = \frac{-2 \cos k_x \sin k_y}{\sqrt{4-4 \sin k_x \sin k_y+(M(t)+2\cos k_z)^2}},\nonumber\\
\dot{y}_{g} = \frac{-2 \sin k_x \cos k_y}{\sqrt{4-4 \sin k_x \sin k_y+(M(t)+2\cos k_z)^2}},\label{eq:rgdotwsm1}\\
\dot{z}_{g} = \frac{-2 \sin k_z (M(t)+2\cos k_z)}{\sqrt{4-4 \sin k_x \sin k_y+(M(t)+2\cos k_z)^2}},\nonumber
\end{align}
and
\begin{align}
\dot{x}_{b} &= \frac{2 M_1^{(t)} (1 - \cos k_y - \sin k_x \sin k_y)}{[4-4 \sin k_x \sin k_y+(M(t)+2\cos k_z)^2]^{3/2}},\nonumber\\
\dot{y}_{b} &=\frac{2 M_1^{(t)} (1 + \cos k_x - \sin k_x \sin k_y)}{[4-4 \sin k_x \sin k_y+(M(t)+2\cos k_z)^2]^{3/2}},\label{eq:rbdotwsm1}\\
\dot{z}_{b} &= 0,\nonumber
\end{align}
where $\mathbf{r}_{g} = (x_g,y_g,z_g)$ and similarly for $\mathbf{r}_b$.
The following important observations can be made from the expressions in Eqs.~\eqref{eq:rgdotwsm1} and \eqref{eq:rbdotwsm1}. 
First we note that $\dot{\mathbf{r}}_b$ is invariant under $k_z\rightarrow -k_z$, which is indicative of the fact that the anomalous Hall velocity is the same for the two Weyl points. 
This is consistent with what was deduced from the linearized regime by the absence of any $\eta$ dependence in Eq.~\eqref{eq:e5omegaqqfin}. It is indeed a signature that the two Weyl nodes experience the effective electric field with different signs. Further, since the anomalous Hall velocity is always perpendicular to the effective electric field, which in our case is along $z$, we have $\dot{z}_{b}=0$. \\

To isolate the effect of $\mathbf{E}_5$ via the wavepacket trajectories, it is necessary to extract the geometric contribution to the wavepacket dynamics.
Thus the probing protocols should be such that the effect of the the group velocity $\dot{\mathbf{r}}_g$ is factored out of the dynamics~\cite{price2012mapping}.
This can be achieved by preparing a wave-packet such that its momentum $k_{c,x}$ minimizes the group velocity contribution.
From Eq.~\eqref{eq:rgdotwsm1}, it can be deduced that if the momentum of the wavepacket is $k_{c,x}=\pi/2$ but $k_{c,y}\neq\pi/2$, then $\dot{x}_{g}=0$. 
Hence the entire contribution to $x_{c}(t)$  is determined by $(\bm{\Omega}_{t\mathbf{k}})^x$, which was shown to encode information about $\mathbf{E}_5$ via the equivalence of Eqs.~\eqref{eq:omegaktwsm1} and \eqref{eq:e5omegaqqfin}. 
The result for such a protocol is shown in Fig.~\ref{fig:e5halldrift}(c). The group velocity contribution $x_g(t)$ stays zero for all times, and $x_c(t)=x_b(t)\neq 0$.
Note that the velocity increases as the Weyl node moves closer to the wavepacket momentum (see Fig.~\ref{fig:e5halldrift}(a)). 
This is indicative of the fact that the response due to $\bm{\Omega}_{t\mathbf{k}}$ is equivalent to the response due to the interplay of the emergent field $\mathbf{E}_5$ and the Berry curvature ${\Omega}_{\vk\vk}$. Although the momentum of the wavepacket $\mathbf{k}_c$ is a constant of motion, the effective Berry curvature experienced by the wavepacket increases in magnitude as the location of the Weyl node moves closer to $\mathbf{k}_c$, and it diverges when they coincide.
A similar protocol can be carried out for $y_c(t)$ with the choice $k_{c,y}=\pi/2$ and $k_{c,x}\neq\pi/2$, the results of which are shown in Fig.~\ref{fig:e5halldrift}(d).
%
%%%%%%%%%%%%%%%%%%%%%%%%%%%%%%%%%%%%%%%%%%%%%%%%%%%%%%%%%%%
%%%%%%%%%%%%%%%%%%%%%%%%%%%%%%%%%%%%%%%%%%%%%%%%%%%%%%%%%%%
\\

Although the above steps isolate the geometric contribution, it might be more desirable from an experimental point of view to have a protocol that does not depend on the preparation of the wavepacket at precise momenta.
To this end, we now discuss a procedure which relies on monitoring the evolution for two distinct situations: 
A differential measurement between the two will cancel the effect of the group velocity and will enable the extraction of the  geometrical contribution~\cite{price2012mapping}.

For concreteness, consider preparing a wavepacket with momentum close to the Weyl node at $k_{W,z}\in[0,\pi]$. 
We will study its time evolution first when $M_1^{(t)}>0$, denoting the corresponding trajectories as $\mathbf{r}_{c1}(t)=\mathbf{r}_{g1}(t)+\mathbf{r}_{b1}(t)$
and second when the sign of $M(t)$ is flipped, {\textit{i.e.}} $M_0\rightarrow -M_0$ and $M_1^{(t)}\rightarrow -M_1^{(t)}$, labelling the trajectories as $\mathbf{r}_{c2}(t)=\mathbf{r}_{g2}(t)+\mathbf{r}_{b2}(t)$. Note that in the second case, the Weyl node then shifts to $\pi-k_{W,z}$, hence the wavepacket is then prepared with $\pi-k_{c,z}$.
The two situations are shown in Fig.~\ref{fig:e5halldrift}(a) and (b). 
From Eqs.~\eqref{eq:rgdotwsm1} and \eqref{eq:rbdotwsm1}, it can be deduced that $x_{g1}(t)= x_{g2}(t)$, whereas  $x_{b1}(t)= -x_{b2}(t)$. Hence, the geometric contribution may be isolated by simply subtracting the responses of the two protocols as $x_b(t)=[x_{c1}(t)-x_{c2}(t)]/2$ (the coordinate $y(t)$ follows analogously, unlike $z(t)$). The numerical results in Fig.~\ref{fig:e5halldrift}(e)-(f) confirm that the group velocity contributions are identical (blue circles and orange squares) while the Berry velocity contribution are opposite (green up and red down triangles) and thus a differential measurement will isolate the geometric contribution $\mathbf{r}_b$.

To summarize this subsection, we have shown that the geometric Hall-like response due to $\mathbf{E}_5$ can be interpreted via the generalized semiclassical equations of motion with gradient corrections for the temporal dependence in the Hamiltonian, and the results have been corroborated with exact wavepacket trajectories calculated using a lattice model with discussions of possible experimental protocols to isolate the geometric contribution. 

\subsection{Cyclotron orbits and pseudo-chiral magnetic effect due to $\mathbf{B}_5$}

In this subsection, we consider $M_1^{(x)}\!\ne\! 0$ and $M_1^{(t)}\!=\!0$ such that $\mathbf{B}_5$ is finite and $\mathbf{E}_5=0$; see Eqs.~\eqref{eq:hamwsm1re5b5}. We begin by showing that the semiclassical equations of motion with the gradient corrections, \eqref{eq:sceomr} and \eqref{eq:sceomk}, can be recast in the form of the usual semiclassical equations, \eqref{eq:sceomrbare} and \eqref{eq:sceomkbare}, in the presence of a magnetic field.  
We then study the corresponding cyclotron orbits analytically in the linearized regime and compare them to exact numerical simulations 
of the wave-packet evolution, confirming their validity.
In the process, we discuss the geometrical contribution to the motion of the wave-packet, which is related to the so-called pseudo-chiral magnetic effect~\cite{Zhou2013,grushin2016inhomogeneous,Landsteiner2016}. 
We end the section by discussing the effects due to inhomogeneous profiles of $\mathbf{B}_5$ that are to be expected in realistic experimental situations.

\subsubsection{Analytical description}

We start with a Weyl semimetal Hamiltonian linearized around the two Weyl nodes, \eqref{eq:hamwsm1rlin}, and as in Sec.~\ref{subsec:e5}, we consider a change of reference frame for the momentum:  
\begin{equation}
\mathbf{q} = \mathbf{k}-\eta\mathbf{b}(\mathbf{r}),~~~\eta=\pm1,
\label{eq:refchangeB5}
\end{equation}
where the Weyl nodes are located at $\eta\mathbf{b}$ and  $\mathbf{b}(\mathbf{r})$ is the Weyl node location depending parametrically on the position $\bf{r}$. 
In the absence of an explicit time dependence, the semiclassical equations of motion with the gradient corrections, \eqref{eq:sceomr} and \eqref{eq:sceomk}, can be expressed as 
\begin{align}
\dot{r}_{c,i} &= \frac{\partial \varepsilon_\mathbf{k}}{\partial k_i} - \Omega_{k_i r_j}\dot{r}_{c,j} - \Omega_{k_i k_j}\dot{k}_{c,j}, \label{eq:sdscr}\\
\dot{k}_{c,i} &=- \frac{\partial \varepsilon_\mathbf{k}}{\partial r_{c,i}} + \Omega_{r_i r_j}\dot{r}_{c,j} + \Omega_{r_i k_j}\dot{k}_{c,j}. \label{eq:sdsck}
\end{align}
To make the analogy to the axial magnetic field $\mathbf{B}_5$, we first look at the transformation of the various Berry curvature-like terms under the change of reference 
\eqref{eq:refchangeB5}:
\begin{eqnarray}
\Omega_{k_i k_j} &=& \Omega_{q_i q_j},\label{eq:omegaqqtrans}\\
\Omega_{k_i r_j} &=& i\left[\left\langle\frac{\partial u}{\partial k_i}\right\vert\left.\frac{\partial u}{\partial r_{c,j}}\right\rangle-\left\langle\frac{\partial u}{\partial r_{c,j}}\right\vert\left.\frac{\partial u}{\partial k_i}\right\rangle\right]\nonumber\\
&=& i\left[\left\langle\frac{\partial u}{\partial q_i}\right\vert\left.\frac{\partial u}{\partial q_l}\right\rangle-\left\langle\frac{\partial u}{\partial q_l}\right\vert\left.\frac{\partial u}{\partial q_i}\right\rangle\right]\frac{\partial q_l}{\partial r_{c,j}}\nonumber\\
&=&-\Omega_{q_i q_l}(\partial_{r_{c,j}}b_l)\eta,\label{eq:omegaqrtrans}\\
\Omega_{r_i r_{c,j}}&=&-\Omega_{r_i q_l}(\partial_{r_{c,j}}b_l)\eta.\label{eq:omegarrtrans}
\end{eqnarray}
Using Eqs.~\eqref{eq:omegaqqtrans}-\eqref{eq:omegarrtrans} the equation of motion Eq.~\eqref{eq:sdscr} can be written as
\begin{align}  
\dot{r}_{c,i} &= \frac{\partial \varepsilon}{\partial q_i} +\eta \Omega_{q_i q_l}(\partial_{r_{c,j}} b_l)\dot{r}_{c,j} - \Omega_{q_i q_j}[\dot{q}_{c,j} +\eta\dot{r}_{c,l} \partial_{r_l}b_j]\nonumber\\
&= \frac{\partial \varepsilon}{\partial q_i} - \Omega_{q_i q_j}\dot{q}_{c,j} \label{eq:sdscrt}
\end{align}
Similarly, the equation of motion for the momentum, \eqref{eq:sdsck}, can be recast as
\begin{align}
\dot{k}_{c,i} &= - \frac{\partial \varepsilon}{\partial q_j}\frac{\partial q_j}{\partial r_{c,i}} -\eta\Omega_{r_i q_l}(\partial_{r_{c,j}}b_l)\dot{r}_{c,j} \nonumber\\&~~~~~~~~~~~~~~+\Omega_{r_i q_j}[\dot{q}_{c,j}+\eta\dot{r}_{c,l}\partial_{r_l}b_j]\nonumber\\
&= \eta\frac{\partial \varepsilon}{\partial q_j}\frac{\partial b_j}{\partial r_{c,i}}  -\eta\Omega_{q_l q_j}\dot{q}_{c,j}\partial_{r_{c,i}}b_l=\eta\dot{r}_{c,l}\partial_{r_{c,i}}b_l.\label{eq:sdsckttemp}
\end{align}
Using the relation $\dot{k}_i = \dot{q_i}+\eta\dot{r}_l\partial_{r_{c,l}}b_i$ from Eq.~\eqref{eq:refchangeB5}, one obtains
\begin{align}
\dot{q}_{c,i}& = \eta\dot{r}_{c,l}(\partial_{r_{c,i}}b_l - \partial_{r_{c,l}}b_i)\nonumber\\
&= \eta[\dot{\mathbf{r}}_c\times(\bm{\nabla}\times\mathbf{b})]_i = \eta[\dot{\mathbf{r}}_c\times\mathbf{B}_5]_i \label{eq:sdsckt}
\end{align}
Comparing with Eqs.~\eqref{eq:sceomrbare} and \eqref{eq:sceomkbare}, it is apparent that, in terms of the shifted momentum $\mathbf{q}$, Eqs.~\eqref{eq:sdscrt} and \eqref{eq:sdsckt} take the form of the usual semiclassical equations~\cite{xiao2010berry} where the role of the magnetic field $\mathbf{B}$ is played by the axial magnetic field $\mathbf{B}_5$.
Together with the equivalence of Eqs.~\eqref{eq:omegaktwsm1} and \eqref{eq:e5omegaqqfin} regarding the axial electric field, we have established that the moving frame, defined in Eqs.~\eqref{eq:sceomrbare} and \eqref{eq:sceomkbare}, is convenient to describe axial gauge fields.

Having established that a spatial variation in the Weyl node separation does indeed lead to an effective axial magnetic field, we expect cyclotron orbits to occur.
To study these, we discuss the semiclassical equations of motion analytically in the linearized regime which will serve to analyze the exact 
numerical simulations of the wave-packet evolution.
We will discuss as well the observable imprints of a pseudo-chiral magnetic effect~\cite{Zhou2013,grushin2016inhomogeneous,Landsteiner2016}.

We start with the linearized Hamiltonian \eqref{eq:hamwsm1rlin}, also expanded to leading order in $M_1^{(x)}$ as
\begin{align}
\mathcal{H}_\eta=(&k^\prime_x-k^\prime_y)\sigma^x-(k^\prime_x+k^\prime_y)\sigma^y\nonumber\\
&-\eta v[k_z-\eta(\beta_0+\beta_1x)]\sigma^z,\label{eq:hamwsm1rlinb5}
\end{align}
where $v=\sqrt{4-M_0^2}$, $\beta_0=\cos^{-1}(-M_0/2)$, and $\beta_1=M_1^{(x)}/\sqrt{4-M_0^2}$.
Using Eq.~\eqref{eq:e5b5definition} the effective magnetic field is along $y$ and it is given by $\mathbf{B}_5 = \beta_1 \hat{y}$

The motion of the wavepacket in the presence of such axial magentic field separates into a cyclotron orbit in the $(x,z)$ plane,
and an unusual motion along the axial magnetic field direction ($\hat{y}$).
The latter has a  trivial contribution due to the band velocity, but also a geometrical contribution that we will shortly associate to a pseudo-chiral magnetic effect. 
To factor out the trivial band structure contribution 
we use a similar protocol as in the previous section; we set the wavepackets' $k_y^\prime=0$ so as to nullify the group velocity along $y$. 
The inspection of the equations of motion (\eqref{eq:sdscrt} and \eqref{eq:sdsckttemp}) reveals that $\dot{k}_y=0=\dot{k}_z$, 
which is consistent with the fact that since translation symmetry is broken only along $x$, only $k_x$ ceases to be a constant of motion. 
Choosing the initial conditions as $\mathbf{r}_c(t=0)=0$ and $k_{c,x}(t=0)=k_{x,0}$, the equations of motion can be solved analytically to obtain the cyclotron orbits as
\begin{align}
v^2\left[x_c(t)-\eta\frac{k_{c,z}-\eta\beta_0}{\beta_1}\right]^2+2\left[z_c(t)+\eta\frac{k_{x,0}^\prime}{\beta_1}\right]^2=\nonumber\\
\frac{1}{\beta_1^2}\left[2k_{x,0}^{\prime 2}+v^2(k_{c,z}-\eta\beta_0)^2\right].
\label{eq:cyclob}
\end{align}
Note that the cyclotron orbits have opposite chiralities for the two Weyl nodes which is indicative of the fact that the two Weyl nodes feel an opposite effective magnetic field. Also the anisotropic group velocities of the Weyl node lead to elliptical orbits with the semi-major and semi-minor axes proportional to $1/\beta_1\propto 1/M_1^{(x)}$; this is consistent with the fact that increasing effective magnetic field causes the cyclotron orbits to become smaller. 

The solution to the equations of motion also show a ballistic motion of the wavepacket along $y$ as
\begin{equation}
y_c(t) = \frac{2 v\beta_1t}{v^2(k_{c,z}-\eta\beta_0)^2+2k_{x,0}^{\prime 2}}.
\label{eq:pcme}
\end{equation}
Unlike the motion of the wave-packet along a magnetic field $\bf{B}$, the motion along the direction of the 
axial field is independent of the node's chirality $\eta$ and is solely due to $\mathbf{B}_5=\beta_1\hat{y}$.
Such wavepacket motion is inherited from the axial field analog of the chiral magnetic effect~\cite{Kharzeev2014,Zyuzin2012a,Fukushima2008,Zyuzin2012,Zhou2013,Grushin2012,Goswami:2013jp,Landsteiner2016}, which has been termed the pseudo-chiral magnetic effect~\cite{Zhou2013,grushin2016inhomogeneous,Landsteiner2016}.
For each node the axial field induces a current parallel and proportional to it ($\mathbf{j}\parallel\mathbf{B}_5$).
Physically, this contribution can be reinterpreted as a magnetization current~\cite{grushin2016inhomogeneous,Huang2017} and is consistent with the fact that $\bf{b}$ itself breaks time-reversal symmetry; note it enters as a Zeeman magnetization coupling in Eq.~\eqref{eq:hamweyllowenergy}). Thus $\bf{B}_5 = \nabla\times\bf{b}$ is physically the curl of a magnetization which is by definition a magnetic (or bound) current~\cite{grushin2016inhomogeneous}.

Although the focus of the present work is on the realization of pseudo electric and mangetic fields, it is important to stress that the results in this section highlight how the dynamical nature of experiments with ultracold atoms is well suited to probe also the chiral magnetic effect originating from a magnetic field $\bf{B}$. The difficulty of probing this effect in a solid-state set-up is that it vanishes in equilibrium~\cite{Vazifeh:2013fe,Land14}, unlike the pseudo chiral magnetic effect which is a magnetization current~\cite{grushin2016inhomogeneous}. Thus, our results suggest that the chiral magnetic effect can be probed by studying the cyclotron orbits of a wavepacket in the presence of a synthetic magnetic field, in an analogous way to the results presented in this section.

\begin{figure}
\includegraphics[width=\columnwidth]{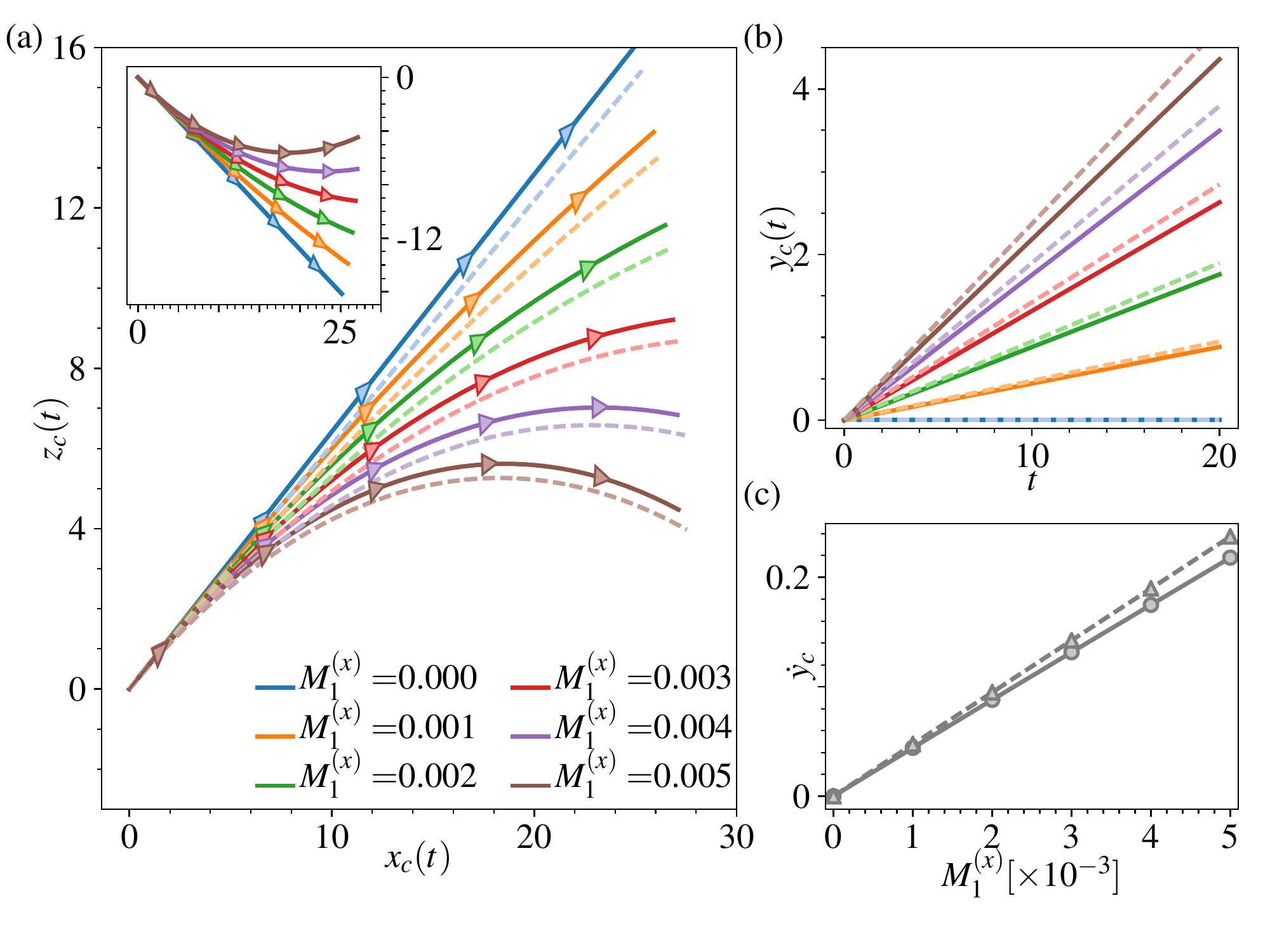}
\caption{{\bf{Cyclotron orbits and chiral pseudo-magnetic effect due to $\mathbf{B}_5$.}} (a) Cyclotron orbits in the $(x,z)$ plane. The solid lines correspond to the exact results with the arrows representing the direction of time. The corresponding dashed show the trajectories from the linearized result \eqref{eq:cyclob}. On increasing the effective $\mathbf{B}_5$, the cyclotron orbits shrink. The main figure corresponds to $\eta=1$, where as the inset shows $\eta=-1$. (b) Manifestation of the chiral magnetic effect due to $\mathbf{B}_5$ represented by a ballistic motion of the wavepacket along $y$. The corresponding velocities, which grow linearly with $M_1{(x)}$, are shown in (c). The dashed lines correspond to the linearized result \eqref{eq:pcme}. The parameters for the plots are $k_{x,0}=1.7,~k_y=\pi/2,~k_z=1.1$, and $M_0=-1$.}
\label{fig:cyclo}
\end{figure}

\subsubsection{Numerical analysis: constant $\bf{B}_5$}

We corroborate the above results by solving the equations of motion numerically for the full lattice model beyond the linearized regime. 
The results are shown in Fig.~\ref{fig:cyclo}. Although there are slight deviations due to the non-linear effects of the lattice, the qualitative behavior is rather similar suggesting that the validity of the interpretations beyond the linearized regime. 
The essential features of the dynamics of the wavepacket remain the same, namely the cyclotron orbits become smaller on increasing $M_1^{(x)}$ and there is ballistic motion along $y$, the velocity of which also linearly increases with $M_1^{(x)}$. 
Also note that, between Fig.~\ref{fig:cyclo}(a) and its inset, the wavepacket is prepared with $k_{c,z}=\eta(\beta_0+\delta k_z)$ but the same $k_{x,0}$. Hence, from the Eq.~\ref{eq:cyclob}, one expects that the $x$ component of the trajectories are the same between the two, whereas the $z$ component is opposite. This leads to a different chirality of the cyclotron orbit between the two Weyl nodes, as seen in Fig.~\ref{fig:cyclo}(a). 

\subsubsection{Numerical analysis: inhomogeneous $\bf{B}_5$}

So far, our discussion assumed that $\mathbf{B}_5$ was taken to be finite and constant (to linear order) everywhere within the sample. 
However $\mathbf{B}_5$ is by construction a bounded field and thus every region with $\mathbf{B}_5>0$ must be compensated with regions where $\mathbf{B}_5<0$, even at linear order \cite{Landsteiner2016}.
This implies that, if $\mathbf{B}_5$ is taken as a positive constant in the bulk, 
as in our previous considerations, the boundaries of the sample must be compensate with $\mathbf{B}_5<0$.
This fact has been used recently to re-interpret the topological surface states of Weyl semimetals, the Fermi arcs, as zeroth pseudo-Landau levels of $\bf{B}_{5}$.
Our results above are therefore valid for wavepackets that have an average center of mass where $\bf{B}_{5}$ is a constant, so that edge-effects and inhomogeneous contributions to 
$\bf{B}_5$ can be safely disregarded.  

\begin{figure}
\includegraphics[width=\columnwidth]{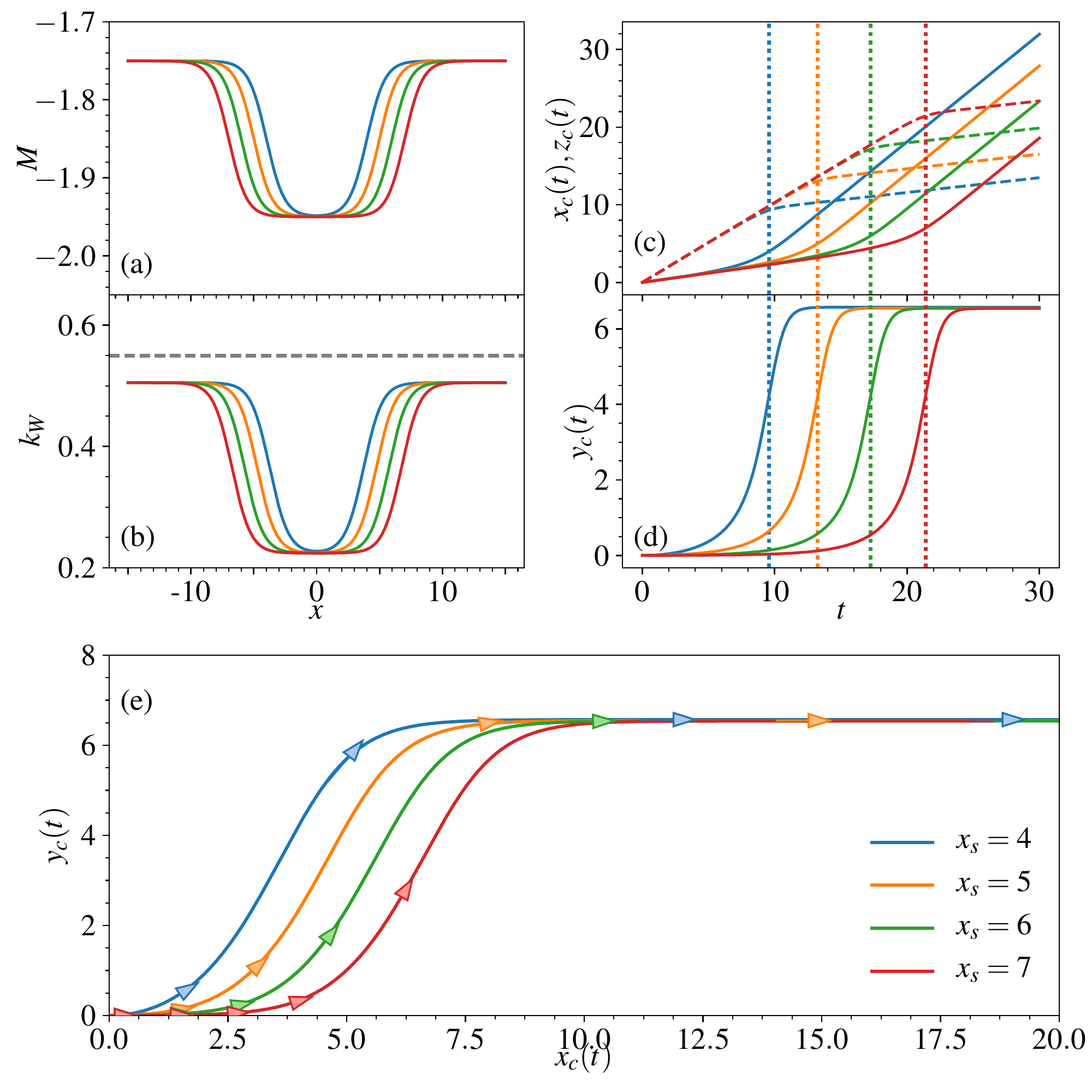}
\caption{{\bf{Configurations for inhomogeneous $\mathbf{B}_5$ and resulting wavepacket trajectories. }}(a) Profile of $M(x)$ for different values of $x_s$. Throughout the figure, data with the same color corresponds to the same value of $x_s$.   (b) Location of Weyl node as a function of $x$. The horizontal dashed corresponds to $k_{c,z}$ of the wavepacket. (c) Drift of the wavepacket along $x$ (solid) and $z$ (dashed) as a function of time. Vertical lines represent times of maximum velocity along $y$. (d) Drift of the wavepacket along $y$ as a function of time. Vertical lines are the same as (c) which denote the times where $x_c(t)=x_s$. (e) Trajectories of the wavepacket with the arrows showing the direction of time. The parameters are $k_{x,0}=1.6,~k_y=\pi/2,~k_z=0.55~, M_0=-1.95,~f=0.1$, and $\alpha=0.7$. }
\label{fig:cpb5}
\end{figure}

However, in realistic experimental set-ups the vector $\mathbf{b}$ will be finite only within a local spatial region and zero otherwise.
This implies that $\bf{B}_5$ will be localized at specific, narrow regions of the system.
In solid state systems it is the boundary with vacuum which will impose such discontinuity in $\mathbf{b}$.
In cold atomic systems the atomic trap potential can act as a boundary, but it will typically impose a smooth, step-like profile 
of the Weyl node separation.
Within the two-band model \eqref{eq:hamwsm1rb} such profile in $\mathbf{b}$ can be modelled by 
\begin{equation}
M(x)\equiv M_0+ f [\tanh(\alpha (x - x_s))- \tanh(\alpha (x +x_s))],
\label{eq:mxtanh}
\end{equation}
and is plotted in Fig.~\ref{fig:cpb5}(a). Its corresponding Weyl node separation is given in Fig.~\ref{fig:cpb5} (b). 
Here, $\pm x_s$, $\alpha$ and $f$  control the location, sharpness and height of the step.

To study the effect of profile Eq.~\eqref{eq:mxtanh} on the wavepacket dynamics numerically, 
we consider a wavepacket with $k_{c,y}=\pi/2$ so as to avoid the effect of the band group velocity along $y$, as described in previous sections. 
It is illustrative to describe what is expected for a wavepacket released from an initial position $x_i$ with $M(x_i) \sim M_0$ and thus $\mathbf{B}_5$ equal to zero. 
Intuitively, this amounts to releasing a wave packet from the left of Fig.~\ref{fig:cpb5}(a) and monitoring its evolution as it encounters the step at $M(x_s)$ due to the profile Eq.~\eqref{eq:mxtanh}.
Before the wavepacket reaches the vicinity of $\pm x_s$, there is no drift of the wavepacket along the $y$ direction and its velocity along the $x$ and $z$ are simply given by the group velocities, which remain approximately constant in the vicinity of the Weyl nodes. 
As the wave packet reaches the vicinity of $\pm x_s$, 
the inhomogeneity of $M(x)$, or equivalently a finite $\bf{B}_5$ field along $y$, will affect its motion. 
Semiclassically, this induces a velocity along the $y$ direction by making the terms $\dot{k}_x$ and $\Omega_{k_y x}$ (see Eqs.~\eqref{eq:sdscr} and \eqref{eq:sdsck}) finite, as those are terms which rely on the derivative with respect to $x$ being finite.
During the course of the evolution we also expect that the velocities along $x$ and $z$ change as they reach $x_s$. 

As the wavepacket moves away from $x_s$, the drift along $y$ due to $\mathbf{B}_5$ stops and the wavepacket moves only in the $(x,z)$ plane. 

Equivalently, such description of the dynamics can also be phrased in terms of the pseudo-Landau level emerging due to $\mathbf{B}_5$~\cite{grushin2016inhomogeneous}. 
As the wavepacket reaches the vicinity of $x_s$, it finds a Landau level in the spectrum which disperses parallel to $B_5$ along $y$, imprinting a finite drift 
along the $y$ direction.
Following Ref.~\cite{grushin2016inhomogeneous}, the motion of the wave packet along $\hat{y}$ is thus the motion along the Fermi arc occurring between a system with $\bf{b}\neq 0$ and a system with $\bf{b}=0$. \\

\begin{figure}
\includegraphics[width=\columnwidth]{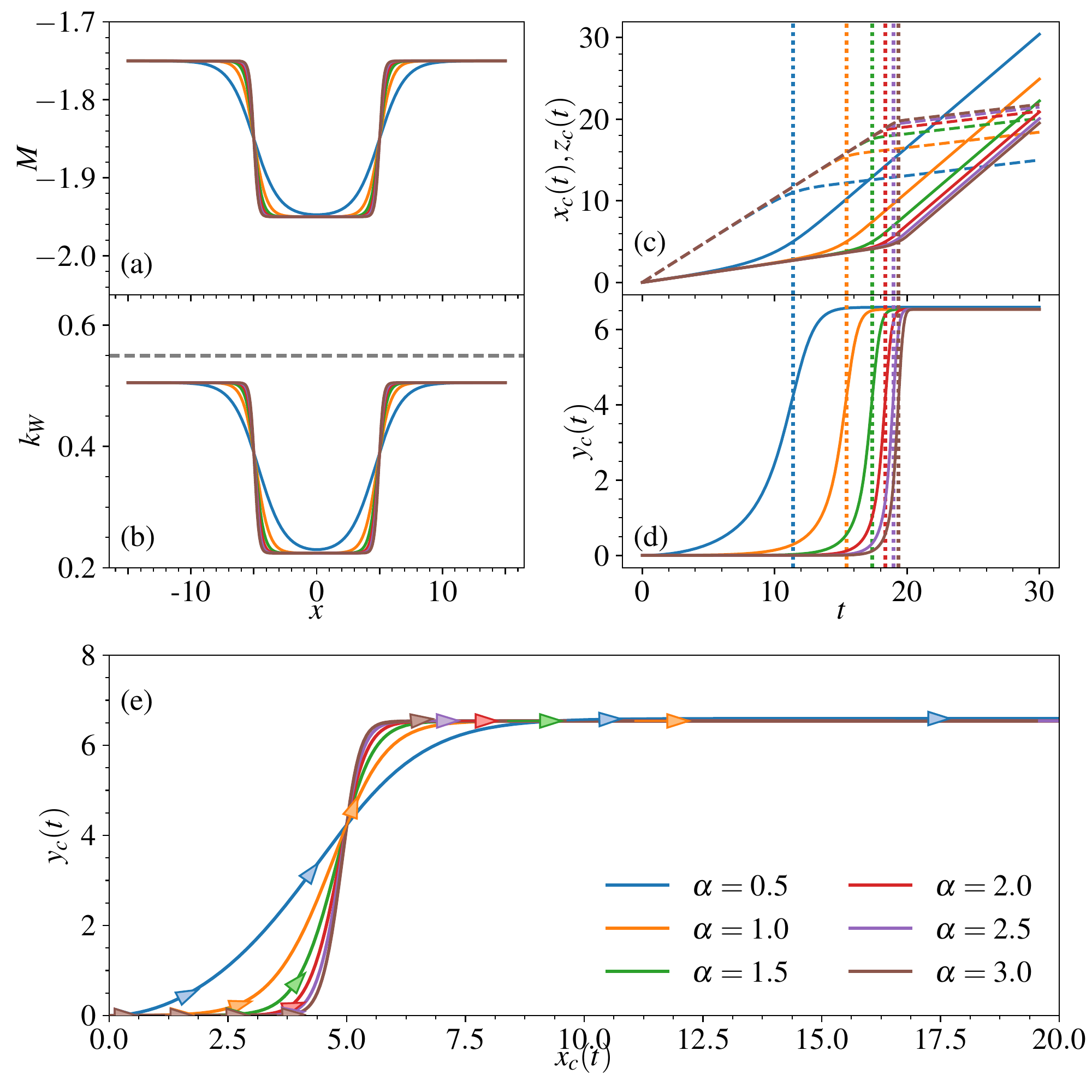}
\caption{Same as Fig.~\ref{fig:cpb5} except now $x_s=5$ where as the different colors represent different values of $\alpha$. Note that increasing the sharpness of the imhomogeneity leads to the $y$ drift being sharper and happening sharply at the time when the $x$ coordinate of the wavepacket is closer to $x_s=5$.}
\label{fig:csb5}
\end{figure}

To confirm and extend this picture we have performed a numerical analysis and study the motion of the wave packet as a function of the step parameters $x_s$, $\alpha$ and $f$. 
The semiclassical trajectories for different step positions, set by $x_s$ are shown in Fig.~\ref{fig:cpb5}. 
The wavepacket starts without any velocity along $y$ and shows no drift until it reaches the vicinity of $x_s$. This can be seen directly from Fig.~\ref{fig:cpb5}(c). The wavepacket drift along the $y$ direction starts to be finite for later times for larger values of $x_s$. In (c) and (d), the vertical lines corresponds to the times where the velocity along the $y$ is the maximum, and from the horizontal lines in (d) it can be seen that at these times, the $x$ component of the wavepacket position is precisely $x_s$. Also at these times, the velocity along $x$ and $z$ also show a change as expected. These observations are further corroborated by the projection of the trajectory of the wavepacket on the $(x,y)$ plane as shown in (e). Note that here also, for smaller values of $x_s$ the wavepacket drifts along the $y$ direction earlier. Due to the motion along $x$, as the wavepacket moves past the region of inhomogeneity, the velocity along $y$ again goes to zero and the $y$ coordinate of the wavepacket flattens out with time.

Next we consider the trajectories for different values of $\alpha$, which controls the width or sharpness of the profile (see Fig.~\ref{fig:csb5} (a),(b)). 
The results are shown in Fig.~\ref{fig:csb5} (c)-(e); as $\alpha$ is increased, the wavepacket drift along $y$ becomes sharper as apparent in Fig.~\ref{fig:csb5}(c). 
From Fig.~\ref{fig:csb5}(e), as the profile of $M(x)$ converges towards a step function, the jump in the $y$ coordinate of the wavepacket also progressively moves closer to $x=x_s$ which is set to $x_s=5$.

\begin{figure}
\includegraphics[width=\columnwidth]{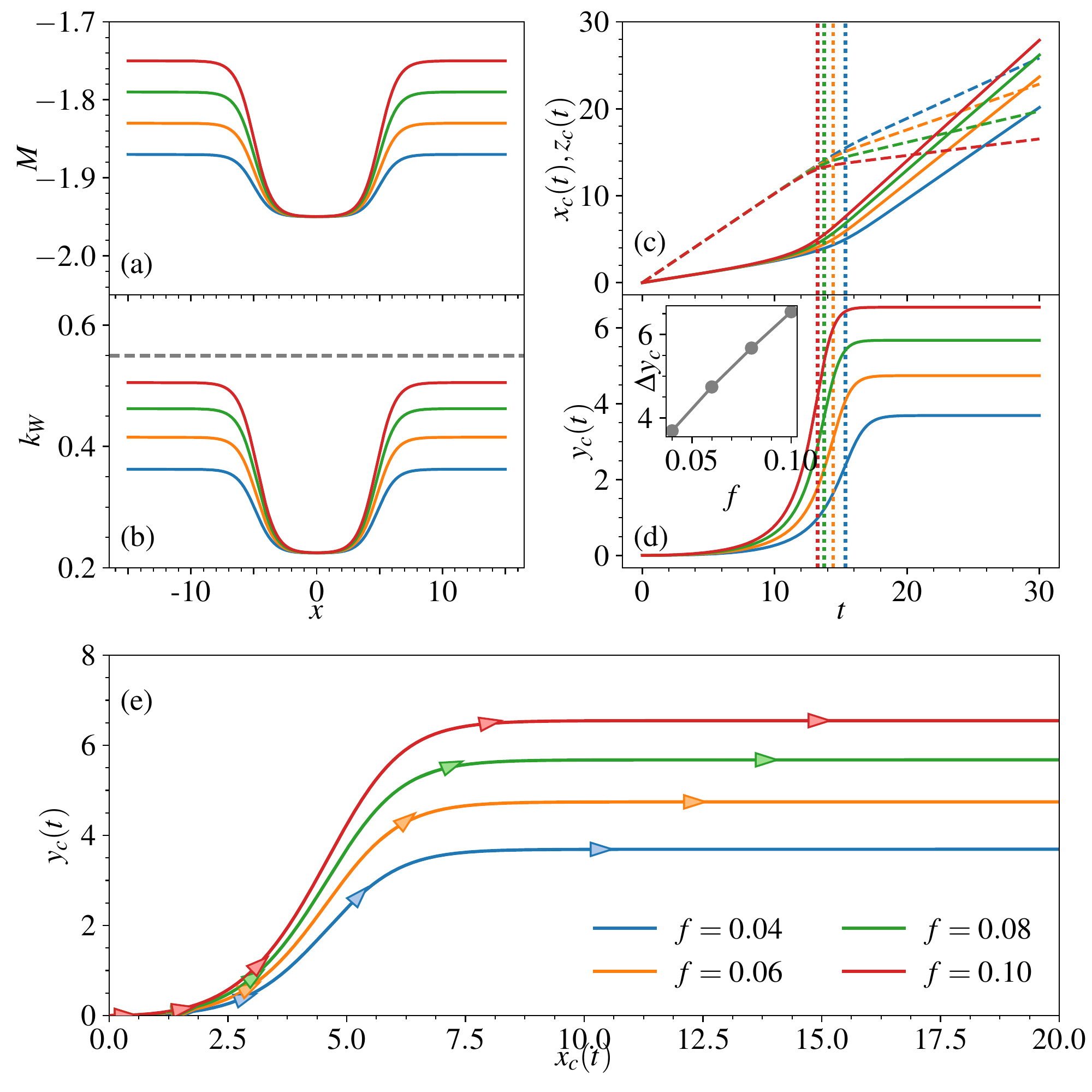}
\caption{Same as Fig.~\ref{fig:cpb5} except now $x_0=5$ where as the different colors represent different values of $f$. The inset in (d) shows the net displacement along $y$ as a function of $f\propto B_5$. The approximate linear behavior numerically corroborates the prediction of Eq.~\eqref{eq:dlybx}.}
\label{fig:cfb5}
\end{figure}

From Figs.~\ref{fig:cpb5} and \ref{fig:csb5} it is apparent that the net $y$-drift of the wavepacket, the total drift once the wave-packet has moved past the region of inhomogeneity along $x$ is the same irrespective of the details of the profile of $\mathbf{b}(x)$. 
This is a consequence of the fact that the net change in $\mathbf{b}$ which is controlled by the step height $f$ is, for all previous cases, the same across the inhomogeneity.
The effect of varying the parameter $f$ in Eq.~\eqref{eq:mxtanh} is shown in Fig.~\ref{fig:cfb5}.
The figure shows that the net $y$ drift is proportional to $f$. 
These observations suggests that if the wavepacket is prepared at a momentum where the group velocity along $y$ is zero and travels from $\mathbf{r}_i$ to $\mathbf{r}_f$
then the net $y$ drift is given by
%
%Say the wavepacket is prepared at a momentum where the group velocity along $y$ is zero and travels from $\mathbf{r}_i$ to $\mathbf{r}_f$. Then, we claim that 
%
\begin{equation}
	y_f - y_i \propto \int_{x_i}^{x_f}dx~ (\mathbf{B}_5(x))_y = b_z(x_f)-b_z(x_i).
	\label{eq:dlybx}
\end{equation}
which we now justify.

Consider the case where $b_z$ takes two different constant values in two spatial regions separated by the region of inhomogeneity, such that $b_z(x)$ is some smooth function interpolating between the two. 
As $b_z(x)$ is smooth, we will approximate by a series of linear segments of a width that we will take to be zero at the end of the calculation. 
Then, in order to show that Eq.~\eqref{eq:dlybx} indeed holds, we need to show that it holds for each of the linear segments. 
With the initial condition $\mathbf{r}_0=0$, the trajectory along $x$ from the solutions of Eqs.~\eqref{eq:sdscr} and \eqref{eq:sdsck} is given by
\begin{equation}
	x_c(t) = \frac{(k_{c,z} - \eta\beta_0) - (k_{c,z} -\eta\beta_0) \cos(\omega t) +\sqrt{2} \frac{k_{x,0}^\prime}{v} \sin(\omega t)}{\eta \beta_1},
	\label{eq:linearxt}
\end{equation}
where $\omega=\sqrt{2}\eta v\beta_1/\sqrt{v^2(k_{c,z}-\eta\beta_0)^2+2k_{x,0}^{\prime 2}}$.
The above equation indicates that the dynamics for an infinitesimally small linear segment is equivalent to considering very short time dynamics. 
Thus we expand the trajectory to linear order in $t$, resulting in
\begin{equation}
	x_c(t)\approx \frac{2k_{x,0}^\prime t}{\sqrt{v^2(k_{c,z}-\eta\beta_0)^2+2k_{x,0}^{\prime 2}}}.
\end{equation}
From time $t=0$ to $t$, the net $y$-drift from Eq.~\eqref{eq:pcme} is 
\begin{equation}
	y_c(t) = \frac{v\beta_1 x_c(t)}{k^\prime_{x,0}\sqrt{v^2(k_{c,z}-\eta\beta_0)^2+2k_{x,0}^{\prime 2}}}.
\end{equation}
This may be recast as
\begin{equation}
	y_c(t) = \frac{v [b_z(x_c(t))-b_z(x_c(t=0))]}{k^\prime_{x,0}\sqrt{v^2(k_{c,z}-\eta\beta_0)^2+2k_{x,0}^{\prime 2}}}.
\end{equation}
showing that Eq.~\eqref{eq:dlybx} holds for infinitesimal segments. 
Applying this argument sequentially results in Eq.~\eqref{eq:dlybx}. 

The short time approximation can be reinterpreted as the requirement that the length scale of the perturbation is short compared to the cyclotron radius. 
Note that in our units the cyclotron radius in Eq.~\eqref{eq:cyclob} is equivalent to the inverse cyclotron frequency, which in turn is the time scale entering the sines and cosines in the trajectory equations of Eq.~\eqref{eq:linearxt}. 
Therefore, the expansion in $t$ is only valid for times much smaller than the inverse cyclotron frequency. 
Alternatively, this statement can also be reinterpreted in the position language: the expansion is valid for an $x$-displacement that covers length scales 
much smaller than the cyclotron radius.

\section{\label{sec:conclusions}Conclusions}

In conclusion, we have presented evidence that cold atomic systems are ideal platforms for creating and probing axial gauge fields for engineered Weyl semimetals. We have demonstrated two realistic models that enable this. We have further shown that semiclassical wavepacket dynamics in these models exhibit a variety of features characteristic of the geometric properties of the Weyl spectrum. While some of these, such as the anomalous Hall responses and the cyclotron orbits, are two-dimensional responses embedded in the three-dimensional system, responses like the chiral pseudo-magnetic effect are exclusive to three dimensions. We particularly note that, while some of these predictions are known to be difficult to experimentally realize in condensed matter platforms, we propose how they may be readily probed by using the fundamentally dynamical nature of ultracold atoms.
For instance, the chiral magnetic effect~\cite{Kharzeev2014,Zyuzin2012a,Fukushima2008,Zyuzin2012,Zhou2013,Grushin2012,Goswami:2013jp,Landsteiner2016} vanishes in equilibrium~\cite{Vazifeh:2013fe,Land14} making the proposed synthetic platforms suitable for its direct detection.

Our work opens a number of future research directions. The tunability of these cold atom realizations should enable access to non-linear effects of the axial gauge fields, such as the chiral anomaly. Given the explicit experimental realizability of these models, they should serve as an ideal platform for helping to address experimentally how anomalies occur in the lattice and in particular the differences between the consistent and covariant anomaly representations~\cite{Land14,Landsteiner2016,Gorbar2017,Gorbar2017a}. 
Furthermore, unlike in condensed matter, both the real and the axial gauge fields may be taken to be strongly varying over the length of a unit cell. While this work has only addressed the semiclassical response, for which weak (in the Hofstadter sense) and slowly varying fields are assumed, recent work has shown that the chiral anomaly has a fractal nature inherited from the Hofstadter butterfly when these systems are placed in large electric and magnetic fields~\cite{Roy2016}.  
In addition, axial gauge fields naturally occur in interface regions between topological phases~\cite{grushin2016inhomogeneous}. The example we considered in the main text, which lead to cyclotron orbits in regions of finite axial magnetic field, could be generalized to more intricate scenarios where topological phases of different kind meet~\cite{Grushin2015,Tchoumakov2017}. 
Finally, the controllability of both disorder and interactions in ultracold atomic and molecular systems will yield the interesting experimental possibility of studying the effects of interaction and disorder on anomalous responses, which can help isolate and discern between intrinsic and extrinsic signatures on the negative magnetoresistance measurements in condensed matter Weyl semimetals~\cite{Arnold2016,Reis2016}.
In short, our theoretical work is provides a realistic basis from which one may explore multiple avenues towards realizing and probing axial gauge fields effects in synthetic systems. 
We thus expect it to be a good starting point to explore a rich phenomenology of novel physical effects in synthetic Weyl semimetals. 

\acknowledgments
We thank J. H. Bardarson, J. Behrends, R. Ilan, and D. Pikulin for enlightening discussions.
AGG was supported by the Marie Curie Programme under EC Grant agreement No. 653846. MK was
supported by Laboratory directed Research and Development
(LDRD) funding from Berkeley Laboratory, provided
by the Director, Office of Science, of the U.S.
Department of Energy under Contract No. DEAC02-
05CH11231, and from the U.S. DOE, Office
of Science, Basic Energy Sciences (BES) as part of the
TIMES initiative. NG is supported by the ERC Starting Grant TopoCold and by the FRS-FNRS (Belgium).

\appendix
\section{Anomalous Hall response due to $\mathbf{E}_5$ from semiclassical equations}\label{sec:app1}
In this Appendix, we show how the response due to $\bm{\Omega}_{t\mathbf k}$ in Eq.~\eqref{eq:sceomr} is equivalent to an anomalous Hall response due to $\mathbf{E}_5$ for a generic Weyl semimemtal model in the linearized regime.
Following Eq.~\eqref{eq:hamweyllowenergy}, we start with a generic low-energy effective model described by
\begin{equation}
\mathcal{H}_{\mathrm{WSM}}^{\text{eff}}  = \sum_{\eta=\pm 1} \left[ \sum_{i,j=x,y,z}\mathcal{D}^{(\eta)j}_{i}\sigma^i(k_j-\eta b_j(t))\right],
\end{equation}
where the time-dependence in $\mathbf{b}(t)$ leads to an $\mathbf{E}_5=\partial_t\mathbf{b}$. The Hamiltonian for the Weyl node corresponding to $\eta$ can be expressed as $\sum_{i=x,y,z}d^{(\eta)}_i\sigma^i$, where
$d^{(\eta)}_i = \mathcal{D}^{(\eta)j}_{i}(k_j-\eta b_j(t))$. Expressing the eigenspinors $\vert u\rangle$ in Eq.~\eqref{eq:omegatk} in terms of $\mathbf{d}^{(\eta)}$, one finds that 
\begin{equation}
(\bm{\Omega}_{t\mathbf k})^i = \eta\epsilon^{lmn}d_l\mathcal{D}_m^{(\eta)i}\mathcal{D}_n^{(\eta)j}\partial_t{b}_j/\vert\mathbf{d}^{(\eta)}\vert^3.
\label{eq:app1}
\end{equation}

Now, consider the change of reference for momenta as
\[\mathbf{q} = \mathbf{k}-\eta\mathbf{b}(t).\]
In this modified reference frame, the anomalous Hall response is $-\Omega_{\mathbf{q}\mathbf{q}}\dot{\mathbf{q}}$, which can be expessed in terms of $\mathbf{d}^{(\eta)}$ as
\begin{eqnarray}
-(\Omega_{\mathbf q \mathbf q}\dot{\mathbf{q}})^i &=& -\epsilon^{lmn}d_l(\partial_{q_i} d_m)(\partial_{q_j} d_n)\dot{q}_{j}/\vert\mathbf{d}\vert^3\nonumber\\
&=&-\epsilon^{lmn}d_l\mathcal{D}_m^{(\eta)i}\mathcal{D}_n^{(\eta)j}\dot{q_j}/\vert\mathbf{d}^{(\eta)}\vert^3,
\end{eqnarray}
which is indeed identical to the response due to $(\bm{\Omega}_{t\mathbf k})^i$ as calculated in Eq.~\eqref{eq:app1}.

While $\mathbf{k}$ is a constant of motion, $\mathbf{q}$ has an explicit time-dependence via $\mathbf{b}(t)$ which can be interpreted as coming from the axial electric field $\eta \mathbf{E}_5$ as $\dot{\mathbf{q}} = -\eta\partial_t\mathbf{b} = \eta \mathbf{E}_5$. 

Hence the response due to $\bm{\Omega}_{t\mathbf k}$ is shown to be equal to an anomalous Hall response due to $\mathbf{E}_5$.

\bibliography{refs}

\begin{thebibliography}{88}%
\makeatletter
\providecommand \@ifxundefined [1]{%
 \@ifx{#1\undefined}
}%
\providecommand \@ifnum [1]{%
 \ifnum #1\expandafter \@firstoftwo
 \else \expandafter \@secondoftwo
 \fi
}%
\providecommand \@ifx [1]{%
 \ifx #1\expandafter \@firstoftwo
 \else \expandafter \@secondoftwo
 \fi
}%
\providecommand \natexlab [1]{#1}%
\providecommand \enquote  [1]{``#1''}%
\providecommand \bibnamefont  [1]{#1}%
\providecommand \bibfnamefont [1]{#1}%
\providecommand \citenamefont [1]{#1}%
\providecommand \href@noop [0]{\@secondoftwo}%
\providecommand \href [0]{\begingroup \@sanitize@url \@href}%
\providecommand \@href[1]{\@@startlink{#1}\@@href}%
\providecommand \@@href[1]{\endgroup#1\@@endlink}%
\providecommand \@sanitize@url [0]{\catcode `\\12\catcode `\$12\catcode
  `\&12\catcode `\#12\catcode `\^12\catcode `\_12\catcode `\%12\relax}%
\providecommand \@@startlink[1]{}%
\providecommand \@@endlink[0]{}%
\providecommand \url  [0]{\begingroup\@sanitize@url \@url }%
\providecommand \@url [1]{\endgroup\@href {#1}{\urlprefix }}%
\providecommand \urlprefix  [0]{URL }%
\providecommand \Eprint [0]{\href }%
\providecommand \doibase [0]{http://dx.doi.org/}%
\providecommand \selectlanguage [0]{\@gobble}%
\providecommand \bibinfo  [0]{\@secondoftwo}%
\providecommand \bibfield  [0]{\@secondoftwo}%
\providecommand \translation [1]{[#1]}%
\providecommand \BibitemOpen [0]{}%
\providecommand \bibitemStop [0]{}%
\providecommand \bibitemNoStop [0]{.\EOS\space}%
\providecommand \EOS [0]{\spacefactor3000\relax}%
\providecommand \BibitemShut  [1]{\csname bibitem#1\endcsname}%
\let\auto@bib@innerbib\@empty
%</preamble>
\bibitem [{\citenamefont {{Polini}}\ \emph {et~al.}(2013)\citenamefont
  {{Polini}}, \citenamefont {{Guinea}}, \citenamefont {{Lewenstein}},
  \citenamefont {{Manoharan}},\ and\ \citenamefont {{Pellegrini}}}]{PGL13}%
  \BibitemOpen
  \bibfield  {author} {\bibinfo {author} {\bibfnamefont {M.}~\bibnamefont
  {{Polini}}}, \bibinfo {author} {\bibfnamefont {F.}~\bibnamefont {{Guinea}}},
  \bibinfo {author} {\bibfnamefont {M.}~\bibnamefont {{Lewenstein}}}, \bibinfo
  {author} {\bibfnamefont {H.~C.}\ \bibnamefont {{Manoharan}}}, \ and\ \bibinfo
  {author} {\bibfnamefont {V.}~\bibnamefont {{Pellegrini}}},\ }\bibfield
  {title} {\enquote {\bibinfo {title} {{Artificial honeycomb lattices for
  electrons, atoms and photons}},}\ }\href
  {http://adsabs.harvard.edu/abs/2013NatNa...8..625P} {\bibfield  {journal}
  {\bibinfo  {journal} {Nature Nanotechnology}\ }\textbf {\bibinfo {volume}
  {8}},\ \bibinfo {pages} {625--633} (\bibinfo {year} {2013})}\BibitemShut
  {NoStop}%
\bibitem [{\citenamefont {{Goldman}}\ \emph {et~al.}(2016)\citenamefont
  {{Goldman}}, \citenamefont {{Budich}},\ and\ \citenamefont
  {{Zoller}}}]{GBZ16}%
  \BibitemOpen
  \bibfield  {author} {\bibinfo {author} {\bibfnamefont {N.}~\bibnamefont
  {{Goldman}}}, \bibinfo {author} {\bibfnamefont {J.~C.}\ \bibnamefont
  {{Budich}}}, \ and\ \bibinfo {author} {\bibfnamefont {P.}~\bibnamefont
  {{Zoller}}},\ }\bibfield  {title} {\enquote {\bibinfo {title} {{Topological
  quantum matter with ultracold gases in optical lattices}},}\ }\href@noop {}
  {\bibfield  {journal} {\bibinfo  {journal} {Nature Physics}\ }\textbf
  {\bibinfo {volume} {12}},\ \bibinfo {pages} {639--645} (\bibinfo {year}
  {2016})}\BibitemShut {NoStop}%
\bibitem [{\citenamefont {Dalibard}\ \emph {et~al.}(2011)\citenamefont
  {Dalibard}, \citenamefont {Gerbier}, \citenamefont {Juzeli{\=u}nas},\ and\
  \citenamefont {{\"O}hberg}}]{dalibard2011colloquium}%
  \BibitemOpen
  \bibfield  {author} {\bibinfo {author} {\bibfnamefont {J.}~\bibnamefont
  {Dalibard}}, \bibinfo {author} {\bibfnamefont {F.}~\bibnamefont {Gerbier}},
  \bibinfo {author} {\bibfnamefont {G.}~\bibnamefont {Juzeli{\=u}nas}}, \ and\
  \bibinfo {author} {\bibfnamefont {P.}~\bibnamefont {{\"O}hberg}},\ }\bibfield
   {title} {\enquote {\bibinfo {title} {Colloquium: Artificial gauge potentials
  for neutral atoms},}\ }\href@noop {} {\bibfield  {journal} {\bibinfo
  {journal} {Reviews of Modern Physics}\ }\textbf {\bibinfo {volume} {83}},\
  \bibinfo {pages} {1523} (\bibinfo {year} {2011})}\BibitemShut {NoStop}%
\bibitem [{\citenamefont {Goldman}\ \emph {et~al.}(2014)\citenamefont
  {Goldman}, \citenamefont {Juzeli{\=u}nas}, \citenamefont {{\"O}hberg},\ and\
  \citenamefont {Spielman}}]{goldman2014light}%
  \BibitemOpen
  \bibfield  {author} {\bibinfo {author} {\bibfnamefont {N.}~\bibnamefont
  {Goldman}}, \bibinfo {author} {\bibfnamefont {G.}~\bibnamefont
  {Juzeli{\=u}nas}}, \bibinfo {author} {\bibfnamefont {P.}~\bibnamefont
  {{\"O}hberg}}, \ and\ \bibinfo {author} {\bibfnamefont {I.~B.}\ \bibnamefont
  {Spielman}},\ }\bibfield  {title} {\enquote {\bibinfo {title} {Light-induced
  gauge fields for ultracold atoms},}\ }\href@noop {} {\bibfield  {journal}
  {\bibinfo  {journal} {Reports on Progress in Physics}\ }\textbf {\bibinfo
  {volume} {77}},\ \bibinfo {pages} {126401} (\bibinfo {year}
  {2014})}\BibitemShut {NoStop}%
\bibitem [{\citenamefont {Aidelsburger}\ \emph {et~al.}(2017)\citenamefont
  {Aidelsburger}, \citenamefont {Nascimbene},\ and\ \citenamefont
  {Goldman}}]{aidelsburger2017artificial}%
  \BibitemOpen
  \bibfield  {author} {\bibinfo {author} {\bibfnamefont {M}~\bibnamefont
  {Aidelsburger}}, \bibinfo {author} {\bibfnamefont {S}~\bibnamefont
  {Nascimbene}}, \ and\ \bibinfo {author} {\bibfnamefont {N}~\bibnamefont
  {Goldman}},\ }\bibfield  {title} {\enquote {\bibinfo {title} {Artificial
  gauge fields in materials and engineered systems},}\ }\href@noop {}
  {\bibfield  {journal} {\bibinfo  {journal} {arXiv preprint arXiv:1710.00851}\
  } (\bibinfo {year} {2017})}\BibitemShut {NoStop}%
\bibitem [{\citenamefont {Lu}\ \emph {et~al.}(2014)\citenamefont {Lu},
  \citenamefont {Joannopoulos},\ and\ \citenamefont
  {Solja{\v{c}}i{\'c}}}]{Lu2014topological}%
  \BibitemOpen
  \bibfield  {author} {\bibinfo {author} {\bibfnamefont {L.}~\bibnamefont
  {Lu}}, \bibinfo {author} {\bibfnamefont {J.~D.}\ \bibnamefont
  {Joannopoulos}}, \ and\ \bibinfo {author} {\bibfnamefont {M.}~\bibnamefont
  {Solja{\v{c}}i{\'c}}},\ }\bibfield  {title} {\enquote {\bibinfo {title}
  {Topological photonics},}\ }\href@noop {} {\bibfield  {journal} {\bibinfo
  {journal} {Nature Photonics}\ }\textbf {\bibinfo {volume} {8}},\ \bibinfo
  {pages} {821--829} (\bibinfo {year} {2014})}\BibitemShut {NoStop}%
\bibitem [{\citenamefont {Hafezi}(2014)}]{hafezi2014synthetic}%
  \BibitemOpen
  \bibfield  {author} {\bibinfo {author} {\bibfnamefont {M.}~\bibnamefont
  {Hafezi}},\ }\bibfield  {title} {\enquote {\bibinfo {title} {Synthetic gauge
  fields with photons},}\ }\href@noop {} {\bibfield  {journal} {\bibinfo
  {journal} {International Journal of Modern Physics B}\ }\textbf {\bibinfo
  {volume} {28}},\ \bibinfo {pages} {1441002} (\bibinfo {year}
  {2014})}\BibitemShut {NoStop}%
\bibitem [{\citenamefont {Amorim}\ \emph {et~al.}(2016)\citenamefont {Amorim},
  \citenamefont {Cortijo}, \citenamefont {de~Juan}, \citenamefont {Grushin},
  \citenamefont {Guinea}, \citenamefont {Gutiérrez-Rubio}, \citenamefont
  {Ochoa}, \citenamefont {Parente}, \citenamefont {Roldán}, \citenamefont
  {San-Jose}, \citenamefont {Schiefele}, \citenamefont {Sturla},\ and\
  \citenamefont {Vozmediano}}]{amorim2016}%
  \BibitemOpen
  \bibfield  {author} {\bibinfo {author} {\bibfnamefont {B.}~\bibnamefont
  {Amorim}}, \bibinfo {author} {\bibfnamefont {A.}~\bibnamefont {Cortijo}},
  \bibinfo {author} {\bibfnamefont {F.}~\bibnamefont {de~Juan}}, \bibinfo
  {author} {\bibfnamefont {A.G.}\ \bibnamefont {Grushin}}, \bibinfo {author}
  {\bibfnamefont {F.}~\bibnamefont {Guinea}}, \bibinfo {author} {\bibfnamefont
  {A.}~\bibnamefont {Gutiérrez-Rubio}}, \bibinfo {author} {\bibfnamefont
  {H.}~\bibnamefont {Ochoa}}, \bibinfo {author} {\bibfnamefont
  {V.}~\bibnamefont {Parente}}, \bibinfo {author} {\bibfnamefont
  {R.}~\bibnamefont {Roldán}}, \bibinfo {author} {\bibfnamefont
  {P.}~\bibnamefont {San-Jose}}, \bibinfo {author} {\bibfnamefont
  {J.}~\bibnamefont {Schiefele}}, \bibinfo {author} {\bibfnamefont
  {M.}~\bibnamefont {Sturla}}, \ and\ \bibinfo {author} {\bibfnamefont
  {M.A.H.}\ \bibnamefont {Vozmediano}},\ }\bibfield  {title} {\enquote
  {\bibinfo {title} {Novel effects of strains in graphene and other two
  dimensional materials},}\ }\href
  {http://www.sciencedirect.com/science/article/pii/S0370157315005402}
  {\bibfield  {journal} {\bibinfo  {journal} {Physics Reports}\ }\textbf
  {\bibinfo {volume} {617}},\ \bibinfo {pages} {1 -- 54} (\bibinfo {year}
  {2016})}\BibitemShut {NoStop}%
\bibitem [{\citenamefont {Aidelsburger}\ \emph {et~al.}(2013)\citenamefont
  {Aidelsburger}, \citenamefont {Atala}, \citenamefont {Lohse}, \citenamefont
  {Barreiro}, \citenamefont {Paredes},\ and\ \citenamefont
  {Bloch}}]{aidelsburger2013realization}%
  \BibitemOpen
  \bibfield  {author} {\bibinfo {author} {\bibfnamefont {M.}~\bibnamefont
  {Aidelsburger}}, \bibinfo {author} {\bibfnamefont {M.}~\bibnamefont {Atala}},
  \bibinfo {author} {\bibfnamefont {M.}~\bibnamefont {Lohse}}, \bibinfo
  {author} {\bibfnamefont {J.~T.}\ \bibnamefont {Barreiro}}, \bibinfo {author}
  {\bibfnamefont {B.}~\bibnamefont {Paredes}}, \ and\ \bibinfo {author}
  {\bibfnamefont {I.}~\bibnamefont {Bloch}},\ }\bibfield  {title} {\enquote
  {\bibinfo {title} {Realization of the hofstadter hamiltonian with ultracold
  atoms in optical lattices},}\ }\href {\doibase
  10.1103/PhysRevLett.111.185301} {\bibfield  {journal} {\bibinfo  {journal}
  {Phys. Rev. Lett.}\ }\textbf {\bibinfo {volume} {111}},\ \bibinfo {pages}
  {185301} (\bibinfo {year} {2013})}\BibitemShut {NoStop}%
\bibitem [{\citenamefont {Miyake}\ \emph {et~al.}(2013)\citenamefont {Miyake},
  \citenamefont {Siviloglou}, \citenamefont {Kennedy}, \citenamefont {Burton},\
  and\ \citenamefont {Ketterle}}]{miyake2013realizing}%
  \BibitemOpen
  \bibfield  {author} {\bibinfo {author} {\bibfnamefont {H.}~\bibnamefont
  {Miyake}}, \bibinfo {author} {\bibfnamefont {G.~A.}\ \bibnamefont
  {Siviloglou}}, \bibinfo {author} {\bibfnamefont {C.~J.}\ \bibnamefont
  {Kennedy}}, \bibinfo {author} {\bibfnamefont {W.~C.}\ \bibnamefont {Burton}},
  \ and\ \bibinfo {author} {\bibfnamefont {W.}~\bibnamefont {Ketterle}},\
  }\bibfield  {title} {\enquote {\bibinfo {title} {Realizing the harper
  hamiltonian with laser-assisted tunneling in optical lattices},}\ }\href
  {\doibase 10.1103/PhysRevLett.111.185302} {\bibfield  {journal} {\bibinfo
  {journal} {Phys. Rev. Lett.}\ }\textbf {\bibinfo {volume} {111}},\ \bibinfo
  {pages} {185302} (\bibinfo {year} {2013})}\BibitemShut {NoStop}%
\bibitem [{\citenamefont {Aidelsburger}\ \emph {et~al.}(2015)\citenamefont
  {Aidelsburger}, \citenamefont {Lohse}, \citenamefont {Schweizer},
  \citenamefont {Atala}, \citenamefont {Barreiro}, \citenamefont {Nascimbene},
  \citenamefont {Cooper}, \citenamefont {Bloch},\ and\ \citenamefont
  {Goldman}}]{aidelsburger2015measuring}%
  \BibitemOpen
  \bibfield  {author} {\bibinfo {author} {\bibfnamefont {M.}~\bibnamefont
  {Aidelsburger}}, \bibinfo {author} {\bibfnamefont {M.}~\bibnamefont {Lohse}},
  \bibinfo {author} {\bibfnamefont {C.}~\bibnamefont {Schweizer}}, \bibinfo
  {author} {\bibfnamefont {M.}~\bibnamefont {Atala}}, \bibinfo {author}
  {\bibfnamefont {J.~T.}\ \bibnamefont {Barreiro}}, \bibinfo {author}
  {\bibfnamefont {S.}~\bibnamefont {Nascimbene}}, \bibinfo {author}
  {\bibfnamefont {N.~R.}\ \bibnamefont {Cooper}}, \bibinfo {author}
  {\bibfnamefont {I.}~\bibnamefont {Bloch}}, \ and\ \bibinfo {author}
  {\bibfnamefont {N.}~\bibnamefont {Goldman}},\ }\bibfield  {title} {\enquote
  {\bibinfo {title} {Measuring the chern number of hofstadter bands with
  ultracold bosonic atoms},}\ }\href@noop {} {\bibfield  {journal} {\bibinfo
  {journal} {Nature Physics}\ }\textbf {\bibinfo {volume} {11}},\ \bibinfo
  {pages} {162--166} (\bibinfo {year} {2015})}\BibitemShut {NoStop}%
\bibitem [{\citenamefont {Kennedy}\ \emph {et~al.}(2015)\citenamefont
  {Kennedy}, \citenamefont {Burton}, \citenamefont {Chung},\ and\ \citenamefont
  {Ketterle}}]{KBCK2015}%
  \BibitemOpen
  \bibfield  {author} {\bibinfo {author} {\bibfnamefont {C.~J.}\ \bibnamefont
  {Kennedy}}, \bibinfo {author} {\bibfnamefont {W.~C.}\ \bibnamefont {Burton}},
  \bibinfo {author} {\bibfnamefont {W.~C.}\ \bibnamefont {Chung}}, \ and\
  \bibinfo {author} {\bibfnamefont {W.}~\bibnamefont {Ketterle}},\ }\bibfield
  {title} {\enquote {\bibinfo {title} {Observation of {B}ose-{E}instein
  condensation in a strong synthetic magnetic field},}\ }\href@noop {}
  {\bibfield  {journal} {\bibinfo  {journal} {Nat Phys}\ }\textbf {\bibinfo
  {volume} {11}},\ \bibinfo {pages} {859} (\bibinfo {year} {2015})}\BibitemShut
  {NoStop}%
\bibitem [{\citenamefont {Jotzu}\ \emph {et~al.}(2014)\citenamefont {Jotzu},
  \citenamefont {Messer}, \citenamefont {Desbuquois}, \citenamefont {Lebrat},
  \citenamefont {Uehlinger}, \citenamefont {Greif},\ and\ \citenamefont
  {Esslinger}}]{jotzu2014experimental}%
  \BibitemOpen
  \bibfield  {author} {\bibinfo {author} {\bibfnamefont {G.}~\bibnamefont
  {Jotzu}}, \bibinfo {author} {\bibfnamefont {M.}~\bibnamefont {Messer}},
  \bibinfo {author} {\bibfnamefont {R.}~\bibnamefont {Desbuquois}}, \bibinfo
  {author} {\bibfnamefont {M.}~\bibnamefont {Lebrat}}, \bibinfo {author}
  {\bibfnamefont {T.}~\bibnamefont {Uehlinger}}, \bibinfo {author}
  {\bibfnamefont {D.}~\bibnamefont {Greif}}, \ and\ \bibinfo {author}
  {\bibfnamefont {T.}~\bibnamefont {Esslinger}},\ }\bibfield  {title} {\enquote
  {\bibinfo {title} {Experimental realization of the topological haldane model
  with ultracold fermions},}\ }\href@noop {} {\bibfield  {journal} {\bibinfo
  {journal} {Nature}\ }\textbf {\bibinfo {volume} {515}},\ \bibinfo {pages}
  {237--240} (\bibinfo {year} {2014})}\BibitemShut {NoStop}%
\bibitem [{\citenamefont {Wu}\ \emph {et~al.}(2016)\citenamefont {Wu},
  \citenamefont {Zhang}, \citenamefont {Sun}, \citenamefont {Xu}, \citenamefont
  {Wang}, \citenamefont {Ji}, \citenamefont {Deng}, \citenamefont {Chen},
  \citenamefont {Liu},\ and\ \citenamefont {Pan}}]{wu2016realization}%
  \BibitemOpen
  \bibfield  {author} {\bibinfo {author} {\bibfnamefont {Z.}~\bibnamefont
  {Wu}}, \bibinfo {author} {\bibfnamefont {L.}~\bibnamefont {Zhang}}, \bibinfo
  {author} {\bibfnamefont {W.}~\bibnamefont {Sun}}, \bibinfo {author}
  {\bibfnamefont {X.-T.}\ \bibnamefont {Xu}}, \bibinfo {author} {\bibfnamefont
  {B.-Z.}\ \bibnamefont {Wang}}, \bibinfo {author} {\bibfnamefont {S.-C.}\
  \bibnamefont {Ji}}, \bibinfo {author} {\bibfnamefont {Y.}~\bibnamefont
  {Deng}}, \bibinfo {author} {\bibfnamefont {S.}~\bibnamefont {Chen}}, \bibinfo
  {author} {\bibfnamefont {X.-J.}\ \bibnamefont {Liu}}, \ and\ \bibinfo
  {author} {\bibfnamefont {J.-W.}\ \bibnamefont {Pan}},\ }\bibfield  {title}
  {\enquote {\bibinfo {title} {Realization of two-dimensional spin-orbit
  coupling for bose-einstein condensates},}\ }\href@noop {} {\bibfield
  {journal} {\bibinfo  {journal} {Science}\ }\textbf {\bibinfo {volume}
  {354}},\ \bibinfo {pages} {83--88} (\bibinfo {year} {2016})}\BibitemShut
  {NoStop}%
\bibitem [{\citenamefont {Fl{\"a}schner}\ \emph {et~al.}(2016)\citenamefont
  {Fl{\"a}schner}, \citenamefont {Vogel}, \citenamefont {Tarnowski},
  \citenamefont {Rem}, \citenamefont {L{\"u}hmann}, \citenamefont {Heyl},
  \citenamefont {Budich}, \citenamefont {Mathey}, \citenamefont {Sengstock},\
  and\ \citenamefont {Weitenberg}}]{flaschner2016observation}%
  \BibitemOpen
  \bibfield  {author} {\bibinfo {author} {\bibfnamefont {N.}~\bibnamefont
  {Fl{\"a}schner}}, \bibinfo {author} {\bibfnamefont {D.}~\bibnamefont
  {Vogel}}, \bibinfo {author} {\bibfnamefont {M.}~\bibnamefont {Tarnowski}},
  \bibinfo {author} {\bibfnamefont {B.~S.}\ \bibnamefont {Rem}}, \bibinfo
  {author} {\bibfnamefont {D.-S.}\ \bibnamefont {L{\"u}hmann}}, \bibinfo
  {author} {\bibfnamefont {M.}~\bibnamefont {Heyl}}, \bibinfo {author}
  {\bibfnamefont {J.~C.}\ \bibnamefont {Budich}}, \bibinfo {author}
  {\bibfnamefont {L.}~\bibnamefont {Mathey}}, \bibinfo {author} {\bibfnamefont
  {K.}~\bibnamefont {Sengstock}}, \ and\ \bibinfo {author} {\bibfnamefont
  {C.}~\bibnamefont {Weitenberg}},\ }\bibfield  {title} {\enquote {\bibinfo
  {title} {Observation of a dynamical topological phase transition},}\
  }\href@noop {} {\bibfield  {journal} {\bibinfo  {journal} {arXiv preprint
  arXiv:1608.05616}\ } (\bibinfo {year} {2016})}\BibitemShut {NoStop}%
\bibitem [{\citenamefont {Tai}\ \emph {et~al.}(2017)\citenamefont {Tai},
  \citenamefont {Lukin}, \citenamefont {Rispoli}, \citenamefont {Schittko},
  \citenamefont {Menke}, \citenamefont {Borgnia}, \citenamefont {Preiss},
  \citenamefont {Grusdt}, \citenamefont {Kaufman},\ and\ \citenamefont
  {Greiner}}]{tai2017microscopy}%
  \BibitemOpen
  \bibfield  {author} {\bibinfo {author} {\bibfnamefont {M.~E.}\ \bibnamefont
  {Tai}}, \bibinfo {author} {\bibfnamefont {A.}~\bibnamefont {Lukin}}, \bibinfo
  {author} {\bibfnamefont {M.}~\bibnamefont {Rispoli}}, \bibinfo {author}
  {\bibfnamefont {R.}~\bibnamefont {Schittko}}, \bibinfo {author}
  {\bibfnamefont {T.}~\bibnamefont {Menke}}, \bibinfo {author} {\bibfnamefont
  {D.}~\bibnamefont {Borgnia}}, \bibinfo {author} {\bibfnamefont {P.~M.}\
  \bibnamefont {Preiss}}, \bibinfo {author} {\bibfnamefont {F.}~\bibnamefont
  {Grusdt}}, \bibinfo {author} {\bibfnamefont {A.~M.}\ \bibnamefont {Kaufman}},
  \ and\ \bibinfo {author} {\bibfnamefont {M.}~\bibnamefont {Greiner}},\
  }\bibfield  {title} {\enquote {\bibinfo {title} {{Microscopy of the
  interacting Harper--Hofstadter model in the two-body limit}},}\ }\href@noop
  {} {\bibfield  {journal} {\bibinfo  {journal} {Nature}\ }\textbf {\bibinfo
  {volume} {546}},\ \bibinfo {pages} {519--523} (\bibinfo {year}
  {2017})}\BibitemShut {NoStop}%
\bibitem [{\citenamefont {{Tran}}\ \emph {et~al.}(2017)\citenamefont {{Tran}},
  \citenamefont {{Dauphin}}, \citenamefont {{Grushin}}, \citenamefont
  {{Zoller}},\ and\ \citenamefont {{Goldman}}}]{TDG17}%
  \BibitemOpen
  \bibfield  {author} {\bibinfo {author} {\bibfnamefont {D.~T.}\ \bibnamefont
  {{Tran}}}, \bibinfo {author} {\bibfnamefont {A.}~\bibnamefont {{Dauphin}}},
  \bibinfo {author} {\bibfnamefont {A.~G.}\ \bibnamefont {{Grushin}}}, \bibinfo
  {author} {\bibfnamefont {P.}~\bibnamefont {{Zoller}}}, \ and\ \bibinfo
  {author} {\bibfnamefont {N.}~\bibnamefont {{Goldman}}},\ }\bibfield  {title}
  {\enquote {\bibinfo {title} {{Probing topology by ``heating'': Quantized
  circular dichroism in ultracold atoms}},}\ }\href@noop {} {\bibfield
  {journal} {\bibinfo  {journal} {ArXiv e-prints}\ } (\bibinfo {year}
  {2017})},\ \Eprint {http://arxiv.org/abs/1704.01990} {arXiv:1704.01990
  [cond-mat.quant-gas]} \BibitemShut {NoStop}%
\bibitem [{\citenamefont {Tian}\ \emph {et~al.}(2015)\citenamefont {Tian},
  \citenamefont {Endres},\ and\ \citenamefont {Pekker}}]{Tian2015Landau}%
  \BibitemOpen
  \bibfield  {author} {\bibinfo {author} {\bibfnamefont {B.}~\bibnamefont
  {Tian}}, \bibinfo {author} {\bibfnamefont {M.}~\bibnamefont {Endres}}, \ and\
  \bibinfo {author} {\bibfnamefont {D.}~\bibnamefont {Pekker}},\ }\bibfield
  {title} {\enquote {\bibinfo {title} {Landau levels in strained optical
  lattices},}\ }\href {\doibase 10.1103/PhysRevLett.115.236803} {\bibfield
  {journal} {\bibinfo  {journal} {Phys. Rev. Lett.}\ }\textbf {\bibinfo
  {volume} {115}},\ \bibinfo {pages} {236803} (\bibinfo {year}
  {2015})}\BibitemShut {NoStop}%
\bibitem [{\citenamefont {Cortijo}\ \emph {et~al.}(2015)\citenamefont
  {Cortijo}, \citenamefont {Ferreir\'os}, \citenamefont {Landsteiner},\ and\
  \citenamefont {Vozmediano}}]{CFL15}%
  \BibitemOpen
  \bibfield  {author} {\bibinfo {author} {\bibfnamefont {A.}~\bibnamefont
  {Cortijo}}, \bibinfo {author} {\bibfnamefont {Y.}~\bibnamefont
  {Ferreir\'os}}, \bibinfo {author} {\bibfnamefont {K.}~\bibnamefont
  {Landsteiner}}, \ and\ \bibinfo {author} {\bibfnamefont {M.~A.~H.}\
  \bibnamefont {Vozmediano}},\ }\bibfield  {title} {\enquote {\bibinfo {title}
  {Elastic gauge fields in {W}eyl semimetals},}\ }\href {\doibase
  10.1103/PhysRevLett.115.177202} {\bibfield  {journal} {\bibinfo  {journal}
  {Phys. Rev. Lett.}\ }\textbf {\bibinfo {volume} {115}},\ \bibinfo {pages}
  {177202} (\bibinfo {year} {2015})}\BibitemShut {NoStop}%
\bibitem [{\citenamefont {Pikulin}\ \emph {et~al.}(2016)\citenamefont
  {Pikulin}, \citenamefont {Chen},\ and\ \citenamefont {Franz}}]{Pikulin2016}%
  \BibitemOpen
  \bibfield  {author} {\bibinfo {author} {\bibfnamefont {D. I.}\ \bibnamefont
  {Pikulin}}, \bibinfo {author} {\bibfnamefont {A.}~\bibnamefont {Chen}}, \
  and\ \bibinfo {author} {\bibfnamefont {M.}~\bibnamefont {Franz}},\ }\bibfield
   {title} {\enquote {\bibinfo {title} {{Chiral Anomaly from Strain-Induced
  Gauge Fields in Dirac and Weyl Semimetals}},}\ }\href {\doibase
  10.1103/PhysRevX.6.041021} {\bibfield  {journal} {\bibinfo  {journal}
  {Physical Review X}\ }\textbf {\bibinfo {volume} {6}},\ \bibinfo {pages}
  {041021} (\bibinfo {year} {2016})}\BibitemShut {NoStop}%
\bibitem [{\citenamefont {Grushin}\ \emph {et~al.}(2016)\citenamefont
  {Grushin}, \citenamefont {Venderbos}, \citenamefont {Vishwanath},\ and\
  \citenamefont {Ilan}}]{grushin2016inhomogeneous}%
  \BibitemOpen
  \bibfield  {author} {\bibinfo {author} {\bibfnamefont {A.~G.}\ \bibnamefont
  {Grushin}}, \bibinfo {author} {\bibfnamefont {J.~W.~F.}\ \bibnamefont
  {Venderbos}}, \bibinfo {author} {\bibfnamefont {A.}~\bibnamefont
  {Vishwanath}}, \ and\ \bibinfo {author} {\bibfnamefont {R.}~\bibnamefont
  {Ilan}},\ }\bibfield  {title} {\enquote {\bibinfo {title} {Inhomogeneous weyl
  and dirac semimetals: Transport in axial magnetic fields and fermi arc
  surface states from pseudo-landau levels},}\ }\href {\doibase
  10.1103/PhysRevX.6.041046} {\bibfield  {journal} {\bibinfo  {journal} {Phys.
  Rev. X}\ }\textbf {\bibinfo {volume} {6}},\ \bibinfo {pages} {041046}
  (\bibinfo {year} {2016})}\BibitemShut {NoStop}%
\bibitem [{\citenamefont {Gorbar}\ \emph
  {et~al.}(2017{\natexlab{a}})\citenamefont {Gorbar}, \citenamefont {Miransky},
  \citenamefont {Shovkovy},\ and\ \citenamefont {Sukhachov}}]{Gorbar2017}%
  \BibitemOpen
  \bibfield  {author} {\bibinfo {author} {\bibfnamefont {E.~V.}\ \bibnamefont
  {Gorbar}}, \bibinfo {author} {\bibfnamefont {V.~A.}\ \bibnamefont
  {Miransky}}, \bibinfo {author} {\bibfnamefont {I.~A.}\ \bibnamefont
  {Shovkovy}}, \ and\ \bibinfo {author} {\bibfnamefont {P.~O.}\ \bibnamefont
  {Sukhachov}},\ }\bibfield  {title} {\enquote {\bibinfo {title} {{Origin of
  the Bardeen-Zumino current in lattice models of Weyl semimetals}},}\ }\href
  {http://arxiv.org/abs/1706.02705} {\bibfield  {journal} {\bibinfo  {journal}
  {arXiv}\ ,\ \bibinfo {pages} {1--13}} (\bibinfo {year}
  {2017}{\natexlab{a}})},\ \Eprint {http://arxiv.org/abs/1706.02705}
  {arXiv:1706.02705} \BibitemShut {NoStop}%
\bibitem [{\citenamefont {Gorbar}\ \emph
  {et~al.}(2017{\natexlab{b}})\citenamefont {Gorbar}, \citenamefont {Miransky},
  \citenamefont {Shovkovy},\ and\ \citenamefont {Sukhachov}}]{Gorbar2017a}%
  \BibitemOpen
  \bibfield  {author} {\bibinfo {author} {\bibfnamefont {E.~V.}\ \bibnamefont
  {Gorbar}}, \bibinfo {author} {\bibfnamefont {V.~A.}\ \bibnamefont
  {Miransky}}, \bibinfo {author} {\bibfnamefont {I.~A.}\ \bibnamefont
  {Shovkovy}}, \ and\ \bibinfo {author} {\bibfnamefont {P.~O.}\ \bibnamefont
  {Sukhachov}},\ }\bibfield  {title} {\enquote {\bibinfo {title} {{Chiral
  response in lattice models of Weyl materials}},}\ }\href
  {http://arxiv.org/abs/1706.09419} {\  (\bibinfo {year}
  {2017}{\natexlab{b}})},\ \Eprint {http://arxiv.org/abs/1706.09419}
  {arXiv:1706.09419} \BibitemShut {NoStop}%
\bibitem [{\citenamefont {Gorbar}\ \emph
  {et~al.}(2017{\natexlab{c}})\citenamefont {Gorbar}, \citenamefont {Miransky},
  \citenamefont {Shovkovy},\ and\ \citenamefont {Sukhachov}}]{Gorbar2017b}%
  \BibitemOpen
  \bibfield  {author} {\bibinfo {author} {\bibfnamefont {E.~V.}\ \bibnamefont
  {Gorbar}}, \bibinfo {author} {\bibfnamefont {V.~A.}\ \bibnamefont
  {Miransky}}, \bibinfo {author} {\bibfnamefont {I.~A.}\ \bibnamefont
  {Shovkovy}}, \ and\ \bibinfo {author} {\bibfnamefont {P.~O.}\ \bibnamefont
  {Sukhachov}},\ }\bibfield  {title} {\enquote {\bibinfo {title}
  {{Pseudomagnetic lens as a valley and chirality splitter in Dirac and Weyl
  materials}},}\ }\href {\doibase 10.1103/PhysRevB.95.241114} {\bibfield
  {journal} {\bibinfo  {journal} {Physical Review B}\ }\textbf {\bibinfo
  {volume} {95}},\ \bibinfo {pages} {241114} (\bibinfo {year}
  {2017}{\natexlab{c}})}\BibitemShut {NoStop}%
\bibitem [{\citenamefont {Gorbar}\ \emph
  {et~al.}(2017{\natexlab{d}})\citenamefont {Gorbar}, \citenamefont {Miransky},
  \citenamefont {Shovkovy},\ and\ \citenamefont {Sukhachov}}]{Gorbar2017c}%
  \BibitemOpen
  \bibfield  {author} {\bibinfo {author} {\bibfnamefont {E.~V.}\ \bibnamefont
  {Gorbar}}, \bibinfo {author} {\bibfnamefont {V.~A.}\ \bibnamefont
  {Miransky}}, \bibinfo {author} {\bibfnamefont {I.~A.}\ \bibnamefont
  {Shovkovy}}, \ and\ \bibinfo {author} {\bibfnamefont {P.~O.}\ \bibnamefont
  {Sukhachov}},\ }\bibfield  {title} {\enquote {\bibinfo {title} {{Consistent
  Chiral Kinetic Theory in Weyl Materials: Chiral Magnetic Plasmons}},}\ }\href
  {\doibase 10.1103/PhysRevLett.118.127601} {\bibfield  {journal} {\bibinfo
  {journal} {Physical Review Letters}\ }\textbf {\bibinfo {volume} {118}},\
  \bibinfo {pages} {127601} (\bibinfo {year} {2017}{\natexlab{d}})}\BibitemShut
  {NoStop}%
\bibitem [{\citenamefont {Gorbar}\ \emph
  {et~al.}(2017{\natexlab{e}})\citenamefont {Gorbar}, \citenamefont {Miransky},
  \citenamefont {Shovkovy},\ and\ \citenamefont {Sukhachov}}]{Gorbar2017d}%
  \BibitemOpen
  \bibfield  {author} {\bibinfo {author} {\bibfnamefont {E.~V.}\ \bibnamefont
  {Gorbar}}, \bibinfo {author} {\bibfnamefont {V.~A.}\ \bibnamefont
  {Miransky}}, \bibinfo {author} {\bibfnamefont {I.~A.}\ \bibnamefont
  {Shovkovy}}, \ and\ \bibinfo {author} {\bibfnamefont {P.~O.}\ \bibnamefont
  {Sukhachov}},\ }\bibfield  {title} {\enquote {\bibinfo {title}
  {{Pseudomagnetic helicons}},}\ }\href {\doibase 10.1103/PhysRevB.95.115422}
  {\bibfield  {journal} {\bibinfo  {journal} {Physical Review B}\ }\textbf
  {\bibinfo {volume} {95}},\ \bibinfo {pages} {115422} (\bibinfo {year}
  {2017}{\natexlab{e}})}\BibitemShut {NoStop}%
\bibitem [{\citenamefont {Liu}\ \emph {et~al.}(2017)\citenamefont {Liu},
  \citenamefont {Pikulin},\ and\ \citenamefont {Franz}}]{Liu2017}%
  \BibitemOpen
  \bibfield  {author} {\bibinfo {author} {\bibfnamefont {T.}~\bibnamefont
  {Liu}}, \bibinfo {author} {\bibfnamefont {D.~I.}\ \bibnamefont {Pikulin}}, \
  and\ \bibinfo {author} {\bibfnamefont {M.}~\bibnamefont {Franz}},\ }\bibfield
   {title} {\enquote {\bibinfo {title} {{Quantum oscillations without magnetic
  field}},}\ }\href {\doibase 10.1103/PhysRevB.95.041201} {\bibfield  {journal}
  {\bibinfo  {journal} {Physical Review B}\ }\textbf {\bibinfo {volume} {95}},\
  \bibinfo {pages} {041201} (\bibinfo {year} {2017})}\BibitemShut {NoStop}%
\bibitem [{\citenamefont {Gorbar}\ \emph
  {et~al.}(2017{\natexlab{f}})\citenamefont {Gorbar}, \citenamefont {Miransky},
  \citenamefont {Shovkovy},\ and\ \citenamefont {Sukhachov}}]{Gorbar2017e}%
  \BibitemOpen
  \bibfield  {author} {\bibinfo {author} {\bibfnamefont {E.~V.}\ \bibnamefont
  {Gorbar}}, \bibinfo {author} {\bibfnamefont {V.~A.}\ \bibnamefont
  {Miransky}}, \bibinfo {author} {\bibfnamefont {I.~A.}\ \bibnamefont
  {Shovkovy}}, \ and\ \bibinfo {author} {\bibfnamefont {P.~O.}\ \bibnamefont
  {Sukhachov}},\ }\bibfield  {title} {\enquote {\bibinfo {title} {{Chiral
  magnetic plasmons in anomalous relativistic matter}},}\ }\href {\doibase
  10.1103/PhysRevB.95.115202} {\bibfield  {journal} {\bibinfo  {journal}
  {Physical Review B}\ }\textbf {\bibinfo {volume} {95}},\ \bibinfo {pages}
  {115202} (\bibinfo {year} {2017}{\natexlab{f}})}\BibitemShut {NoStop}%
\bibitem [{\citenamefont {Gorbar}\ \emph
  {et~al.}(2017{\natexlab{g}})\citenamefont {Gorbar}, \citenamefont {Miransky},
  \citenamefont {Shovkovy},\ and\ \citenamefont {Sukhachov}}]{Gorbar2017f}%
  \BibitemOpen
  \bibfield  {author} {\bibinfo {author} {\bibfnamefont {E.~V.}\ \bibnamefont
  {Gorbar}}, \bibinfo {author} {\bibfnamefont {V.~A.}\ \bibnamefont
  {Miransky}}, \bibinfo {author} {\bibfnamefont {I.~A.}\ \bibnamefont
  {Shovkovy}}, \ and\ \bibinfo {author} {\bibfnamefont {P.~O.}\ \bibnamefont
  {Sukhachov}},\ }\bibfield  {title} {\enquote {\bibinfo {title} {{Second-order
  chiral kinetic theory: Chiral magnetic and pseudomagnetic waves}},}\ }\href
  {\doibase 10.1103/PhysRevB.95.205141} {\bibfield  {journal} {\bibinfo
  {journal} {Physical Review B}\ }\textbf {\bibinfo {volume} {95}},\ \bibinfo
  {pages} {205141} (\bibinfo {year} {2017}{\natexlab{g}})}\BibitemShut
  {NoStop}%
\bibitem [{\citenamefont {Huang}\ \emph {et~al.}(2017)\citenamefont {Huang},
  \citenamefont {Zhou},\ and\ \citenamefont {Shen}}]{Huang2017}%
  \BibitemOpen
  \bibfield  {author} {\bibinfo {author} {\bibfnamefont {Z.-M.}\ \bibnamefont
  {Huang}}, \bibinfo {author} {\bibfnamefont {J.}~\bibnamefont {Zhou}}, \ and\
  \bibinfo {author} {\bibfnamefont {S.-Q.}\ \bibnamefont {Shen}},\ }\bibfield
  {title} {\enquote {\bibinfo {title} {{Topological responses from chiral
  anomaly in multi-Weyl semimetals}},}\ }\href {\doibase
  10.1103/PhysRevB.96.085201} {\bibfield  {journal} {\bibinfo  {journal}
  {Physical Review B}\ }\textbf {\bibinfo {volume} {96}},\ \bibinfo {pages}
  {085201} (\bibinfo {year} {2017})}\BibitemShut {NoStop}%
\bibitem [{\citenamefont {Guan}\ \emph {et~al.}(2017)\citenamefont {Guan},
  \citenamefont {Yu}, \citenamefont {Liu}, \citenamefont {Liu}, \citenamefont
  {Dong}, \citenamefont {Lu}, \citenamefont {Yao},\ and\ \citenamefont
  {Yang}}]{Guan2017}%
  \BibitemOpen
  \bibfield  {author} {\bibinfo {author} {\bibfnamefont {S.}~\bibnamefont
  {Guan}}, \bibinfo {author} {\bibfnamefont {Z.-M.}\ \bibnamefont {Yu}},
  \bibinfo {author} {\bibfnamefont {Y.}~\bibnamefont {Liu}}, \bibinfo {author}
  {\bibfnamefont {G.-B.}\ \bibnamefont {Liu}}, \bibinfo {author} {\bibfnamefont
  {L.}~\bibnamefont {Dong}}, \bibinfo {author} {\bibfnamefont {Y.}~\bibnamefont
  {Lu}}, \bibinfo {author} {\bibfnamefont {Y.}~\bibnamefont {Yao}}, \ and\
  \bibinfo {author} {\bibfnamefont {S.~A.}\ \bibnamefont {Yang}},\ }\bibfield
  {title} {\enquote {\bibinfo {title} {{Artificial gravity field, astrophysical
  analogues, and topological phase transitions in strained topological
  semimetals}},}\ }\href {\doibase 10.1038/s41535-017-0026-7} {\bibfield
  {journal} {\bibinfo  {journal} {npj Quantum Materials}\ }\textbf {\bibinfo
  {volume} {2}},\ \bibinfo {pages} {23} (\bibinfo {year} {2017})}\BibitemShut
  {NoStop}%
\bibitem [{\citenamefont {Volovik}(2003)}]{V03}%
  \BibitemOpen
  \bibfield  {author} {\bibinfo {author} {\bibfnamefont {G.~E.}\ \bibnamefont
  {Volovik}},\ }\href@noop {} {\emph {\bibinfo {title} {{The Universe in Helium
  Droplet}}}}\ (\bibinfo  {publisher} {{Clarendon, Oxford, UK}},\ \bibinfo
  {year} {2003})\BibitemShut {NoStop}%
\bibitem [{\citenamefont {Turner}\ and\ \citenamefont
  {Vishwanath}()}]{Turner:2013tf}%
  \BibitemOpen
  \bibfield  {author} {\bibinfo {author} {\bibfnamefont {A.~M.}\ \bibnamefont
  {Turner}}\ and\ \bibinfo {author} {\bibfnamefont {A.}~\bibnamefont
  {Vishwanath}},\ }\bibfield  {title} {\enquote {\bibinfo {title} {{Beyond Band
  Insulators: Topology of Semi-metals and Interacting Phases}},}\ }\href@noop
  {} {\bibinfo  {journal} {arXiv:1301.0330}\ }\BibitemShut {NoStop}%
\bibitem [{\citenamefont {Hosur}\ and\ \citenamefont
  {Qi}(2013)}]{hosur2013recent}%
  \BibitemOpen
\bibfield  {journal} {  }\bibfield  {author} {\bibinfo {author} {\bibfnamefont
  {Pavan}\ \bibnamefont {Hosur}}\ and\ \bibinfo {author} {\bibfnamefont
  {Xiaoliang}\ \bibnamefont {Qi}},\ }\bibfield  {title} {\enquote {\bibinfo
  {title} {{Recent developments in transport phenomena in Weyl semimetals}},}\
  }\href@noop {} {\bibfield  {journal} {\bibinfo  {journal} {Comptes Rendus
  Physique}\ }\textbf {\bibinfo {volume} {14}},\ \bibinfo {pages} {857--870}
  (\bibinfo {year} {2013})}\BibitemShut {NoStop}%
\bibitem [{\citenamefont {{Armitage}}\ \emph {et~al.}(2017)\citenamefont
  {{Armitage}}, \citenamefont {{Mele}},\ and\ \citenamefont
  {{Vishwanath}}}]{AMV17}%
  \BibitemOpen
  \bibfield  {author} {\bibinfo {author} {\bibfnamefont {N.~P.}\ \bibnamefont
  {{Armitage}}}, \bibinfo {author} {\bibfnamefont {E.~J.}\ \bibnamefont
  {{Mele}}}, \ and\ \bibinfo {author} {\bibfnamefont {A.}~\bibnamefont
  {{Vishwanath}}},\ }\bibfield  {title} {\enquote {\bibinfo {title} {{Weyl and
  Dirac Semimetals in Three Dimensional Solids}},}\ }\href@noop {} {\bibfield
  {journal} {\bibinfo  {journal} {ArXiv e-prints}\ } (\bibinfo {year}
  {2017})},\ \Eprint {http://arxiv.org/abs/1705.01111} {arXiv:1705.01111
  [cond-mat.str-el]} \BibitemShut {NoStop}%
\bibitem [{\citenamefont {Nielsen}\ and\ \citenamefont
  {Ninomiya}(1981{\natexlab{a}})}]{NielNino81a}%
  \BibitemOpen
  \bibfield  {author} {\bibinfo {author} {\bibfnamefont {H.B.}\ \bibnamefont
  {Nielsen}}\ and\ \bibinfo {author} {\bibfnamefont {M.}~\bibnamefont
  {Ninomiya}},\ }\bibfield  {title} {\enquote {\bibinfo {title} {Absence of
  neutrinos on a lattice: {(I)}. {P}roof by homotopy theory},}\ }\href
  {\doibase 10.1016/0550-3213(81)90361-8} {\bibfield  {journal} {\bibinfo
  {journal} {Nucl. Phys. B}\ }\textbf {\bibinfo {volume} {185}},\ \bibinfo
  {pages} {20 -- 40} (\bibinfo {year} {1981}{\natexlab{a}})}\BibitemShut
  {NoStop}%
\bibitem [{\citenamefont {Nielsen}\ and\ \citenamefont
  {Ninomiya}(1981{\natexlab{b}})}]{NielNino81b}%
  \BibitemOpen
  \bibfield  {author} {\bibinfo {author} {\bibfnamefont {H.B.}\ \bibnamefont
  {Nielsen}}\ and\ \bibinfo {author} {\bibfnamefont {M.}~\bibnamefont
  {Ninomiya}},\ }\bibfield  {title} {\enquote {\bibinfo {title} {Absence of
  neutrinos on a lattice: ({II}). {I}ntuitive topological proof},}\ }\href
  {\doibase 10.1016/0550-3213(81)90524-1} {\bibfield  {journal} {\bibinfo
  {journal} {Nucl. Phys. B}\ }\textbf {\bibinfo {volume} {193}},\ \bibinfo
  {pages} {173 -- 194} (\bibinfo {year} {1981}{\natexlab{b}})}\BibitemShut
  {NoStop}%
\bibitem [{\citenamefont {Nielsen}\ and\ \citenamefont
  {Ninomiya}(1981{\natexlab{c}})}]{NielNino81c}%
  \BibitemOpen
  \bibfield  {author} {\bibinfo {author} {\bibfnamefont {H.B.}\ \bibnamefont
  {Nielsen}}\ and\ \bibinfo {author} {\bibfnamefont {M.}~\bibnamefont
  {Ninomiya}},\ }\bibfield  {title} {\enquote {\bibinfo {title} {A no-go
  theorem for regularizing chiral fermions},}\ }\href {\doibase
  http://dx.doi.org/10.1016/0370-2693(81)91026-1} {\bibfield  {journal}
  {\bibinfo  {journal} {Phys. Lett. B}\ }\textbf {\bibinfo {volume} {105}},\
  \bibinfo {pages} {219 -- 223} (\bibinfo {year}
  {1981}{\natexlab{c}})}\BibitemShut {NoStop}%
\bibitem [{\citenamefont {Nielsen}\ and\ \citenamefont
  {Ninomiya}(1983)}]{NielNino83}%
  \BibitemOpen
  \bibfield  {author} {\bibinfo {author} {\bibfnamefont {H.B.}\ \bibnamefont
  {Nielsen}}\ and\ \bibinfo {author} {\bibfnamefont {M.}~\bibnamefont
  {Ninomiya}},\ }\bibfield  {title} {\enquote {\bibinfo {title} {The
  {A}dler-{B}ell-{J}ackiw anomaly and {W}eyl fermions in a crystal},}\
  }\href@noop {} {\bibfield  {journal} {\bibinfo  {journal} {Phys. Lett.}\
  }\textbf {\bibinfo {volume} {130B}},\ \bibinfo {pages} {389--396} (\bibinfo
  {year} {1983})}\BibitemShut {NoStop}%
\bibitem [{\citenamefont {Chernodub}\ \emph {et~al.}(2014)\citenamefont
  {Chernodub}, \citenamefont {Cortijo}, \citenamefont {Grushin}, \citenamefont
  {Landsteiner},\ and\ \citenamefont {Vozmediano}}]{CCG14}%
  \BibitemOpen
  \bibfield  {author} {\bibinfo {author} {\bibfnamefont {Maxim~N.}\
  \bibnamefont {Chernodub}}, \bibinfo {author} {\bibfnamefont {Alberto}\
  \bibnamefont {Cortijo}}, \bibinfo {author} {\bibfnamefont {Adolfo~G.}\
  \bibnamefont {Grushin}}, \bibinfo {author} {\bibfnamefont {Karl}\
  \bibnamefont {Landsteiner}}, \ and\ \bibinfo {author} {\bibfnamefont
  {Mar\'{\i}a A.~H.}\ \bibnamefont {Vozmediano}},\ }\bibfield  {title}
  {\enquote {\bibinfo {title} {Condensed matter realization of the axial
  magnetic effect},}\ }\href {\doibase 10.1103/PhysRevB.89.081407} {\bibfield
  {journal} {\bibinfo  {journal} {Phys. Rev. B}\ }\textbf {\bibinfo {volume}
  {89}},\ \bibinfo {pages} {081407} (\bibinfo {year} {2014})}\BibitemShut
  {NoStop}%
\bibitem [{\citenamefont {Jiang}\ \emph {et~al.}(2015)\citenamefont {Jiang},
  \citenamefont {Jiang}, \citenamefont {Liu}, \citenamefont {Sun},\ and\
  \citenamefont {Xie}}]{Jiang2015}%
  \BibitemOpen
  \bibfield  {author} {\bibinfo {author} {\bibfnamefont {Q.-D.}\ \bibnamefont
  {Jiang}}, \bibinfo {author} {\bibfnamefont {H.}~\bibnamefont {Jiang}},
  \bibinfo {author} {\bibfnamefont {H.}~\bibnamefont {Liu}}, \bibinfo {author}
  {\bibfnamefont {Q.-F.}\ \bibnamefont {Sun}}, \ and\ \bibinfo {author}
  {\bibfnamefont {X.~C.}\ \bibnamefont {Xie}},\ }\bibfield  {title} {\enquote
  {\bibinfo {title} {{Topological Imbert-Fedorov Shift in Weyl Semimetals}},}\
  }\href {\doibase 10.1103/PhysRevLett.115.156602} {\bibfield  {journal}
  {\bibinfo  {journal} {Physical Review Letters}\ }\textbf {\bibinfo {volume}
  {115}},\ \bibinfo {pages} {156602} (\bibinfo {year} {2015})}\BibitemShut
  {NoStop}%
\bibitem [{\citenamefont {Yang}\ \emph {et~al.}(2015)\citenamefont {Yang},
  \citenamefont {Pan},\ and\ \citenamefont {Zhang}}]{Yang2015}%
  \BibitemOpen
  \bibfield  {author} {\bibinfo {author} {\bibfnamefont {S.~A.}\ \bibnamefont
  {Yang}}, \bibinfo {author} {\bibfnamefont {H.}~\bibnamefont {Pan}}, \ and\
  \bibinfo {author} {\bibfnamefont {F.}~\bibnamefont {Zhang}},\ }\bibfield
  {title} {\enquote {\bibinfo {title} {{Chirality-Dependent Hall Effect in Weyl
  Semimetals}},}\ }\href {\doibase 10.1103/PhysRevLett.115.156603} {\bibfield
  {journal} {\bibinfo  {journal} {Physical Review Letters}\ }\textbf {\bibinfo
  {volume} {115}},\ \bibinfo {pages} {156603} (\bibinfo {year}
  {2015})}\BibitemShut {NoStop}%
\bibitem [{\citenamefont {Schuster}\ \emph {et~al.}(2016)\citenamefont
  {Schuster}, \citenamefont {Iadecola}, \citenamefont {Chamon}, \citenamefont
  {Jackiw},\ and\ \citenamefont {Pi}}]{Schuster2016}%
  \BibitemOpen
  \bibfield  {author} {\bibinfo {author} {\bibfnamefont {T.}~\bibnamefont
  {Schuster}}, \bibinfo {author} {\bibfnamefont {T.}~\bibnamefont {Iadecola}},
  \bibinfo {author} {\bibfnamefont {C.}~\bibnamefont {Chamon}}, \bibinfo
  {author} {\bibfnamefont {R.}~\bibnamefont {Jackiw}}, \ and\ \bibinfo {author}
  {\bibfnamefont {S.-Y.}\ \bibnamefont {Pi}},\ }\bibfield  {title} {\enquote
  {\bibinfo {title} {{Dissipationless conductance in a topological coaxial
  cable}},}\ }\href {\doibase 10.1103/PhysRevB.94.115110} {\bibfield  {journal}
  {\bibinfo  {journal} {Physical Review B}\ }\textbf {\bibinfo {volume} {94}},\
  \bibinfo {pages} {115110} (\bibinfo {year} {2016})}\BibitemShut {NoStop}%
\bibitem [{\citenamefont {Cortijo}\ \emph {et~al.}(2016)\citenamefont
  {Cortijo}, \citenamefont {Kharzeev}, \citenamefont {Landsteiner},\ and\
  \citenamefont {Vozmediano}}]{cortijo2016strain}%
  \BibitemOpen
  \bibfield  {author} {\bibinfo {author} {\bibfnamefont {A.}~\bibnamefont
  {Cortijo}}, \bibinfo {author} {\bibfnamefont {D.}~\bibnamefont {Kharzeev}},
  \bibinfo {author} {\bibfnamefont {K.}~\bibnamefont {Landsteiner}}, \ and\
  \bibinfo {author} {\bibfnamefont {M.~A.~H.}\ \bibnamefont {Vozmediano}},\
  }\bibfield  {title} {\enquote {\bibinfo {title} {Strain-induced chiral
  magnetic effect in weyl semimetals},}\ }\href {\doibase
  10.1103/PhysRevB.94.241405} {\bibfield  {journal} {\bibinfo  {journal} {Phys.
  Rev. B}\ }\textbf {\bibinfo {volume} {94}},\ \bibinfo {pages} {241405}
  (\bibinfo {year} {2016})}\BibitemShut {NoStop}%
\bibitem [{\citenamefont {Cortijo}\ and\ \citenamefont
  {Zubkov}(2016)}]{cortijo2016emergent}%
  \BibitemOpen
  \bibfield  {author} {\bibinfo {author} {\bibfnamefont {A.}~\bibnamefont
  {Cortijo}}\ and\ \bibinfo {author} {\bibfnamefont {M.~A.}\ \bibnamefont
  {Zubkov}},\ }\bibfield  {title} {\enquote {\bibinfo {title} {Emergent gravity
  in the cubic tight-binding model of weyl semimetal in the presence of elastic
  deformations},}\ }\href@noop {} {\bibfield  {journal} {\bibinfo  {journal}
  {Annals of Physics}\ }\textbf {\bibinfo {volume} {366}},\ \bibinfo {pages}
  {45--56} (\bibinfo {year} {2016})}\BibitemShut {NoStop}%
\bibitem [{\citenamefont {Bevan}\ \emph {et~al.}(1997)\citenamefont {Bevan},
  \citenamefont {Manninen}, \citenamefont {Cook}, \citenamefont {Hook},
  \citenamefont {Hall}, \citenamefont {Vachaspati},\ and\ \citenamefont
  {Volovik}}]{bevan1997momentum}%
  \BibitemOpen
  \bibfield  {author} {\bibinfo {author} {\bibfnamefont {T.~D.~C.}\
  \bibnamefont {Bevan}}, \bibinfo {author} {\bibfnamefont {A.~J.}\ \bibnamefont
  {Manninen}}, \bibinfo {author} {\bibfnamefont {J.~B.}\ \bibnamefont {Cook}},
  \bibinfo {author} {\bibfnamefont {J.~R.}\ \bibnamefont {Hook}}, \bibinfo
  {author} {\bibfnamefont {H.~E.}\ \bibnamefont {Hall}}, \bibinfo {author}
  {\bibfnamefont {T.}~\bibnamefont {Vachaspati}}, \ and\ \bibinfo {author}
  {\bibfnamefont {G.~E.}\ \bibnamefont {Volovik}},\ }\bibfield  {title}
  {\enquote {\bibinfo {title} {Momentum creation by vortices in superfluid 3he
  as a model of primordial baryogenesis},}\ }\href {\doibase 10.1038/386689a0}
  {\bibfield  {journal} {\bibinfo  {journal} {Nature}\ }\textbf {\bibinfo
  {volume} {386}},\ \bibinfo {pages} {689--692} (\bibinfo {year}
  {1997})}\BibitemShut {NoStop}%
\bibitem [{\citenamefont {Klinkhamer}\ and\ \citenamefont
  {Volovik}(2005)}]{klinkhamer2005emergent}%
  \BibitemOpen
  \bibfield  {author} {\bibinfo {author} {\bibfnamefont {F.~R.}\ \bibnamefont
  {Klinkhamer}}\ and\ \bibinfo {author} {\bibfnamefont {G.~E.}\ \bibnamefont
  {Volovik}},\ }\bibfield  {title} {\enquote {\bibinfo {title} {Emergent cpt
  violation from the splitting of fermi points},}\ }\href@noop {} {\bibfield
  {journal} {\bibinfo  {journal} {International Journal of Modern Physics A}\
  }\textbf {\bibinfo {volume} {20}},\ \bibinfo {pages} {2795--2812} (\bibinfo
  {year} {2005})}\BibitemShut {NoStop}%
\bibitem [{\citenamefont {Son}\ and\ \citenamefont
  {Spivak}(2013)}]{son2013chiral}%
  \BibitemOpen
  \bibfield  {author} {\bibinfo {author} {\bibfnamefont {D.~T.}\ \bibnamefont
  {Son}}\ and\ \bibinfo {author} {\bibfnamefont {B.~Z.}\ \bibnamefont
  {Spivak}},\ }\bibfield  {title} {\enquote {\bibinfo {title} {{Chiral anomaly
  and classical negative magnetoresistance of Weyl metals}},}\ }\href {\doibase
  10.1103/PhysRevB.88.104412} {\bibfield  {journal} {\bibinfo  {journal} {Phys.
  Rev. B}\ }\textbf {\bibinfo {volume} {88}},\ \bibinfo {pages} {104412}
  (\bibinfo {year} {2013})}\BibitemShut {NoStop}%
\bibitem [{\citenamefont {Tchoumakov}\ \emph
  {et~al.}(2017{\natexlab{a}})\citenamefont {Tchoumakov}, \citenamefont
  {Civelli},\ and\ \citenamefont {Goerbig}}]{Tchoumakov2017b}%
  \BibitemOpen
  \bibfield  {author} {\bibinfo {author} {\bibfnamefont {S.}~\bibnamefont
  {Tchoumakov}}, \bibinfo {author} {\bibfnamefont {M.}~\bibnamefont {Civelli}},
  \ and\ \bibinfo {author} {\bibfnamefont {M.~O.}\ \bibnamefont {Goerbig}},\
  }\bibfield  {title} {\enquote {\bibinfo {title} {{Magnetic description of the
  Fermi arc in type-I and type-II Weyl semimetals}},}\ }\href {\doibase
  10.1103/PhysRevB.95.125306} {\bibfield  {journal} {\bibinfo  {journal}
  {Physical Review B}\ }\textbf {\bibinfo {volume} {95}},\ \bibinfo {pages}
  {125306} (\bibinfo {year} {2017}{\natexlab{a}})}\BibitemShut {NoStop}%
\bibitem [{\citenamefont {Grushin}\ \emph {et~al.}(2015)\citenamefont
  {Grushin}, \citenamefont {Venderbos},\ and\ \citenamefont
  {Bardarson}}]{Grushin2015}%
  \BibitemOpen
  \bibfield  {author} {\bibinfo {author} {\bibfnamefont {A.~G.}\ \bibnamefont
  {Grushin}}, \bibinfo {author} {\bibfnamefont {J.~W.~F.}\ \bibnamefont
  {Venderbos}}, \ and\ \bibinfo {author} {\bibfnamefont {J.~H.}\ \bibnamefont
  {Bardarson}},\ }\bibfield  {title} {\enquote {\bibinfo {title} {{Coexistence
  of Fermi arcs with two-dimensional gapless Dirac states}},}\ }\href@noop {}
  {\bibfield  {journal} {\bibinfo  {journal} {Physical Review B}\ }\textbf
  {\bibinfo {volume} {91}} (\bibinfo {year} {2015})}\BibitemShut {NoStop}%
\bibitem [{\citenamefont {Lau}\ \emph {et~al.}(2017)\citenamefont {Lau},
  \citenamefont {Koepernik}, \citenamefont {van~den Brink},\ and\ \citenamefont
  {Ortix}}]{Lau2017}%
  \BibitemOpen
  \bibfield  {author} {\bibinfo {author} {\bibfnamefont {A.}~\bibnamefont
  {Lau}}, \bibinfo {author} {\bibfnamefont {K.}~\bibnamefont {Koepernik}},
  \bibinfo {author} {\bibfnamefont {J.}~\bibnamefont {van~den Brink}}, \ and\
  \bibinfo {author} {\bibfnamefont {C.}~\bibnamefont {Ortix}},\ }\bibfield
  {title} {\enquote {\bibinfo {title} {{Coexistence of Fermi arcs and Dirac
  cones on the surface of time-reversal invariant Weyl semimetals}},}\ }\href
  {http://arxiv.org/abs/1701.01660} {\  (\bibinfo {year} {2017})},\ \Eprint
  {http://arxiv.org/abs/1701.01660} {arXiv:1701.01660} \BibitemShut {NoStop}%
\bibitem [{\citenamefont {Inhofer}\ \emph {et~al.}(2017)\citenamefont
  {Inhofer}, \citenamefont {Tchoumakov}, \citenamefont {Assaf}, \citenamefont
  {F{\`{e}}ve}, \citenamefont {Berroir}, \citenamefont {Jouffrey},
  \citenamefont {Carpentier}, \citenamefont {Goerbig}, \citenamefont
  {Pla{\c{c}}ais}, \citenamefont {Bendias}, \citenamefont {Mahler},
  \citenamefont {Bocquillon}, \citenamefont {Schlereth}, \citenamefont
  {Br{\"{u}}ne}, \citenamefont {Buhmann},\ and\ \citenamefont
  {Molenkamp}}]{Inhofer2017}%
  \BibitemOpen
  \bibfield  {author} {\bibinfo {author} {\bibfnamefont {A.}~\bibnamefont
  {Inhofer}}, \bibinfo {author} {\bibfnamefont {S.}~\bibnamefont {Tchoumakov}},
  \bibinfo {author} {\bibfnamefont {B.~A.}\ \bibnamefont {Assaf}}, \bibinfo
  {author} {\bibfnamefont {G.}~\bibnamefont {F{\`{e}}ve}}, \bibinfo {author}
  {\bibfnamefont {J.~M.}\ \bibnamefont {Berroir}}, \bibinfo {author}
  {\bibfnamefont {V.}~\bibnamefont {Jouffrey}}, \bibinfo {author}
  {\bibfnamefont {D.}~\bibnamefont {Carpentier}}, \bibinfo {author}
  {\bibfnamefont {M.}~\bibnamefont {Goerbig}}, \bibinfo {author} {\bibfnamefont
  {B.}~\bibnamefont {Pla{\c{c}}ais}}, \bibinfo {author} {\bibfnamefont
  {K.}~\bibnamefont {Bendias}}, \bibinfo {author} {\bibfnamefont {D.~M.}\
  \bibnamefont {Mahler}}, \bibinfo {author} {\bibfnamefont {E.}~\bibnamefont
  {Bocquillon}}, \bibinfo {author} {\bibfnamefont {R.}~\bibnamefont
  {Schlereth}}, \bibinfo {author} {\bibfnamefont {C.}~\bibnamefont
  {Br{\"{u}}ne}}, \bibinfo {author} {\bibfnamefont {H.}~\bibnamefont
  {Buhmann}}, \ and\ \bibinfo {author} {\bibfnamefont {L.~W.}\ \bibnamefont
  {Molenkamp}},\ }\bibfield  {title} {\enquote {\bibinfo {title} {{Topological
  confined massive surface states in strained bulk HgTe probed by RF
  compressibility}},}\ }\href {http://arxiv.org/abs/1704.04045} {\  (\bibinfo
  {year} {2017})},\ \Eprint {http://arxiv.org/abs/1704.04045}
  {arXiv:1704.04045} \BibitemShut {NoStop}%
\bibitem [{\citenamefont {Tchoumakov}\ \emph
  {et~al.}(2017{\natexlab{b}})\citenamefont {Tchoumakov}, \citenamefont
  {Jouffrey}, \citenamefont {Inhofer}, \citenamefont {Bocquillon},
  \citenamefont {Pla{\c{c}}ais}, \citenamefont {Carpentier},\ and\
  \citenamefont {Goerbig}}]{Tchoumakov2017}%
  \BibitemOpen
  \bibfield  {author} {\bibinfo {author} {\bibfnamefont {S.}~\bibnamefont
  {Tchoumakov}}, \bibinfo {author} {\bibfnamefont {V.}~\bibnamefont
  {Jouffrey}}, \bibinfo {author} {\bibfnamefont {A.}~\bibnamefont {Inhofer}},
  \bibinfo {author} {\bibfnamefont {E.}~\bibnamefont {Bocquillon}}, \bibinfo
  {author} {\bibfnamefont {B.}~\bibnamefont {Pla{\c{c}}ais}}, \bibinfo {author}
  {\bibfnamefont {D.}~\bibnamefont {Carpentier}}, \ and\ \bibinfo {author}
  {\bibfnamefont {M.~O.}\ \bibnamefont {Goerbig}},\ }\bibfield  {title}
  {\enquote {\bibinfo {title} {{Massive states in topological
  heterojunctions}},}\ }\href {http://arxiv.org/abs/1704.08954} {\  (\bibinfo
  {year} {2017}{\natexlab{b}})},\ \Eprint {http://arxiv.org/abs/1704.08954}
  {arXiv:1704.08954} \BibitemShut {NoStop}%
\bibitem [{\citenamefont {Ho}\ \emph {et~al.}(2017)\citenamefont {Ho},
  \citenamefont {Castro},\ and\ \citenamefont {Cazalilla}}]{Ho2017}%
  \BibitemOpen
  \bibfield  {author} {\bibinfo {author} {\bibfnamefont {Y.-H.}\ \bibnamefont
  {Ho}}, \bibinfo {author} {\bibfnamefont {E.~V.}\ \bibnamefont {Castro}}, \
  and\ \bibinfo {author} {\bibfnamefont {M.~A.}\ \bibnamefont {Cazalilla}},\
  }\bibfield  {title} {\enquote {\bibinfo {title} {{The Haldane model under
  nonuniform strain}},}\ }\href {http://arxiv.org/abs/1705.08600} {\  (\bibinfo
  {year} {2017})},\ \Eprint {http://arxiv.org/abs/1705.08600}
  {arXiv:1705.08600} \BibitemShut {NoStop}%
\bibitem [{\citenamefont {Bertlmann}(2000)}]{bertlmann2000anomalies}%
  \BibitemOpen
  \bibfield  {author} {\bibinfo {author} {\bibfnamefont {R.~A}\ \bibnamefont
  {Bertlmann}},\ }\href@noop {} {\emph {\bibinfo {title} {Anomalies in quantum
  field theory}}},\ Vol.~\bibinfo {volume} {91}\ (\bibinfo  {publisher} {Oxford
  University Press},\ \bibinfo {year} {2000})\BibitemShut {NoStop}%
\bibitem [{\citenamefont {Landsteiner}(2014)}]{Land14}%
  \BibitemOpen
  \bibfield  {author} {\bibinfo {author} {\bibfnamefont {Karl}\ \bibnamefont
  {Landsteiner}},\ }\bibfield  {title} {\enquote {\bibinfo {title} {{Anomalous
  transport of Weyl fermions in Weyl semimetals}},}\ }\href {\doibase
  10.1103/PhysRevB.89.075124} {\bibfield  {journal} {\bibinfo  {journal} {Phys.
  Rev. B}\ }\textbf {\bibinfo {volume} {89}},\ \bibinfo {pages} {75124}
  (\bibinfo {year} {2014})}\BibitemShut {NoStop}%
\bibitem [{\citenamefont {Landsteiner}(2016)}]{Landsteiner2016}%
  \BibitemOpen
  \bibfield  {author} {\bibinfo {author} {\bibfnamefont {Karl}\ \bibnamefont
  {Landsteiner}},\ }\bibfield  {title} {\enquote {\bibinfo {title} {Notes on
  anomaly induced transport},}\ }\href@noop {} {\bibfield  {journal} {\bibinfo
  {journal} {arXiv:1610.04413}\ } (\bibinfo {year} {2016})}\BibitemShut
  {NoStop}%
\bibitem [{\citenamefont {{Krinner}}\ \emph {et~al.}(2017)\citenamefont
  {{Krinner}}, \citenamefont {{Esslinger}},\ and\ \citenamefont
  {{Brantut}}}]{BrantutReview}%
  \BibitemOpen
  \bibfield  {author} {\bibinfo {author} {\bibfnamefont {S.}~\bibnamefont
  {{Krinner}}}, \bibinfo {author} {\bibfnamefont {T.}~\bibnamefont
  {{Esslinger}}}, \ and\ \bibinfo {author} {\bibfnamefont {J.-P.}\ \bibnamefont
  {{Brantut}}},\ }\bibfield  {title} {\enquote {\bibinfo {title} {{Two-terminal
  transport measurements with cold atoms}},}\ }\href@noop {} {\bibfield
  {journal} {\bibinfo  {journal} {ArXiv e-prints}\ } (\bibinfo {year}
  {2017})},\ \Eprint {http://arxiv.org/abs/1706.01085} {arXiv:1706.01085
  [cond-mat.quant-gas]} \BibitemShut {NoStop}%
\bibitem [{\citenamefont {Price}\ and\ \citenamefont
  {Cooper}(2012)}]{price2012mapping}%
  \BibitemOpen
  \bibfield  {author} {\bibinfo {author} {\bibfnamefont {H.~M.}\ \bibnamefont
  {Price}}\ and\ \bibinfo {author} {\bibfnamefont {N.~R.}\ \bibnamefont
  {Cooper}},\ }\bibfield  {title} {\enquote {\bibinfo {title} {Mapping the
  berry curvature from semiclassical dynamics in optical lattices},}\
  }\href@noop {} {\bibfield  {journal} {\bibinfo  {journal} {Physical Review
  A}\ }\textbf {\bibinfo {volume} {85}},\ \bibinfo {pages} {033620} (\bibinfo
  {year} {2012})}\BibitemShut {NoStop}%
\bibitem [{\citenamefont {Dauphin}\ and\ \citenamefont
  {Goldman}(2013)}]{dauphin2013extracting}%
  \BibitemOpen
  \bibfield  {author} {\bibinfo {author} {\bibfnamefont {A.}~\bibnamefont
  {Dauphin}}\ and\ \bibinfo {author} {\bibfnamefont {N.}~\bibnamefont
  {Goldman}},\ }\bibfield  {title} {\enquote {\bibinfo {title} {Extracting the
  chern number from the dynamics of a fermi gas: Implementing a quantum hall
  bar for cold atoms},}\ }\href@noop {} {\bibfield  {journal} {\bibinfo
  {journal} {Physical review letters}\ }\textbf {\bibinfo {volume} {111}},\
  \bibinfo {pages} {135302} (\bibinfo {year} {2013})}\BibitemShut {NoStop}%
\bibitem [{\citenamefont {Duca}\ \emph {et~al.}(2014)\citenamefont {Duca},
  \citenamefont {Li}, \citenamefont {Reitter}, \citenamefont {Bloch},
  \citenamefont {Schleier-Smith},\ and\ \citenamefont
  {Schneider}}]{duca2014aharonov}%
  \BibitemOpen
  \bibfield  {author} {\bibinfo {author} {\bibfnamefont {L.}~\bibnamefont
  {Duca}}, \bibinfo {author} {\bibfnamefont {T.}~\bibnamefont {Li}}, \bibinfo
  {author} {\bibfnamefont {M.}~\bibnamefont {Reitter}}, \bibinfo {author}
  {\bibfnamefont {I.}~\bibnamefont {Bloch}}, \bibinfo {author} {\bibfnamefont
  {M.}~\bibnamefont {Schleier-Smith}}, \ and\ \bibinfo {author} {\bibfnamefont
  {U.}~\bibnamefont {Schneider}},\ }\bibfield  {title} {\enquote {\bibinfo
  {title} {An aharonov-bohm interferometer for determining bloch band
  topology},}\ }\href@noop {} {\bibfield  {journal} {\bibinfo  {journal}
  {Science}\ ,\ \bibinfo {pages} {1259052}} (\bibinfo {year}
  {2014})}\BibitemShut {NoStop}%
\bibitem [{\citenamefont {Price}\ \emph {et~al.}(2016)\citenamefont {Price},
  \citenamefont {Zilberberg}, \citenamefont {Ozawa}, \citenamefont
  {Carusotto},\ and\ \citenamefont {Goldman}}]{price2016measurement}%
  \BibitemOpen
  \bibfield  {author} {\bibinfo {author} {\bibfnamefont {H.~M.}\ \bibnamefont
  {Price}}, \bibinfo {author} {\bibfnamefont {O.}~\bibnamefont {Zilberberg}},
  \bibinfo {author} {\bibfnamefont {T.}~\bibnamefont {Ozawa}}, \bibinfo
  {author} {\bibfnamefont {I.}~\bibnamefont {Carusotto}}, \ and\ \bibinfo
  {author} {\bibfnamefont {N.}~\bibnamefont {Goldman}},\ }\bibfield  {title}
  {\enquote {\bibinfo {title} {Measurement of chern numbers through
  center-of-mass responses},}\ }\href {\doibase 10.1103/PhysRevB.93.245113}
  {\bibfield  {journal} {\bibinfo  {journal} {Phys. Rev. B}\ }\textbf {\bibinfo
  {volume} {93}},\ \bibinfo {pages} {245113} (\bibinfo {year}
  {2016})}\BibitemShut {NoStop}%
\bibitem [{\citenamefont {Vazifeh}\ and\ \citenamefont
  {Franz}(2013)}]{Vazifeh:2013fe}%
  \BibitemOpen
  \bibfield  {author} {\bibinfo {author} {\bibfnamefont {M.~M.}\ \bibnamefont
  {Vazifeh}}\ and\ \bibinfo {author} {\bibfnamefont {M.}~\bibnamefont
  {Franz}},\ }\bibfield  {title} {\enquote {\bibinfo {title} {{Electromagnetic
  Response of Weyl Semimetals}},}\ }\href@noop {} {\bibfield  {journal}
  {\bibinfo  {journal} {Phys. Rev. Lett.}\ }\textbf {\bibinfo {volume} {111}},\
  \bibinfo {pages} {027201} (\bibinfo {year} {2013})}\BibitemShut {NoStop}%
\bibitem [{\citenamefont {Dub\ifmmode~\check{c}\else \v{c}\fi{}ek}\ \emph
  {et~al.}(2015)\citenamefont {Dub\ifmmode~\check{c}\else \v{c}\fi{}ek},
  \citenamefont {Kennedy}, \citenamefont {Lu}, \citenamefont {Ketterle},
  \citenamefont {Solja\ifmmode \check{c}\else
  \v{c}\fi{}i\ifmmode~\acute{c}\else \'{c}\fi{}},\ and\ \citenamefont
  {Buljan}}]{dubeck2015weyl}%
  \BibitemOpen
  \bibfield  {author} {\bibinfo {author} {\bibfnamefont {Tena}\ \bibnamefont
  {Dub\ifmmode~\check{c}\else \v{c}\fi{}ek}}, \bibinfo {author} {\bibfnamefont
  {Colin~J.}\ \bibnamefont {Kennedy}}, \bibinfo {author} {\bibfnamefont {Ling}\
  \bibnamefont {Lu}}, \bibinfo {author} {\bibfnamefont {Wolfgang}\ \bibnamefont
  {Ketterle}}, \bibinfo {author} {\bibfnamefont {Marin}\ \bibnamefont
  {Solja\ifmmode \check{c}\else \v{c}\fi{}i\ifmmode~\acute{c}\else
  \'{c}\fi{}}}, \ and\ \bibinfo {author} {\bibfnamefont {Hrvoje}\ \bibnamefont
  {Buljan}},\ }\bibfield  {title} {\enquote {\bibinfo {title} {Weyl points in
  three-dimensional optical lattices: Synthetic magnetic monopoles in momentum
  space},}\ }\href {\doibase 10.1103/PhysRevLett.114.225301} {\bibfield
  {journal} {\bibinfo  {journal} {Phys. Rev. Lett.}\ }\textbf {\bibinfo
  {volume} {114}},\ \bibinfo {pages} {225301} (\bibinfo {year}
  {2015})}\BibitemShut {NoStop}%
\bibitem [{Note1()}]{Note1}%
  \BibitemOpen
  \bibinfo {note} {Note that, for the pair of Weyl points at $\protect \mathbf
  {k}_{W,\pm }=(-\pi /2,-\pi /2,\pm \protect \qopname \relax
  o{cos}^{-1}(-M/2))$, the analysis remains unchanged except for a flip in the
  chiralities. On the other hand, choosing the pair of Weyl points located at
  the same value of $k_z$, for instance the pair $\protect \mathbf {k}_{W,\pm
  }=(\pm \pi /2,\pm \pi /2,\protect \qopname \relax o{cos}^{-1}(-M/2))$ does
  not result in axial fields since any variation in $M$ leads to an overall
  shift in the Weyl points and not a relative shift between the two. The
  effective low-energy Hamiltonian for such pair does not posses an axial gauge
  potential.}\BibitemShut {Stop}%
\bibitem [{\citenamefont {de~Juan}\ \emph {et~al.}(2012)\citenamefont
  {de~Juan}, \citenamefont {Sturla},\ and\ \citenamefont
  {Vozmediano}}]{DeJuan2012}%
  \BibitemOpen
  \bibfield  {author} {\bibinfo {author} {\bibfnamefont {F.}~\bibnamefont
  {de~Juan}}, \bibinfo {author} {\bibfnamefont {M.}~\bibnamefont {Sturla}}, \
  and\ \bibinfo {author} {\bibfnamefont {M.~A.~H.}\ \bibnamefont
  {Vozmediano}},\ }\bibfield  {title} {\enquote {\bibinfo {title} {{Space
  Dependent Fermi Velocity in Strained Graphene}},}\ }\href {\doibase
  10.1103/PhysRevLett.108.227205} {\bibfield  {journal} {\bibinfo  {journal}
  {Physical Review Letters}\ }\textbf {\bibinfo {volume} {108}},\ \bibinfo
  {pages} {227205} (\bibinfo {year} {2012})}\BibitemShut {NoStop}%
\bibitem [{\citenamefont {Kolovsky}(2011)}]{kolovsky2011creating}%
  \BibitemOpen
  \bibfield  {author} {\bibinfo {author} {\bibfnamefont {A.~R.}\ \bibnamefont
  {Kolovsky}},\ }\bibfield  {title} {\enquote {\bibinfo {title} {Creating
  artificial magnetic fields for cold atoms by photon-assisted tunneling},}\
  }\href@noop {} {\bibfield  {journal} {\bibinfo  {journal} {EPL (Europhysics
  Letters)}\ }\textbf {\bibinfo {volume} {93}},\ \bibinfo {pages} {20003}
  (\bibinfo {year} {2011})}\BibitemShut {NoStop}%
\bibitem [{\citenamefont {Bermudez}\ \emph {et~al.}(2011)\citenamefont
  {Bermudez}, \citenamefont {Schaetz},\ and\ \citenamefont
  {Porras}}]{bermudez2011synthetic}%
  \BibitemOpen
  \bibfield  {author} {\bibinfo {author} {\bibfnamefont {A.}~\bibnamefont
  {Bermudez}}, \bibinfo {author} {\bibfnamefont {T.}~\bibnamefont {Schaetz}}, \
  and\ \bibinfo {author} {\bibfnamefont {D.}~\bibnamefont {Porras}},\
  }\bibfield  {title} {\enquote {\bibinfo {title} {Synthetic gauge fields for
  vibrational excitations of trapped ions},}\ }\href@noop {} {\bibfield
  {journal} {\bibinfo  {journal} {Physical Review Letters}\ }\textbf {\bibinfo
  {volume} {107}},\ \bibinfo {pages} {150501} (\bibinfo {year}
  {2011})}\BibitemShut {NoStop}%
\bibitem [{\citenamefont {Goldman}\ \emph {et~al.}(2015)\citenamefont
  {Goldman}, \citenamefont {Dalibard}, \citenamefont {Aidelsburger},\ and\
  \citenamefont {Cooper}}]{goldman2015periodically}%
  \BibitemOpen
  \bibfield  {author} {\bibinfo {author} {\bibfnamefont {N.}~\bibnamefont
  {Goldman}}, \bibinfo {author} {\bibfnamefont {J.}~\bibnamefont {Dalibard}},
  \bibinfo {author} {\bibfnamefont {M.}~\bibnamefont {Aidelsburger}}, \ and\
  \bibinfo {author} {\bibfnamefont {N.~R.}\ \bibnamefont {Cooper}},\ }\bibfield
   {title} {\enquote {\bibinfo {title} {Periodically driven quantum matter: The
  case of resonant modulations},}\ }\href {\doibase 10.1103/PhysRevA.91.033632}
  {\bibfield  {journal} {\bibinfo  {journal} {Phys. Rev. A}\ }\textbf {\bibinfo
  {volume} {91}},\ \bibinfo {pages} {033632} (\bibinfo {year}
  {2015})}\BibitemShut {NoStop}%
\bibitem [{\citenamefont {Creffield}\ \emph {et~al.}(2016)\citenamefont
  {Creffield}, \citenamefont {Pieplow}, \citenamefont {Sols},\ and\
  \citenamefont {Goldman}}]{creffield2016realization}%
  \BibitemOpen
  \bibfield  {author} {\bibinfo {author} {\bibfnamefont {C.~E.}\ \bibnamefont
  {Creffield}}, \bibinfo {author} {\bibfnamefont {G.}~\bibnamefont {Pieplow}},
  \bibinfo {author} {\bibfnamefont {F.}~\bibnamefont {Sols}}, \ and\ \bibinfo
  {author} {\bibfnamefont {N.}~\bibnamefont {Goldman}},\ }\bibfield  {title}
  {\enquote {\bibinfo {title} {Realization of uniform synthetic magnetic fields
  by periodically shaking an optical square lattice},}\ }\href@noop {}
  {\bibfield  {journal} {\bibinfo  {journal} {New Journal of Physics}\ }\textbf
  {\bibinfo {volume} {18}},\ \bibinfo {pages} {093013} (\bibinfo {year}
  {2016})}\BibitemShut {NoStop}%
\bibitem [{\citenamefont {Goldman}\ \emph {et~al.}(2013)\citenamefont
  {Goldman}, \citenamefont {Gerbier},\ and\ \citenamefont
  {Lewenstein}}]{goldman2013realizing}%
  \BibitemOpen
  \bibfield  {author} {\bibinfo {author} {\bibfnamefont {N.}~\bibnamefont
  {Goldman}}, \bibinfo {author} {\bibfnamefont {F.}~\bibnamefont {Gerbier}}, \
  and\ \bibinfo {author} {\bibfnamefont {M.}~\bibnamefont {Lewenstein}},\
  }\bibfield  {title} {\enquote {\bibinfo {title} {Realizing non-abelian gauge
  potentials in optical square lattices: an application to atomic {C}hern
  insulators},}\ }\href@noop {} {\bibfield  {journal} {\bibinfo  {journal}
  {Journal of Physics B: Atomic, Molecular and Optical Physics}\ }\textbf
  {\bibinfo {volume} {46}},\ \bibinfo {pages} {134010} (\bibinfo {year}
  {2013})}\BibitemShut {NoStop}%
\bibitem [{\citenamefont {Jaksch}\ and\ \citenamefont
  {Zoller}(2003)}]{jaksch2003creation}%
  \BibitemOpen
  \bibfield  {author} {\bibinfo {author} {\bibfnamefont {D.}~\bibnamefont
  {Jaksch}}\ and\ \bibinfo {author} {\bibfnamefont {P.}~\bibnamefont
  {Zoller}},\ }\bibfield  {title} {\enquote {\bibinfo {title} {Creation of
  effective magnetic fields in optical lattices: the hofstadter butterfly for
  cold neutral atoms},}\ }\href@noop {} {\bibfield  {journal} {\bibinfo
  {journal} {New Journal of Physics}\ }\textbf {\bibinfo {volume} {5}},\
  \bibinfo {pages} {56} (\bibinfo {year} {2003})}\BibitemShut {NoStop}%
\bibitem [{\citenamefont {Osterloh}\ \emph {et~al.}(2005)\citenamefont
  {Osterloh}, \citenamefont {Baig}, \citenamefont {Santos}, \citenamefont
  {Zoller},\ and\ \citenamefont {Lewenstein}}]{osterloh2005cold}%
  \BibitemOpen
  \bibfield  {author} {\bibinfo {author} {\bibfnamefont {K.}~\bibnamefont
  {Osterloh}}, \bibinfo {author} {\bibfnamefont {M.}~\bibnamefont {Baig}},
  \bibinfo {author} {\bibfnamefont {L.}~\bibnamefont {Santos}}, \bibinfo
  {author} {\bibfnamefont {P.}~\bibnamefont {Zoller}}, \ and\ \bibinfo {author}
  {\bibfnamefont {M.}~\bibnamefont {Lewenstein}},\ }\bibfield  {title}
  {\enquote {\bibinfo {title} {Cold atoms in non-abelian gauge potentials: from
  the hofstadter" moth" to lattice gauge theory},}\ }\href@noop {} {\bibfield
  {journal} {\bibinfo  {journal} {Physical Review Letters}\ }\textbf {\bibinfo
  {volume} {95}},\ \bibinfo {pages} {010403} (\bibinfo {year}
  {2005})}\BibitemShut {NoStop}%
\bibitem [{\citenamefont {Tarruell}\ \emph {et~al.}(2012)\citenamefont
  {Tarruell}, \citenamefont {Greif}, \citenamefont {Uehlinger}, \citenamefont
  {Jotzu},\ and\ \citenamefont {Esslinger}}]{tarruell2012creating}%
  \BibitemOpen
  \bibfield  {author} {\bibinfo {author} {\bibfnamefont {L.}~\bibnamefont
  {Tarruell}}, \bibinfo {author} {\bibfnamefont {D.}~\bibnamefont {Greif}},
  \bibinfo {author} {\bibfnamefont {T.}~\bibnamefont {Uehlinger}}, \bibinfo
  {author} {\bibfnamefont {G.}~\bibnamefont {Jotzu}}, \ and\ \bibinfo {author}
  {\bibfnamefont {T.}~\bibnamefont {Esslinger}},\ }\bibfield  {title} {\enquote
  {\bibinfo {title} {Creating, moving and merging dirac points with a fermi gas
  in a tunable honeycomb lattice},}\ }\href@noop {} {\bibfield  {journal}
  {\bibinfo  {journal} {Nature}\ }\textbf {\bibinfo {volume} {483}},\ \bibinfo
  {pages} {302--305} (\bibinfo {year} {2012})}\BibitemShut {NoStop}%
\bibitem [{\citenamefont {Goldman}\ \emph {et~al.}(2016)\citenamefont
  {Goldman}, \citenamefont {Jotzu}, \citenamefont {Messer}, \citenamefont
  {G{\"o}rg}, \citenamefont {Desbuquois},\ and\ \citenamefont
  {Esslinger}}]{goldman2016creating}%
  \BibitemOpen
  \bibfield  {author} {\bibinfo {author} {\bibfnamefont {N.}~\bibnamefont
  {Goldman}}, \bibinfo {author} {\bibfnamefont {G.}~\bibnamefont {Jotzu}},
  \bibinfo {author} {\bibfnamefont {M.}~\bibnamefont {Messer}}, \bibinfo
  {author} {\bibfnamefont {F.}~\bibnamefont {G{\"o}rg}}, \bibinfo {author}
  {\bibfnamefont {R.}~\bibnamefont {Desbuquois}}, \ and\ \bibinfo {author}
  {\bibfnamefont {T.}~\bibnamefont {Esslinger}},\ }\bibfield  {title} {\enquote
  {\bibinfo {title} {Creating topological interfaces and detecting chiral edge
  modes in a two-dimensional optical lattice},}\ }\href@noop {} {\bibfield
  {journal} {\bibinfo  {journal} {Physical Review A}\ }\textbf {\bibinfo
  {volume} {94}},\ \bibinfo {pages} {043611} (\bibinfo {year}
  {2016})}\BibitemShut {NoStop}%
\bibitem [{\citenamefont {Xiao}\ \emph {et~al.}(2010)\citenamefont {Xiao},
  \citenamefont {Chang},\ and\ \citenamefont {Niu}}]{xiao2010berry}%
  \BibitemOpen
  \bibfield  {author} {\bibinfo {author} {\bibfnamefont {D.}~\bibnamefont
  {Xiao}}, \bibinfo {author} {\bibfnamefont {M.-C.}\ \bibnamefont {Chang}}, \
  and\ \bibinfo {author} {\bibfnamefont {Q.}~\bibnamefont {Niu}},\ }\bibfield
  {title} {\enquote {\bibinfo {title} {Berry phase effects on electronic
  properties},}\ }\href {\doibase 10.1103/RevModPhys.82.1959} {\bibfield
  {journal} {\bibinfo  {journal} {Rev. Mod. Phys.}\ }\textbf {\bibinfo {volume}
  {82}},\ \bibinfo {pages} {1959--2007} (\bibinfo {year} {2010})}\BibitemShut
  {NoStop}%
\bibitem [{\citenamefont {Sundaram}\ and\ \citenamefont
  {Niu}(1999)}]{sundaram1999wavepacket}%
  \BibitemOpen
  \bibfield  {author} {\bibinfo {author} {\bibfnamefont {Ganesh}\ \bibnamefont
  {Sundaram}}\ and\ \bibinfo {author} {\bibfnamefont {Qian}\ \bibnamefont
  {Niu}},\ }\bibfield  {title} {\enquote {\bibinfo {title} {Wave-packet
  dynamics in slowly perturbed crystals: Gradient corrections and berry-phase
  effects},}\ }\href {\doibase 10.1103/PhysRevB.59.14915} {\bibfield  {journal}
  {\bibinfo  {journal} {Phys. Rev. B}\ }\textbf {\bibinfo {volume} {59}},\
  \bibinfo {pages} {14915--14925} (\bibinfo {year} {1999})}\BibitemShut
  {NoStop}%
\bibitem [{Note2()}]{Note2}%
  \BibitemOpen
  \bibinfo {note} {The parabolic spectrum at $t=0$ is an artifact of our
  initial condition $M_0=2$, and is inconsequential to our results as for any
  $t>0$, the system is a Weyl semimetal.}\BibitemShut {Stop}%
\bibitem [{\citenamefont {Zhou}\ \emph {et~al.}(2013)\citenamefont {Zhou},
  \citenamefont {Jiang}, \citenamefont {Niu},\ and\ \citenamefont
  {Shi}}]{Zhou2013}%
  \BibitemOpen
  \bibfield  {author} {\bibinfo {author} {\bibfnamefont {J.-H.}\ \bibnamefont
  {Zhou}}, \bibinfo {author} {\bibfnamefont {H.}~\bibnamefont {Jiang}},
  \bibinfo {author} {\bibfnamefont {Q.}~\bibnamefont {Niu}}, \ and\ \bibinfo
  {author} {\bibfnamefont {R}~\bibnamefont {Shi}, \bibfnamefont {J}},\
  }\bibfield  {title} {\enquote {\bibinfo {title} {{Topological Invariants of
  Metals and the Related Physical Effects}},}\ }\href@noop {} {\bibfield
  {journal} {\bibinfo  {journal} {Chinese Physics Letters}\ }\textbf {\bibinfo
  {volume} {30}},\ \bibinfo {pages} {027101} (\bibinfo {year}
  {2013})}\BibitemShut {NoStop}%
\bibitem [{\citenamefont {Kharzeev}(2014)}]{Kharzeev2014}%
  \BibitemOpen
  \bibfield  {author} {\bibinfo {author} {\bibfnamefont {D.~E.}\ \bibnamefont
  {Kharzeev}},\ }\bibfield  {title} {\enquote {\bibinfo {title} {{The Chiral
  Magnetic Effect and anomaly-induced transport}},}\ }\href {\doibase
  10.1016/j.ppnp.2014.01.002} {\bibfield  {journal} {\bibinfo  {journal}
  {Progress in Particle and Nuclear Physics}\ }\textbf {\bibinfo {volume}
  {75}},\ \bibinfo {pages} {133--151} (\bibinfo {year} {2014})}\BibitemShut
  {NoStop}%
\bibitem [{\citenamefont {Zyuzin}\ \emph {et~al.}(2012)\citenamefont {Zyuzin},
  \citenamefont {Wu},\ and\ \citenamefont {Burkov}}]{Zyuzin2012a}%
  \BibitemOpen
  \bibfield  {author} {\bibinfo {author} {\bibfnamefont {A.~A.}\ \bibnamefont
  {Zyuzin}}, \bibinfo {author} {\bibfnamefont {S.}~\bibnamefont {Wu}}, \ and\
  \bibinfo {author} {\bibfnamefont {A.~A.}\ \bibnamefont {Burkov}},\ }\bibfield
   {title} {\enquote {\bibinfo {title} {{Weyl semimetal with broken time
  reversal and inversion symmetries}},}\ }\href {\doibase
  10.1103/PhysRevB.85.165110} {\bibfield  {journal} {\bibinfo  {journal}
  {Physical Review B}\ }\textbf {\bibinfo {volume} {85}},\ \bibinfo {pages}
  {165110} (\bibinfo {year} {2012})}\BibitemShut {NoStop}%
\bibitem [{\citenamefont {Fukushima}\ \emph {et~al.}(2008)\citenamefont
  {Fukushima}, \citenamefont {Kharzeev},\ and\ \citenamefont
  {Warringa}}]{Fukushima2008}%
  \BibitemOpen
  \bibfield  {author} {\bibinfo {author} {\bibfnamefont {K.}~\bibnamefont
  {Fukushima}}, \bibinfo {author} {\bibfnamefont {D.~E.}\ \bibnamefont
  {Kharzeev}}, \ and\ \bibinfo {author} {\bibfnamefont {H.~J.}\ \bibnamefont
  {Warringa}},\ }\bibfield  {title} {\enquote {\bibinfo {title} {{Chiral
  magnetic effect}},}\ }\href {\doibase 10.1103/PhysRevD.78.074033} {\bibfield
  {journal} {\bibinfo  {journal} {Physical Review D}\ }\textbf {\bibinfo
  {volume} {78}},\ \bibinfo {pages} {074033} (\bibinfo {year}
  {2008})}\BibitemShut {NoStop}%
\bibitem [{\citenamefont {Zyuzin}\ and\ \citenamefont
  {Burkov}(2012)}]{Zyuzin2012}%
  \BibitemOpen
  \bibfield  {author} {\bibinfo {author} {\bibfnamefont {A.~A.}\ \bibnamefont
  {Zyuzin}}\ and\ \bibinfo {author} {\bibfnamefont {A.~A.}\ \bibnamefont
  {Burkov}},\ }\bibfield  {title} {\enquote {\bibinfo {title} {{Topological
  response in Weyl semimetals and the chiral anomaly}},}\ }\href {\doibase
  10.1103/PhysRevB.86.115133} {\bibfield  {journal} {\bibinfo  {journal}
  {Physical Review B}\ }\textbf {\bibinfo {volume} {86}},\ \bibinfo {pages}
  {115133} (\bibinfo {year} {2012})}\BibitemShut {NoStop}%
\bibitem [{\citenamefont {Grushin}(2012)}]{Grushin2012}%
  \BibitemOpen
  \bibfield  {author} {\bibinfo {author} {\bibfnamefont {Adolfo~G.}\
  \bibnamefont {Grushin}},\ }\bibfield  {title} {\enquote {\bibinfo {title}
  {{Consequences of a condensed matter realization of Lorentz-violating QED in
  Weyl semi-metals}},}\ }\href {\doibase 10.1103/PhysRevD.86.045001} {\bibfield
   {journal} {\bibinfo  {journal} {Phys. Rev. D}\ }\textbf {\bibinfo {volume}
  {86}},\ \bibinfo {pages} {045001} (\bibinfo {year} {2012})}\BibitemShut
  {NoStop}%
\bibitem [{\citenamefont {Goswami}\ and\ \citenamefont
  {Tewari}(2013)}]{Goswami:2013jp}%
  \BibitemOpen
  \bibfield  {author} {\bibinfo {author} {\bibfnamefont {Pallab}\ \bibnamefont
  {Goswami}}\ and\ \bibinfo {author} {\bibfnamefont {Sumanta}\ \bibnamefont
  {Tewari}},\ }\bibfield  {title} {\enquote {\bibinfo {title} {{Axionic field
  theory of (3+1)-dimensional Weyl semimetals}},}\ }\href {\doibase
  10.1103/PhysRevB.88.245107} {\bibfield  {journal} {\bibinfo  {journal} {Phys.
  Rev. B}\ }\textbf {\bibinfo {volume} {88}},\ \bibinfo {pages} {245107}
  (\bibinfo {year} {2013})}\BibitemShut {NoStop}%
\bibitem [{\citenamefont {Roy}\ \emph {et~al.}(2016)\citenamefont {Roy},
  \citenamefont {Kolodrubetz}, \citenamefont {Moore},\ and\ \citenamefont
  {Grushin}}]{Roy2016}%
  \BibitemOpen
  \bibfield  {author} {\bibinfo {author} {\bibfnamefont {S.}~\bibnamefont
  {Roy}}, \bibinfo {author} {\bibfnamefont {M.}~\bibnamefont {Kolodrubetz}},
  \bibinfo {author} {\bibfnamefont {J.E.}\ \bibnamefont {Moore}}, \ and\
  \bibinfo {author} {\bibfnamefont {Adolfo~G.}\ \bibnamefont {Grushin}},\
  }\bibfield  {title} {\enquote {\bibinfo {title} {{Chern numbers and chiral
  anomalies in Weyl butterflies}},}\ }\href {\doibase
  10.1103/PhysRevB.94.161107} {\bibfield  {journal} {\bibinfo  {journal}
  {Physical Review B}\ }\textbf {\bibinfo {volume} {94}} (\bibinfo {year}
  {2016}),\ 10.1103/PhysRevB.94.161107}\BibitemShut {NoStop}%
\bibitem [{\citenamefont {Arnold}\ \emph {et~al.}(2016)\citenamefont {Arnold},
  \citenamefont {Shekhar}, \citenamefont {Wu}, \citenamefont {Sun},
  \citenamefont {{Dos Reis}}, \citenamefont {Kumar}, \citenamefont {Naumann},
  \citenamefont {Ajeesh}, \citenamefont {Schmidt}, \citenamefont {Grushin},
  \citenamefont {Bardarson}, \citenamefont {Baenitz}, \citenamefont {Sokolov},
  \citenamefont {Borrmann}, \citenamefont {Nicklas}, \citenamefont {Felser},
  \citenamefont {Hassinger},\ and\ \citenamefont {Yan}}]{Arnold2016}%
  \BibitemOpen
  \bibfield  {author} {\bibinfo {author} {\bibfnamefont {F.}~\bibnamefont
  {Arnold}}, \bibinfo {author} {\bibfnamefont {C.}~\bibnamefont {Shekhar}},
  \bibinfo {author} {\bibfnamefont {S.-C.}\ \bibnamefont {Wu}}, \bibinfo
  {author} {\bibfnamefont {Y.}~\bibnamefont {Sun}}, \bibinfo {author}
  {\bibfnamefont {R.D.}\ \bibnamefont {{Dos Reis}}}, \bibinfo {author}
  {\bibfnamefont {N.}~\bibnamefont {Kumar}}, \bibinfo {author} {\bibfnamefont
  {M.}~\bibnamefont {Naumann}}, \bibinfo {author} {\bibfnamefont {M.O.}\
  \bibnamefont {Ajeesh}}, \bibinfo {author} {\bibfnamefont {M.}~\bibnamefont
  {Schmidt}}, \bibinfo {author} {\bibfnamefont {Adolfo~G.}\ \bibnamefont
  {Grushin}}, \bibinfo {author} {\bibfnamefont {J.H.}\ \bibnamefont
  {Bardarson}}, \bibinfo {author} {\bibfnamefont {M.}~\bibnamefont {Baenitz}},
  \bibinfo {author} {\bibfnamefont {D.}~\bibnamefont {Sokolov}}, \bibinfo
  {author} {\bibfnamefont {H.}~\bibnamefont {Borrmann}}, \bibinfo {author}
  {\bibfnamefont {M.}~\bibnamefont {Nicklas}}, \bibinfo {author} {\bibfnamefont
  {C.}~\bibnamefont {Felser}}, \bibinfo {author} {\bibfnamefont
  {E.}~\bibnamefont {Hassinger}}, \ and\ \bibinfo {author} {\bibfnamefont
  {B.}~\bibnamefont {Yan}},\ }\bibfield  {title} {\enquote {\bibinfo {title}
  {{Negative magnetoresistance without well-defined chirality in the Weyl
  semimetal TaP}},}\ }\href {\doibase 10.1038/ncomms11615} {\bibfield
  {journal} {\bibinfo  {journal} {Nature Communications}\ }\textbf {\bibinfo
  {volume} {7}} (\bibinfo {year} {2016}),\ 10.1038/ncomms11615}\BibitemShut
  {NoStop}%
\bibitem [{\citenamefont {Reis}\ \emph {et~al.}(2016)\citenamefont {Reis},
  \citenamefont {Ajeesh}, \citenamefont {Kumar}, \citenamefont {Arnold},
  \citenamefont {Shekhar}, \citenamefont {Naumann}, \citenamefont {Schmidt},
  \citenamefont {Nicklas},\ and\ \citenamefont {Hassinger}}]{Reis2016}%
  \BibitemOpen
  \bibfield  {author} {\bibinfo {author} {\bibfnamefont {R.~D.~Dos}\
  \bibnamefont {Reis}}, \bibinfo {author} {\bibfnamefont {M~O}\ \bibnamefont
  {Ajeesh}}, \bibinfo {author} {\bibfnamefont {N}~\bibnamefont {Kumar}},
  \bibinfo {author} {\bibfnamefont {F}~\bibnamefont {Arnold}}, \bibinfo
  {author} {\bibfnamefont {C}~\bibnamefont {Shekhar}}, \bibinfo {author}
  {\bibfnamefont {M}~\bibnamefont {Naumann}}, \bibinfo {author} {\bibfnamefont
  {M}~\bibnamefont {Schmidt}}, \bibinfo {author} {\bibfnamefont
  {M}~\bibnamefont {Nicklas}}, \ and\ \bibinfo {author} {\bibfnamefont
  {E}~\bibnamefont {Hassinger}},\ }\bibfield  {title} {\enquote {\bibinfo
  {title} {{On the search for the chiral anomaly in Weyl semimetals: The
  negative longitudinal magnetoresistance}},}\ }\href {\doibase
  10.1088/1367-2630/18/8/085006} {\bibfield  {journal} {\bibinfo  {journal}
  {New Journal of Physics}\ }\textbf {\bibinfo {volume} {18}} (\bibinfo {year}
  {2016}),\ 10.1088/1367-2630/18/8/085006},\ \Eprint
  {http://arxiv.org/abs/1606.03389} {arXiv:1606.03389} \BibitemShut {NoStop}%
\end{thebibliography}%

\end{document}